\documentclass[twocolumn]{aastex63}

\shorttitle{MAPS X: deuterated molecules}
\shortauthors{Cataldi et al.}
\graphicspath{{./}{figures/}}

\usepackage[version=4]{mhchem} 
\usepackage{xcolor}
\usepackage{xspace}
\newcommand{\nthp}{\ce{N2H+}\xspace}
\newcommand{\ntdp}{\ce{N2D+}\xspace}
\begin{document}

\title{Molecules with ALMA at Planet-forming Scales (MAPS). X. Studying deuteration at high angular resolution toward protoplanetary disks}

\correspondingauthor{Gianni Cataldi}
\email{gianni.cataldi@nao.ac.jp}

\author[0000-0002-2700-9676]{Gianni Cataldi}
\affil{National Astronomical Observatory of Japan, Osawa 2-21-1, Mitaka, Tokyo 181-8588, Japan}
\affil{Department of Astronomy, Graduate School of Science, The University of Tokyo, Tokyo 113-0033, Japan}

\author[0000-0003-4099-6941]{Yoshihide Yamato}
\affil{Department of Astronomy, Graduate School of Science, The University of Tokyo, Tokyo 113-0033, Japan}

\author[0000-0003-3283-6884]{Yuri Aikawa}
\affil{Department of Astronomy, Graduate School of Science, The University of Tokyo, Tokyo 113-0033, Japan}

\author[0000-0002-8716-0482]{Jennifer B. Bergner}
\altaffiliation{NASA Hubble Fellowship Program Sagan Fellow}
\affiliation{University of Chicago Department of the Geophysical Sciences, Chicago, IL 60637, USA}

\author[0000-0002-2026-8157]{Kenji Furuya}
\affiliation{National Astronomical Observatory of Japan, Osawa 2-21-1, Mitaka, Tokyo 181-8588, Japan}

\author[0000-0003-4784-3040]{Viviana V. Guzm\'{a}n}
\affiliation{Instituto de Astrof\'isica, Pontificia Universidad Cat\'olica de Chile, Av. Vicu\~na Mackenna 4860, 7820436 Macul, Santiago, Chile}

\author[0000-0001-6947-6072]{Jane Huang}
\altaffiliation{NASA Hubble Fellowship Program Sagan Fellow}
\affiliation{Department of Astronomy, University of Michigan, 323 West Hall, 1085 S. University Avenue, Ann Arbor, MI 48109, USA}
\affiliation{Center for Astrophysics \textbar\ Harvard \& Smithsonian, 60 Garden St., Cambridge, MA 02138, USA}

\author[0000-0002-8932-1219]{Ryan A. Loomis}
\affiliation{National Radio Astronomy Observatory, 520 Edgemont Rd., Charlottesville, VA 22903, USA}

\author[0000-0001-8642-1786]{Chunhua Qi}
\affiliation{Center for Astrophysics \textbar\ Harvard \& Smithsonian, 60 Garden St., Cambridge, MA 02138, USA}

\author[0000-0003-2253-2270]{Sean M. Andrews}
\affiliation{Center for Astrophysics \textbar\ Harvard \& Smithsonian, 60 Garden St., Cambridge, MA 02138, USA}

\author[0000-0003-4179-6394]{Edwin A. Bergin}
\affiliation{Department of Astronomy, University of Michigan, 323 West Hall, 1085 S. University Avenue, Ann Arbor, MI 48109, USA}

\author[0000-0003-2014-2121]{Alice S. Booth}
\affiliation{Leiden Observatory, Leiden University, 2300 RA Leiden, the Netherlands}
\affiliation{School of Physics \& Astronomy, University of Leeds, Leeds LS2 9JT, UK}

\author[0000-0003-4001-3589]{Arthur D. Bosman}
\affiliation{Department of Astronomy, University of Michigan, 323 West Hall, 1085 S. University Avenue, Ann Arbor, MI 48109, USA}

\author[0000-0003-2076-8001]{L. Ilsedore Cleeves}
\affiliation{Department of Astronomy, University of Virginia, 530 McCormick Rd, Charlottesville, VA 22904}

\author[0000-0002-1483-8811]{Ian Czekala}
\altaffiliation{NASA Hubble Fellowship Program Sagan Fellow}
\affiliation{Department of Astronomy and Astrophysics, 525 Davey Laboratory, The Pennsylvania State University, University Park, PA 16802, USA}
\affiliation{Center for Exoplanets and Habitable Worlds, 525 Davey Laboratory, The Pennsylvania State University, University Park, PA 16802, USA}
\affiliation{Center for Astrostatistics, 525 Davey Laboratory, The Pennsylvania State University, University Park, PA 16802, USA}
\affiliation{Institute for Computational \& Data Sciences, The Pennsylvania State University, University Park, PA 16802, USA}
\affiliation{Department of Astronomy, 501 Campbell Hall, University of California, Berkeley, CA 94720-3411, USA}

\author[0000-0003-1008-1142]{John~D.~Ilee}
\affil{School of Physics \& Astronomy, University of Leeds, Leeds LS2 9JT, UK}

\author[0000-0003-1413-1776]{Charles J. Law}
\affiliation{Center for Astrophysics \textbar\ Harvard \& Smithsonian, 60 Garden St., Cambridge, MA 02138, USA}

\author[0000-0003-1837-3772]{Romane Le Gal}
\affiliation{Center for Astrophysics \textbar\ Harvard \& Smithsonian, 60 Garden St., Cambridge, MA 02138, USA}
\affiliation{IRAP, Universit\'{e} de Toulouse, CNRS, CNES, UT3, 31400 Toulouse, France}
\affiliation{Univ. Grenoble Alpes, CNRS, IPAG, F-38000 Grenoble, France}
\affiliation{IRAM, 300 rue de la piscine, F-38406 Saint-Martin d'H\`{e}res, France}

\author[0000-0002-7616-666X]{Yao Liu}
\affiliation{Purple Mountain Observatory \& Key Laboratory for Radio Astronomy, Chinese Academy of Sciences, Nanjing 210023, China}

\author[0000-0002-7607-719X]{Feng Long}
\affiliation{Center for Astrophysics \textbar\ Harvard \& Smithsonian, 60 Garden St., Cambridge, MA 02138, USA}

\author[0000-0002-1637-7393]{Fran\c cois M\'enard}
\affiliation{Univ. Grenoble Alpes, CNRS, IPAG, F-38000 Grenoble, France}

\author[0000-0002-7058-7682]{Hideko Nomura}
\affiliation{National Astronomical Observatory of Japan, Osawa 2-21-1, Mitaka, Tokyo 181-8588, Japan}

\author[0000-0001-8798-1347]{Karin I. \"Oberg}
\affiliation{Center for Astrophysics \textbar\ Harvard \& Smithsonian, 60 Garden St., Cambridge, MA 02138, USA}

\author[0000-0002-6429-9457]{Kamber R. Schwarz}
\altaffiliation{NASA Hubble Fellowship Program Sagan Fellow}
\affiliation{Lunar and Planetary Laboratory, University of Arizona, 1629 E. University Blvd, Tucson, AZ 85721, USA}

\author[0000-0003-1534-5186]{Richard Teague}
\affiliation{Center for Astrophysics \textbar\ Harvard \& Smithsonian, 60 Garden St., Cambridge, MA 02138, USA}

\author[0000-0002-6034-2892]{Takashi Tsukagoshi}
\affiliation{National Astronomical Observatory of Japan, Osawa 2-21-1, Mitaka, Tokyo 181-8588, Japan}

\author[0000-0001-6078-786X]{Catherine Walsh}
\affiliation{School of Physics \& Astronomy, University of Leeds, Leeds LS2 9JT, UK}

\author[0000-0003-1526-7587]{David J. Wilner}
\affiliation{Center for Astrophysics \textbar\ Harvard \& Smithsonian, 60 Garden St., Cambridge, MA 02138, USA}

\author[0000-0002-0661-7517]{Ke Zhang}
\altaffiliation{NASA Hubble Fellow}
\affiliation{Department of Astronomy, University of Wisconsin-Madison, 475 N Charter St, Madison, WI 53706}
\affiliation{Department of Astronomy, University of Michigan, 
323 West Hall, 1085 S. University Avenue, 
Ann Arbor, MI 48109, USA}


\begin{abstract}
Deuterium fractionation is dependent on various physical and chemical parameters. Thus, the formation location and thermal history of material in the solar system is often studied by measuring its D/H ratio. This requires knowledge about the deuteration processes operating during the planet formation era. We aim to study these processes by radially resolving the DCN/HCN (at 0.3$\arcsec$ resolution) and \ce{N2D+}/\ce{N2H+} ($\sim$0.3--0.9$\arcsec$) column density ratios toward the five protoplanetary disks observed by the Molecules with ALMA at Planet-forming scales (MAPS) Large Program. DCN is detected in all five sources, with one newly reported detection. \ce{N2D+} is detected in four sources, two of which are newly reported detections. We derive column density profiles that allow us to study the spatial variation of the DCN/HCN and \ce{N2D+}/\ce{N2H+} ratios at high resolution. DCN/HCN varies considerably for different parts of the disks, ranging from $10^{-3}$ to $10^{-1}$. In particular, the inner disk regions generally show significantly lower HCN deuteration compared with the outer disk. In addition, our analysis confirms that two deuterium fractionation channels are active, which can alter the D/H ratio within the pool of organic molecules. \ce{N2D+} is found in the cold outer regions beyond $\sim$50\,au, with \ce{N2D+}/\ce{N2H+} ranging between $10^{-2}$ and 1 across the disk sample. This is consistent with the theoretical expectation that \ce{N2H+} deuteration proceeds via the low-temperature channel only. This paper is part of the MAPS special issue of the Astrophysical Journal Supplement.
\end{abstract}


\section{Introduction} \label{sec:intro}
It is clear that the physical and chemical properties of a protoplanetary disk heavily influence the properties and evolution of the bodies (planets, asteroids, and comets) emerging from it. In particular, the composition of nascent planets is set by the composition of the disk at the location of formation \citep[e.g.][]{Oberg21}. Varying physical parameters such as temperature, gas density, or UV exposure result in a radially and vertically varying composition of the disk \citep[e.g.][]{Henning13}. Spatially resolving the chemical composition of protoplanetary disks is thus a cornerstone of planet formation studies. In this paper, we consider the radial distribution of two deuterated molecules: DCN and \ce{N2D+}. We are particularly interested in the radial variation of the deuteration fraction: DCN/HCN and \ce{N2D+}/\ce{N2H+}. This will allow us to connect deuterium chemistry in disks to measurements of deuterium fractionation in solar system bodies.

In the interstellar medium (ISM), the elemental D/H ratio (by number) is of the order of 1.5$\times10^{-5}$ \citep[e.g.][and references therein]{Hebrard05,Linsky06}, consistent with predictions of big bang nucleosynthesis \citep[e.g.][]{Cyburt16}. However, the deuteration fraction, that is, the ratio of a deuterated molecule to its non-deuterated isotopologue, can exceed the elemental ratio by several orders of magnitude. This phenomenon is known as deuterium fractionation and depends on chemical and physical parameters. For example, the degree of fractionation is generally enhanced at low temperatures. As a consequence, the deuteration fraction carries information about the physical properties of the environment in which the molecules were formed and is often used to infer the formation location and thermal history of material in the ISM or the solar system. For example, comparison of the deuteration fractions of the water in Earth's oceans and in solar system asteroids or comets is used to study the possibility of asteroidal or cometary water delivery to Earth during terrestrial planet formation \citep[e.g.][]{Alexander17,OBrien18,Lis19}.

As in molecular clouds, deuteration in protoplanetary disks starts from a number of exchange reactions involving the main reservoir of deuterium: HD \citep[e.g.][]{Millar89,Turner01}. At low temperature ($\lesssim30$\,K), the most relevant reaction is
\begin{equation}\label{eq:HD_exchange_reaction_lowT}
    \ce{HD + H_3^+ <=> H2 + H2D+}.
\end{equation}
\ce{H2D+} can then propagate the D to other molecules such as \ce{N2D+} and DCN by proton transfer and subsequent recombination:  N$_2$ + H$_2$D$^+$ $\rightarrow$ N$_2$D$^+$ + H$_2$, HNC + H$_2$D$^+$ $\rightarrow$ DCNH$^+$ + H$_2$, and DCNH$^+$ + \ce{e-} $\rightarrow$ DCN + H. Reaction (\ref{eq:HD_exchange_reaction_lowT}) is exothermic in the forward direction ($\Delta E\approx 230$\,K). Therefore, at gas temperatures below $\sim$30\,K, the inverse reaction is suppressed and significant fractionation can occur. In addition, at these low temperatures, molecules that can destroy \ce{H2D+} are frozen out, further enhancing the fractionation \citep[e.g.][]{Roberts00}. It is also important to consider the spin state of \ce{H2}. Ortho-H$_2$ has a higher internal energy than para-H$_2$, meaning that it can more easily drive the reverse reaction. Thus, a higher ortho-to-para ratio means less efficient deuteration \citep[e.g.][]{Willacy15}. Compared to molecular clouds, the ortho/para ratio of H$_2$ is more easily thermalized by ion-molecule reactions and grain-surface conversions in protoplanetary disks \citep{Aikawa15,Furuya19}. The strong thermal gradients in protoplanetary disks then likely lead to a gradient of the ortho-para ratio, with a higher abundance of ortho-H$_2$ in the warm inner disk.

Reaction (\ref{eq:HD_exchange_reaction_lowT}) is often referred to as the low temperature deuteration channel. At temperatures $\gtrsim30$\,K, the most relevant fractionation reactions are instead expected to be \citep{Millar89}
\begin{align}
    \ce{HD + CH_3^+ &<=> H2 + CH2D+},\label{eq:HD_exchange_reaction_highT1}\\
    \ce{HD + C2H_2^+ &<=> H2 + C2HD+}.\label{eq:HD_exchange_reaction_highT2}
\end{align}
These reactions are more exothermic than reaction (\ref{eq:HD_exchange_reaction_lowT}), that is, $\Delta E\approx 500$\,K \citep[e.g.][]{Roberts00,Roueff13,Nyman19}. Thus, they could dominate the fractionation for temperatures $\gtrsim 30$ K and are often called the high-temperature deuteration channel. DCN, for example, can be formed by the reaction of an N atom with CHD, which is formed by the dissociative recombination of CH$_2$D$^+$. The different deuteration pathways are therefore expected to operate in different parts of the disk \citep{Aikawa18}: in the midplane of the cold outer disk, fractionation should mainly be initiated by the low-temperature channel, while in the warmer inner region and the disk atmosphere, the high-temperature channel should dominate.

While HCN is mainly destroyed by photodissociation and various ion-molecule reactions \citep[e.g.][]{Aikawa99,Willacy00}, \ce{N2H+} is destroyed by CO and is thus expected to be abundant in cold regions beyond the CO snow line \citep[e.g.][]{Qi13}. Therefore, \ce{N2H+} deuteration should proceed via the low-temperature channel only \citep{Millar89}. On the other hand, both the high- and the low-temperature channels are expected to contribute to DCN formation \citep[e.g.][]{Huang17,Salinas17,Aikawa18}. This picture can be observationally tested by radially resolving the distribution of DCN and \ntdp \citep[e.g.][]{Salinas17,Huang17,Oberg21_TWHya}. DCO$^+$ can be formed by both pathways and thus is also used to study the two deuteration pathways \citep[e.g.][]{Oberg12,Salinas17,Carney18,Oberg21_TWHya}. In this work, we build on these previous efforts by using high-resolution DCN and N$_2$D$^+$ data obtained by the Molecules with ALMA at Planet-forming Scales (MAPS) Large Program\footnote{\url{https://www.alma-maps.info}} \citep[][]{Oberg21_MAPS}. MAPS targeted five protoplanetary disks \citep{Oberg21_MAPS}: three disks around T~Tauri stars (IM~Lup, GM~Aur, AS~209) and two disks around Herbig~Ae stars (HD~163296 and MWC~480). These systems cover a range of stellar masses from 1.1\,M$_\odot$ for IM~Lup and GM~Aur \citep{Teague21_MAPS} to 2.1\,M$_\odot$ for MWC~480 \citep{Simon19}, a range of stellar luminosities from 1.2\,L$_\odot$ for GM~Aur \citep{Macias18} to 21.9\,L$_\odot$ for MWC~480 \citep{Montesinos09}, and a range of ages from $\sim$1\,Myr for IM~Lup \citep{Mawet12} and AS~209 \citep{Andrews18} to $\sim$7\,Myr for MWC~480 \citep{Simon00,Montesinos09}. All disks have dust substructures in the form of rings and gaps \citep{Andrews18,Huang18,Long18,Huang20}. GM~Aur is the only disk with a central dust cavity. The $^{12}$CO 2--1 disk size, which encloses 90\% of the flux \citep{Law21_MAPS_radial_profiles} is largest for IM~Lup ($\sim$480\,au) and smallest for AS~209 ($\sim$200\,au). Low signal-to-noise ratio (SNR) $^{12}$CO 2--1 emission often extends to considerably larger radii \citep[e.g.\ out to $\sim$800--900\,au for IM~Lup,][]{Law21_MAPS_radial_profiles}. A more detailed description of the MAPS targets is given by \citet{Oberg21_MAPS}. The MAPS data allow us to study the deuteration at an unprecedented spatial resolution and sensitivity. We aim to compare radial deuteration profiles to model predictions and to study the temperature dependence of the deuteration fraction.

Besides studying deuteration, we aim to investigate the relation of the CO snow line with N$_2$D$^+$. As mentioned above, the abundance of both \nthp and \ntdp is expected to anticorrelate with the CO abundance, because \nthp destruction is enhanced in the presence of CO \citep[e.g.][]{Bergin01,Qi13}. This was used by \citet{Qi13,Qi15,Qi19} to estimate the radius of the CO snow line by matching it to the inner edge of the observed \nthp emission. But theoretical models show that, besides the midplane, \nthp can be moderately abundant in the warm molecular layer as well \citep[e.g.][]{vantHoff17}. This can complicate the inference of the CO snow line. On the other hand, \ntdp is expected to mainly trace the midplane \citep{Aikawa18}. Therefore, we will test whether \ntdp can be used as an alternative to \nthp to trace the CO snow line.

In summary, we observed DCN and N$_2$D$^+$ with high sensitivity and angular resolution to study the deuteration chemistry of five protoplanetary disks. In Section \ref{sec:observations}, we give an overview of the data. In Section \ref{sec:observational_results}, we discuss the emission morphology inferred from zeroth moment maps and radial emission profiles. We then proceed to derive radial column density profiles and deuteration profiles in Section \ref{sec:analysis}. We discuss the implications of our results in Section \ref{sec:discussion} and summarize our conclusions in Section \ref{sec:summary}.

\section{Observations}\label{sec:observations}
\subsection{MAPS data: HCN, \texorpdfstring{H$^{13}$CN}{H13CN}, DCN, \texorpdfstring{\ce{N2D+}}{N2D+}}\label{sec:MAPS_data_description}
The MAPS Large Program (project code 2018.1.01055.L) used four spectral setups: two in ALMA Band 3 at a wavelength of $\sim$3\,mm and two in ALMA Band 6 at $\sim$1.3\,mm \citep{Oberg21_MAPS}. Two array configurations were used: a short baseline configuration and a long baseline configuration, resulting in baselines between 15 and 3638\,m. The details of the data calibration are described in \citet{Oberg21_MAPS}.

\subsubsection{CLEANing and JvM correction}\label{sec:CLEANING_JvM}
As described in detail in \citet{Czekala21_MAPS}, the MAPS collaboration produced images from the calibrated visibilities using the CLEAN deconvolution algorithm \citep{Hogbom74} implemented in the \texttt{CASA} task \textit{tclean}, where the "multiscale" version of the algorithm was used (\texttt{deconvolver=``multiscale"}). Furthermore, MAPS employed a flux scale correction to the CLEANed images that was first discussed by \citet[][hereafter JvM correction]{Jorsater95}. A brief summary of the correction is given in Appendix \ref{appendix:JvM_correction_summary}, while the full details can be found in \citet{Czekala21_MAPS}.

\subsubsection{Adopted MAPS imaging products}\label{sec:adopted_MAPS_imaging_products}
In this paper, we use the following lines observed by MAPS: the $J=3-2$ transitions of HCN, DCN, and \ntdp in Band 6, and the $J=1-0$ transitions of HCN and H$^{13}$CN in Band 3. For each emission line, images with several different beam sizes were produced by MAPS. For our Band 3 data, we use the MAPS fiducial images with a circular 0.3$\arcsec$ beam. For the Band 6 data, we use images tapered to the same circular 0.3$\arcsec$ beam \citep{Czekala21_MAPS} for the weak DCN and \ntdp lines to improve the sensitivity. For the strong HCN 3-2 line in Band 6, we use both the images with 0.15$\arcsec$ and 0.3$\arcsec$ circular beams, as listed in Table \ref{tab:HCN_image_params}. In particular, when fitting for the column densities, we use the image cubes with 0.3$\arcsec$ beams in order to have uniform spatial resolution for all the emission lines included in the fit. Since MAPS provided individual HCN and H$^{13}$CN image cubes with spectral axes centered onto individual hyperfine components \citep{Oberg21_MAPS}, we used the \texttt{CASA} tasks \textit{regrid} and \textit{imageconcat} to combine these cubes into a single cube. The basic properties of the MAPS data cubes we used are listed in Tables \ref{tab:HCN_image_params} (for HCN, DCN, and H$^{13}$CN) and \ref{tab:N2Hp_image_params} (for \ntdp).

\begin{table*}
\centering
\caption{Overview of properties of HCN, DCN, and H$^{13}$CN data cubes provided by MAPS and used in this study.}\label{tab:HCN_image_params}
\begin{tabular}{ccccccc}
\hline
line & rms\tablenotemark{a} & JvM $\epsilon$\tablenotemark{b} & spectral resolution & channel width & beam size & usage\tablenotemark{c}\\
& [mJy\,beam$^{-1}$] & & [km\,s$^{-1}$] &[km\,s$^{-1}$] & &\\
\hline\hline
\multicolumn{7}{c}{IM~Lup}\\
HCN 1--0 & 0.7 & 0.72 & 0.24 & 0.5 & $0.30\arcsec\times0.30\arcsec$ &\\
HCN 3--2 & 0.3 & 0.25 & 0.16 & 0.2 & $0.15\arcsec\times0.15\arcsec$& MOM0, PROF \\
HCN 3--2 & 0.6 & 0.49 & 0.16 & 0.2 & $0.30\arcsec\times0.30\arcsec$ & FLUX, AASPEC, COLDENS\\
DCN 3--2 & 1.1 & 0.79 & 0.20& 0.2 & $0.30\arcsec\times0.30\arcsec$ & \\
H$^{13}$CN 1--0& 0.7 & 0.73 & 0.49 & 0.5 & $0.30\arcsec\times0.30\arcsec$&\\
\hline
\multicolumn{7}{c}{GM~Aur}\\
HCN 1--0 & 1.8 & 1.00 & 0.24 & 0.5 & $0.31\arcsec\times0.31\arcsec$ &\\
HCN 3--2 & 0.8 & 0.57 & 0.16 & 0.2 & $0.15\arcsec\times0.15\arcsec$& MOM0, PROF \\
HCN 3--2 & 1.0 & 0.74 & 0.16 & 0.2 & $0.30\arcsec\times0.30\arcsec$ & FLUX, AASPEC, COLDENS\\
DCN 3--2 & 0.7 & 0.63 & 0.20 & 0.2 & $0.30\arcsec\times0.30\arcsec$&\\
H$^{13}$CN 1--0 & 1.5 & 1.00 & 0.49 & 0.5 & $0.33\arcsec\times0.33\arcsec$&\\
\hline
\multicolumn{7}{c}{AS~209}\\
HCN 1--0 & 0.8 & 0.83 & 0.24 & 0.5 & $0.30\arcsec\times0.29\arcsec$ &\\
HCN 3--2 & 0.5 & 0.28 & 0.16 & 0.2 & $0.15\arcsec\times0.15\arcsec$& MOM0, PROF \\
HCN 3--2 & 0.8 & 0.51 & 0.16 & 0.2 & $0.30\arcsec\times0.30\arcsec$ & FLUX, AASPEC, COLDENS\\
DCN 3--2 & 0.7 & 0.58 & 0.20 & 0.2 & $0.30\arcsec\times0.30\arcsec$&\\
H$^{13}$CN 1--0& 0.8 & 0.90 & 0.49 & 0.5 & $0.30\arcsec\times0.30\arcsec$&\\
\hline
\multicolumn{7}{c}{HD~163296}\\
HCN 1--0 & 0.8 & 0.93 & 0.24 & 0.5 & $0.30\arcsec\times0.30\arcsec$ &\\
HCN 3--2 & 0.4 & 0.30 & 0.16 & 0.2 & $0.15\arcsec\times0.15\arcsec$& MOM0, PROF \\
HCN 3--2 & 0.7 & 0.53 & 0.16 & 0.2 & $0.30\arcsec\times0.30\arcsec$ & FLUX, AASPEC, COLDENS\\
DCN 3--2 & 0.9 & 0.72 & 0.20 & 0.2 & $0.30\arcsec\times0.30\arcsec$&\\
H$^{13}$CN 1--0 & 0.7 & 0.95 & 0.49 & 0.5 & $0.30\arcsec\times0.30\arcsec$&\\
\hline
\multicolumn{7}{c}{MWC~480}\\
HCN 1--0 & 1.7 & 0.99 & 0.24 & 0.5 & $0.31\arcsec\times0.31\arcsec$ &\\
HCN 3--2 & 0.8 & 0.54 & 0.16 & 0.2 & $0.15\arcsec\times0.15\arcsec$& MOM0, PROF \\
HCN 3--2 & 1.0 & 0.74 & 0.16 & 0.2 & $0.30\arcsec\times0.30\arcsec$ & FLUX, AASPEC, COLDENS\\
DCN 3--2 & 0.8 & 0.73 & 0.20 & 0.2 & $0.30\arcsec\times0.30\arcsec$&\\
H$^{13}$CN 1--0& 1.5 & 1.00 & 0.49 & 0.5 & $0.31\arcsec\times0.31\arcsec$&\\
\hline
\end{tabular}
\tablenotetext{a}{Measured in a emission-free region of the non-primary beam corrected data cube.}
\tablenotetext{b}{Ratio of the CLEAN beam area to the dirty beam area used in the JvM correction. See Appendix \ref{appendix:JvM_correction_summary} and \citet{Czekala21_MAPS} for details.}
\tablenotetext{c}{For HCN 3--2, the following flags indicate the usage: FLUX: disk-integrated flux; MOM0: zeroth moment; PROF: radial emission profile; AASPEC: azimuthally averaged spectra; COLDENS: column density calculation. For the moment 0 maps, the radial emission profiles, and the disk-integrated fluxes, we used the non-primary beam corrected images, while primary beam corrected images are used for all other analyses.}
\end{table*}

\subsection{\texorpdfstring{\ce{N2H+}}{N2H+} archival data}\label{sec:archival_N2H+_data}
No transitions of \nthp were observed by MAPS. Fortunately, the $J=3-2$ transition of this molecule has been targeted by previous observations in ALMA Band 7 ($\lambda\sim1$\,mm) for all five of our targets. For IM~Lup, GM~Aur, and AS~209, we use data presented in \citet[][project code 2015.1.00678.S, $\sim$0.3--0.4$\arcsec$ resolution]{Qi19}. For HD~163296, we use the data by \citet[][project code 2012.1.00681.S, $\sim$0.5$\arcsec$ resolution]{Qi15}. Finally, for MWC~480, we use data presented in \citet[][project code 2015.1.00657.S, $\sim$1$\arcsec$ resolution]{Loomis20}. The calibration of the data is described in the original papers. Using \texttt{CASA} 5.6, we imaged the visibilities in the same way as for the MAPS data: first, Keplerian masks (i.e.\ masks that select the regions of the image cube where emission is expected based on the Keplerian rotation of the disk) were constructed following the procedure described in \citet{Czekala21_MAPS}, taking into account the hyperfine components listed in Table \ref{tab:atomic_parameters_N2H+}. Using the Keplerian masks, the data were imaged with the multiscale CLEAN deconvolver implemented in the \texttt{CASA} task \textit{tclean} with scales of [0, 5, 15, 25] pixels and a Briggs parameter of 0.5. Finally, to yield the correct flux, the JvM correction was applied to all images following \citet{Czekala21_MAPS}. For most sources, this process results in \nthp image cubes with fairly circular beams. The exception is MWC~480 where the beam is $0.93\arcsec\times0.48\arcsec$. However, the beam sizes of the resulting \ce{N2H+} images do not match the beam sizes of the MAPS \ce{N2D+} images. Thus, we produced additional \ce{N2H+} and \ce{N2D+} data cubes with matching beams using the \texttt{CASA} task \textit{imsmooth}. These additional cubes are used for the derivation of azimuthally averaged spectra and column densities. For IM~Lup, HD~163296, and MWC~480, we only had to smooth the \ce{N2D+} data. For AS~209 and GM~Aur, both the \ce{N2H+} and \ce{N2D+} data required smoothing to achieve matching beam sizes. Table \ref{tab:N2Hp_image_params} lists the properties and the usage of the various \nthp and \ntdp image cubes.

\begin{table*}
\footnotesize
\centering
\caption{Overview of properties and usage of \nthp and \ntdp data cubes.}\label{tab:N2Hp_image_params}
\begin{tabular}{ccccccccc}
\hline
line & rms\tablenotemark{a} & JvM $\epsilon$\tablenotemark{b} & $\delta v$\tablenotemark{c} & channel width & beam size & beam PA\tablenotemark{d} & data source\tablenotemark{e} & usage\tablenotemark{f}\\
& [mJy\,beam$^{-1}$] & & [km\,s$^{-1}$] & [km\,s$^{-1}$] & & [\arcdeg] & &\\
\hline\hline
\multicolumn{9}{c}{IM~Lup}\\
\ce{N2H+} 3--2 & 3.6 & 0.71 & 0.15 & 0.15 & $0.42\arcsec\times0.35\arcsec$& $76$ & Qi19 & \\
\ce{N2D+} 3--2 & 1.6 & 0.78 & 0.09 & 0.2 & $0.3\arcsec\times0.3\arcsec$ & N/A & \"Oberg21 & FLUX, MOM0, PROF\\
\ce{N2D+} 3--2 & 1.8 & 0.78 & 0.09 & 0.2 & $0.42\arcsec\times0.35\arcsec$ & $76$ & \"Oberg21 & AASPEC, COLDENS\\
\hline
\multicolumn{9}{c}{GM~Aur}\\
\ce{N2H+} 3--2 & 2.4 & 0.75 & 0.15 & 0.15 & $0.31\arcsec\times0.20\arcsec$ & $-2$ & Qi19 & FLUX, MOM0, PROF\\
\ce{N2H+} 3--2 & 2.4 & 0.75 & 0.15 & 0.15 & $0.32\arcsec\times0.32\arcsec$ & N/A & Qi19 & AASPEC, COLDENS\\
\ce{N2D+} 3--2 & 1.2 & 0.73 & 0.09 & 0.2 & $0.3\arcsec\times0.3\arcsec$ & N/A & \"Oberg21 & FLUX, MOM0, PROF\\
\ce{N2D+} 3--2 & 1.3 & 0.73 & 0.09 & 0.2 & $0.32\arcsec\times0.32\arcsec$ & N/A & \"Oberg21 & AASPEC, COLDENS\\
\hline
\multicolumn{9}{c}{AS~209}\\
\ce{N2H+} 3--2 & 2.1 & 0.67 & 0.15 & 0.15 & $0.34\arcsec\times0.24\arcsec$ & $-74$ & Qi19 & FLUX, MOM0, PROF\\
\ce{N2H+} 3--2 & 2.1 & 0.67 & 0.15 & 0.15 & $0.35\arcsec\times0.35\arcsec$ & N/A & Qi19 & AASPEC, COLDENS\\
\ce{N2D+} 3--2 & 1.1 & 0.58 & 0.09 & 0.2 & $0.3\arcsec\times0.3\arcsec$ & N/A & \"Oberg21 & FLUX, MOM0, PROF\\
\ce{N2D+} 3--2 & 1.2 & 0.58 & 0.09 & 0.2 & $0.35\arcsec\times0.35\arcsec$ & N/A & \"Oberg21 & AASPEC, COLDENS\\
\hline
\multicolumn{9}{c}{HD~163296}\\
\ce{N2H+} 3--2 & 2.8 & 0.84 & 0.26 & 0.2 & $0.45\arcsec\times0.35\arcsec$ & $-88$ & Qi15 & \\
\ce{N2D+} 3--2 & 1.2 & 0.69 & 0.09 & 0.2 & $0.3\arcsec\times0.3\arcsec$ & N/A & \"Oberg21 &  FLUX, MOM0, PROF\\
\ce{N2D+} 3--2 & 1.4 & 0.69 & 0.09 & 0.2 & $0.45\arcsec\times0.35\arcsec$ & $-88$ & \"Oberg21 & AASPEC, COLDENS\\
\hline
\multicolumn{9}{c}{MWC~480}\\
\ce{N2H+} 3--2 & 1.8 & 0.95 & 1.21 & 1.05 & $0.93\arcsec\times0.48\arcsec$ & $-29$ & Loomis20\\
\ce{N2D+} 3--2 & 1.3 & 0.74 & 0.09 & 0.2 & $0.3\arcsec\times0.3\arcsec$ & N/A & \"Oberg21 & FLUX, MOM0, PROF\\
\ce{N2D+} 3--2 & 2.2 & 0.74 & 0.09 & 0.2 & $0.93\arcsec\times0.48\arcsec$ & $-29$ & \"Oberg21 & AASPEC, COLDENS\\
\hline
\end{tabular}
\tablenotetext{a}{Measured in a emission-free region of the non-primary beam corrected data cube.}
\tablenotetext{b}{Ratio of the CLEAN beam area to the dirty beam area used in the JvM correction. See Appendix \ref{appendix:JvM_correction_summary} and \citet{Czekala21_MAPS} for details.}
\tablenotetext{c}{Spectral resolution.}
\tablenotetext{d}{Position angle of the beam major axis, measured North through East. For circular beams, the position angle is undefined.}
\tablenotetext{e}{References: Qi19: \citet{Qi19}, \"Oberg21: \citet{Oberg21_MAPS}, Qi15: \citet{Qi15}, Loomis20: \citet{Loomis20}}
\tablenotetext{f}{For lines with several image cubes, the following flags indicate their usage: FLUX: disk-integrated flux; MOM0: zeroth moment; PROF: radial emission profile; AASPEC: azimuthally averaged spectra; COLDENS: column  density calculation. For the moment 0 maps, the radial emission profiles, and the disk-integrated fluxes, we used the non-primary beam corrected images, while primary beam corrected images are used for all other analyses.}
\end{table*}

\section{Observational results}\label{sec:observational_results}

\subsection{Disk-integrated fluxes and zeroth moment maps}\label{sec:disk_integrated_flux_mom0}
Disk-integrated fluxes of all the emission lines considered in this work are presented in Table \ref{tab:disk_integrated_fluxes}. These were calculated by integrating the flux within a Keplerian mask. The outer radii of the masks were determined from visual inspection of the radial emission profiles presented in Sections \ref{sec:HCN_radial_emission_profiles} and \ref{sec:N2H+_radial_emission_profiles}. If the radial profile was too noisy to determine an outer edge, we looked for guidance from stronger lines: for the H$^{13}$CN 1--0 lines, we used the same outer radii as for HCN 1--0, and for \ntdp 3--2 toward GM~Aur, we used the same radius as for \nthp 3--2. In addition, for \ce{N2D+} 3--2 toward HD~163296 and MWC~480, a nonzero inner radius of the mask was set such that the negative emission in the inner region is avoided (see Section \ref{sec:N2H+_radial_emission_profiles} and Appendix \ref{appendix:negative_flux}). HCN 1--0 and 3--2 have a significant fraction of their flux in hyperfine components. For these lines, we constructed the mask as a superpostion of Keplerian masks, each centered at one of the hyperfine components listed in Table \ref{tab:atomic_parameters_HCN}. Errors were calculated by repeating the flux measurement procedure at off-source positions and taking the standard deviation of the off-source fluxes. A 10\% flux calibration error \citep{ALMA_THB_Cycle8} was added in quadrature. If the resulting signal-to-noise ratio (SNR) was smaller than 3, Table \ref{tab:disk_integrated_fluxes} reports the 3$\sigma$ upper limit. To calculate these disk-integrated fluxes and errors, we used the image cubes that are not corrected for the primary beam \citep[see][section 3.5, for a discussion of the primary beam correction]{ALMA_THB_Cycle8}. This is because for primary beam corrected images, the noise increases toward the edges of the image, making our approach to estimate the error from off-source positions invalid. However, we verified that the difference to fluxes extracted from primary beam corrected images is negligible. Non-primary beam corrected images were also used for the moment 0 maps and the radial emission profiles. For all other analyses presented in this paper, we used the primary beam corrected images.

The HCN lines as well as DCN are detected in all five targets. DCN is detected for the first time toward GM~Aur. H$^{13}$CN 1--0 is not detected in any of the disks, but a matched filter analysis in the $uv$ plane \citep{Loomis18} described in Appendix \ref{appendix:matched_filter} yields a tentative detection ($\sim$3$\sigma$) for GM~Aur. Although undetected (or tentatively detected in the case of GM~Aur), the H$^{13}$CN 1--0 data are useful to constrain the HCN column density. \nthp 3--2 is detected in all five sources. \ntdp 3--2 is firmly detected toward IM~Lup, AS~209, HD~163296, and MWC~480, with the detections toward IM~Lup and MWC~480 reported for the first time. Toward one source, GM~Aur, the SNR of the integrated \ntdp 3--2 flux is only 3.4. The matched filter analysis (Appendix \ref{appendix:matched_filter}) and the disk-integrated spectrum (Appendix \ref{appendix:az_averaged_spectra_gallery}) also do not provide a definitive detection. Thus, we consider \ntdp 3--2 toward GM~Aur tentatively detected.

\begin{rotatetable*}
\begin{deluxetable*}{cCCCCCCCCCCCCCCC}
\tabletypesize{\scriptsize}
\tablecaption{Disk-integrated fluxes (derived from non primary beam corrected images). Upper limits are at 3$\sigma$ significance.}
\label{tab:disk_integrated_fluxes}
\tablehead{
& \multicolumn{3}{c}{IM Lup} & \multicolumn{3}{c}{GM Aur} & \multicolumn{3}{c}{AS 209} & \multicolumn{3}{c}{HD 163296}& \multicolumn{3}{c}{MWC 480}\\
& \colhead{$r_\mathrm{min}$\tablenotemark{a}} & \colhead{$r_\mathrm{max}$\tablenotemark{b}} & \colhead{flux} & \colhead{$r_\mathrm{min}$} & \colhead{$r_\mathrm{max}$} & \colhead{flux} & \colhead{$r_\mathrm{min}$} & \colhead{$r_\mathrm{max}$} & \colhead{flux} & \colhead{$r_\mathrm{min}$} & \colhead{$r_\mathrm{max}$} & \colhead{flux} & \colhead{$r_\mathrm{min}$} & \colhead{$r_\mathrm{max}$} & \colhead{flux} \\ 
& \colhead{[au]}& \colhead{[au]}& \colhead{[mJy\,km\,s$^{-1}$]}& \colhead{[au]}& \colhead{[au]}& \colhead{[mJy\,km\,s$^{-1}$]}& \colhead{[au]}& \colhead{[au]}& \colhead{[mJy\,km\,s$^{-1}$]}& \colhead{[au]}& \colhead{[au]}& \colhead{[mJy\,km\,s$^{-1}$]}& \colhead{[au]}& \colhead{[au]}& \colhead{[mJy\,km\,s$^{-1}$]}
}
\startdata 
HCN 1--0 & 0 & 600 & 240 \pm 27 & 0 & 400 & 125 \pm 16 & 0 & 250 & 198 \pm 21 & 0 & 500 & 658 \pm 68 & 0 & 300 & 155 \pm 18\\
HCN 3--2 & 0 & 600 & 2409 \pm 242 & 0 & 400 & 1792 \pm 180 & 0 & 300 & 2965 \pm 297 & 0 & 500 & 7346 \pm 736 & 0 & 300 & 2469 \pm 248\\
DCN 3--2 & 0 & 500 & 88 \pm 13 & 0 & 400 & 37 \pm 6 & 0 & 200 & 235 \pm 24 & 0 & 400 & 115 \pm 15 & 0 & 400 & 74 \pm 10\\
H$^{13}$CN 1--0 & 0 & 600 & <16 & 0 & 400 & <36 & 0 & 250 & <13 & 0 & 500 & <15 & 0 & 300 & <20 \\
\nthp 3--2 & 0 & 500 & 1483 \pm 150 & 0 & 400 & 1025 \pm 103 & 0 & 300 & 698 \pm 71 & 0 & 350 & 450 \pm 51 & 0 & 350 & 279 \pm 31\\
\ntdp 3--2 & 0 & 400 & 96 \pm 14 & 0 & 400 & 24 \pm 7 & 0 & 250 & 74 \pm 9 & 50 & 250 & 135 \pm 15 & 60 & 250 & 28 \pm 6 \\
\enddata
\tablenotetext{a}{Minimum radius of the Keplerian mask.}
\tablenotetext{b}{Maximum radius of the Keplerian mask.}
\end{deluxetable*}
\end{rotatetable*}

The MAPS collaboration produced zeroth moment maps by applying a Keplerian mask to the data cubes and integrating over the velocity axis. The Keplerian masks take into account the hyperfine structure of the emission lines. These maps were used for all scientific analyses (in particular to derive radial emission profiles). To mitigate arc-like artefacts in these maps and better visualize radial structures, MAPS also produced ``hybrid'' zeroth moment maps. These were produced by combining a Keplerian mask and a smoothed $\sigma$-clip mask \citep[thus their name ``hybrid'', see][]{Law21_MAPS_radial_profiles}. We emphasize that by using a clipping mask, some emission is inevitably lost if the clipping threshold is larger than 0$\sigma$. Therefore, the hybrid zeroth moment maps are for presentational purposes only and are not used for any quantitative analysis. Note also that the use of masks implies that the noise level is not constant over the map. This is because in general, each pixel in the zeroth moment map is calculated by integrating over a different number of channels \citep[see Figure 2 in ][]{Law21_MAPS_radial_profiles}.

\begin{figure*}
\plotone{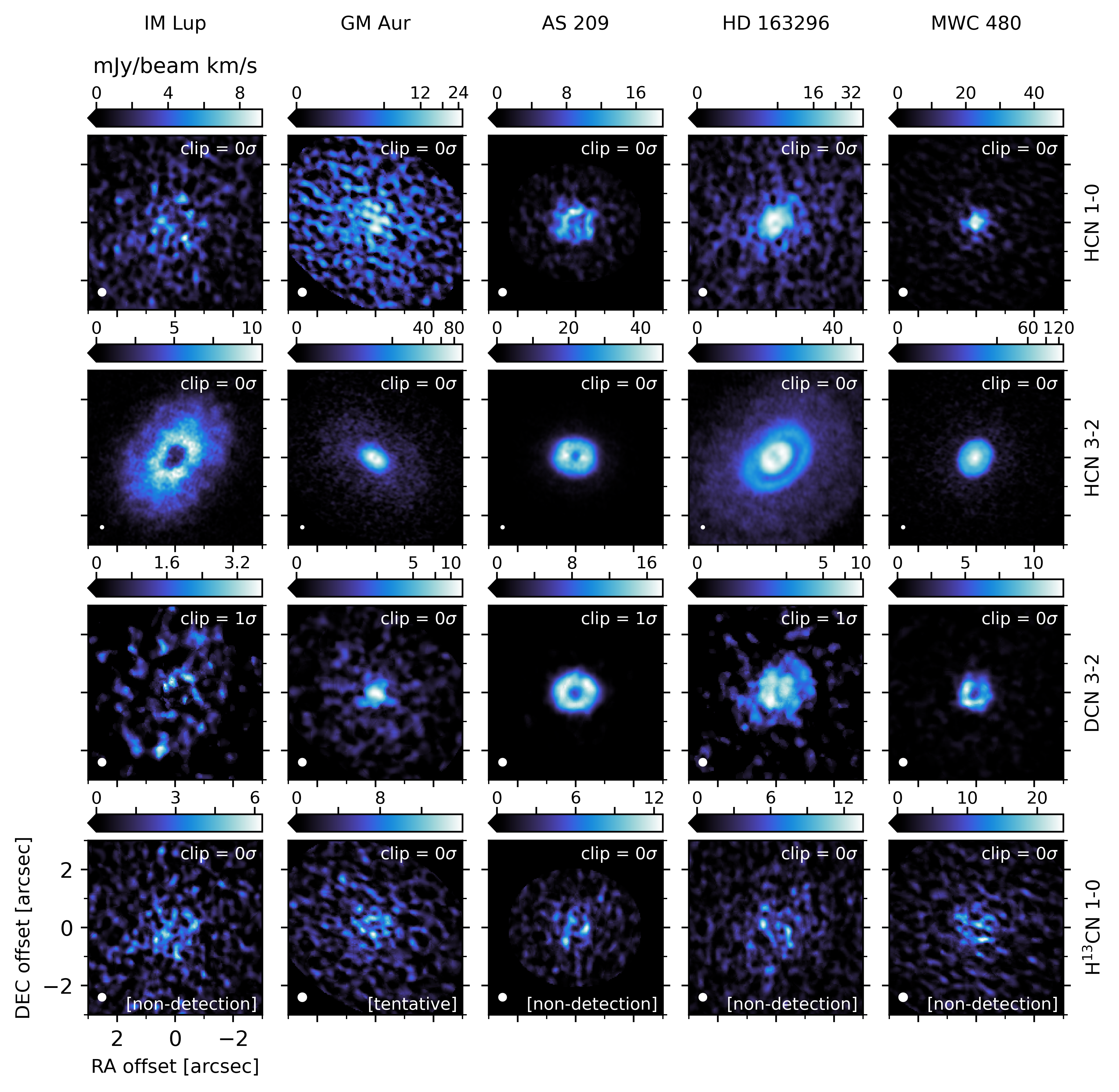}
\caption{Gallery of hybrid zeroth moment maps for HCN 1--0, HCN 3--2, DCN 3--2, and H$^{13}$CN 1--0, generated by combining a Keplerian mask and a smoothed $\sigma$-clip mask \citep{Law21_MAPS_radial_profiles}. The use of a $\sigma$-clip mask means that some emission is inevitably lost, except for a threshold of $0\sigma$. Thus, the flux scale can be unreliable, and these maps should be used for presentational purposes only. The clip values employed are indicated in the upper right of each map. The color scales employ either linear or arcsinh stretches, with the lower end saturating at 0\,mJy\,beam$^{-1}$\,km\,s$^{-1}$. The beam is shown by the white ellipse. Note that the noise is not uniform over these maps due to the use of masks. Text in the lower right of the panels marks lines not or only tentatively detected in total flux or a matched $uv$-plane filter.\label{fig:HCN_mom0_gallery}}
\end{figure*}

\begin{figure*}
\plotone{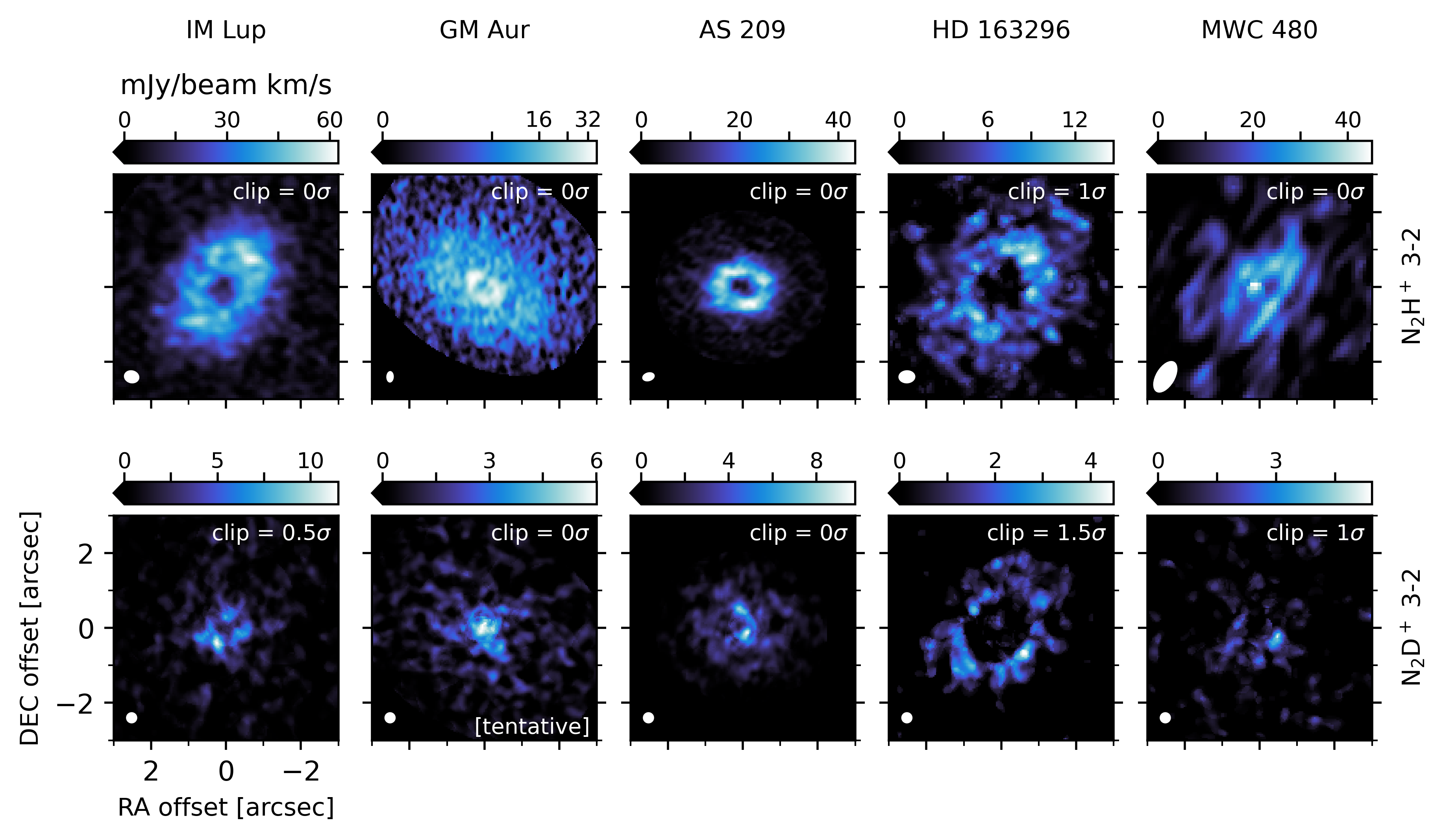}
\caption{Gallery of hybrid zeroth moment maps for N$_2$H$^+$ 3--2 and N$_2$D$^+$ 3--2, generated by combining a Keplerian mask and a smoothed $\sigma$-clip mask \citep{Law21_MAPS_radial_profiles}. The use of a $\sigma$ clip mask means that some emission is inevitably lost, except for a threshold of $0\sigma$. Thus, the flux scale can be unreliable, and these maps should be used for presentational purposes only. The clip values employed are indicated in the upper right of each map. The color scales saturate at  0\,mJy\,beam$^{-1}$\,km\,s$^{-1}$ at the lower end and employ a linear stretch, except for \nthp toward GM~Aur (arcsinh stretch). The ellipses in the lower left indicate the beam size. N$_2$D$^+$ 3--2 toward GM~Aur is only tentatively detected in total flux as well as with a matched filter analysis in the $uv$ plane. \label{fig:N2H+_mom0_gallery}}
\end{figure*}

In Figures \ref{fig:HCN_mom0_gallery} and \ref{fig:N2H+_mom0_gallery}, we show the hybrid zeroth moment maps produced by the MAPS collaboration for all the emission lines. Several values of the $\sigma$ clip were tested, and the final value was chosen by visual inspection of the maps. The maps for the archival \nthp 3--2 data were generated in the same way.

While the hybrid zeroth moment maps show various substructures, the radial profiles presented in the next Sections \ref{sec:HCN_radial_emission_profiles} and \ref{sec:N2H+_radial_emission_profiles} show these substructures more clearly. Thus, we concentrate on discussing the radial profiles in the following.

\subsection{Radial emission profiles of HCN, DCN, and \texorpdfstring{H$^{13}$CN}{H13CN}}\label{sec:HCN_radial_emission_profiles}
\begin{figure*}
\plotone{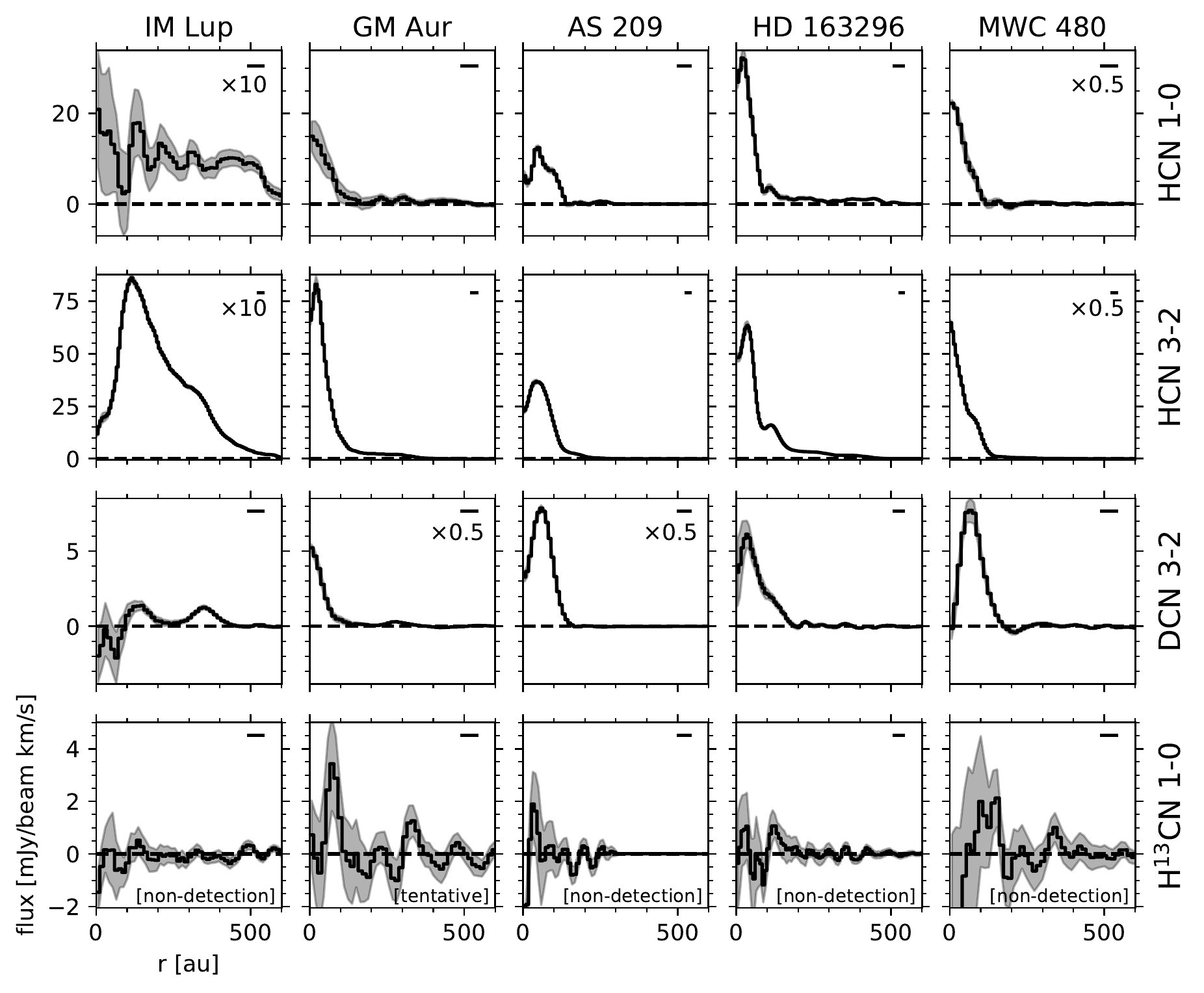}
\caption{Radial emission profiles of HCN 1--0, HCN 3--2, DCN 3--2 and H$^{13}$CN 1--0. The horizontal dashed line marks the zero flux level. The shaded area shows the $\pm1\sigma$ error. If the profile has been scaled by a constant factor, the scaling is indicated in the upper right of the panel. The beam major axis is shown as a horizontal line in the upper right. H$^{13}$CN 1--0 is marked as undetected or tentatively detected based on the total flux measurement and the matched filter analysis. \label{fig:HCN_radial_profiles_gallery}}
\end{figure*}

Figure \ref{fig:HCN_radial_profiles_gallery} shows deprojected radial emission profiles produced by the MAPS collaboration for the HCN, DCN, and H$^{13}$CN emission lines. These were produced by azimuthally averaging a zeroth moment map that was produced with a Keplerian mask only (i.e.\ without a $\sigma$-clipping mask). We use the profiles derived by averaging over the full azimuth of the zeroth moment map in order to maximize the SNR. The uncertainty in each radial bin is estimated as the standard error on the mean in the annulus over which the emission was averaged \citep{Law21_MAPS_radial_profiles}. 

The HCN 1--0 and 3--2 emission shows varied radial emission morphologies. Centrally peaked emission is seen in MWC~480, while the other disks show a central depression. Various rings and shoulders are also observed. The morphology of HCN is discussed in more detail in \citet{Guzman21_MAPS} and \citet{Bergner21_MAPS}. For DCN 3--2, we identify three distinct morphologies:
\begin{enumerate}
    \item Two DCN rings for IM~Lup, centered at $\sim$140\,au and $\sim$350\,au, respectively. These rings are also seen in the zeroth moment map (Figure \ref{fig:HCN_mom0_gallery}).
    \item Centrally peaked DCN emission and a weak outer ring at $\sim$280\,au for GM~Aur.
    \item A single ring, centered at $\sim$60\,au for AS~209, 30\,au for HD~163296, and 70\,au for MWC~480. The ring in HD~163296 has a shoulder at $\sim$100\,au.
\end{enumerate}
The relation of the outer DCN rings of IM~Lup and GM~Aur to other deuterated molecules (\ntdp and \ce{DCO+}) is further discussed in Section \ref{sec:DCN_N2D+_DCO+_comparison}.

The DCN structure identified here is generally consistent with the findings by \citet{Law21_MAPS_radial_profiles}. However, by employing an azimuthal wedge along the disk major axis (instead of the full azimuth) when calculating the radial profile, they find that the shoulder in the HD~163296 profile is in fact a distinct ring at 118\,au. Furthermore, by looking at the higher-resolution (0.15$\arcsec$) image, they identify a ring at 16\,au for GM~Aur, instead of a centrally peaked profile. For a detailed characterization of the substructures (precise ring centers with uncertainties, gap widths, etc.) see \citet{Law21_MAPS_radial_profiles}.

Taking into account the additional information gained from the high-resolution profiles by \citet{Law21_MAPS_radial_profiles} described in the previous paragraph, there is generally good agreement between the structures seen in the HCN and DCN radial profiles. For IM~Lup, the inner DCN ring corresponds to a ring in HCN 3--2, while the outer DCN ring corresponds to a shoulder in the HCN 3--2 profile. For GM~Aur, both HCN and DCN show a bright inner ring and a faint outer ring. For AS~209, both HCN and DCN are in a ring, while for HD~163296, both species show a double ring. MWC~480 is a notable exception in that HCN and DCN show different morphologies: while HCN is centrally peaked, DCN is in a ring.

There are also a few associations of the HCN and DCN line emission structure with dust substructures \citep{Law21_MAPS_radial_profiles}. For example, the inner HCN and DCN rings in IM~Lup are associated with a continuum ring-gap structure at $\sim$125\,au. The most prominent feature is the association of the outer DCN rings in IM~Lup and GM~Aur with the edge of the dust continuum disk, as discussed in Section \ref{sec:DCN_N2D+_DCO+_comparison}. For a detailed analysis of the relationships between the radial structures seen in the MAPS data for HCN, DCN, as well as other molecules and the dust, see \citet{Law21_MAPS_radial_profiles}.

\subsection{Radial emission profiles of \texorpdfstring{\ce{N2H+}}{N2H+} and \texorpdfstring{\ce{N2D+}}{N2D+}}\label{sec:N2H+_radial_emission_profiles}
\begin{figure*}
\plotone{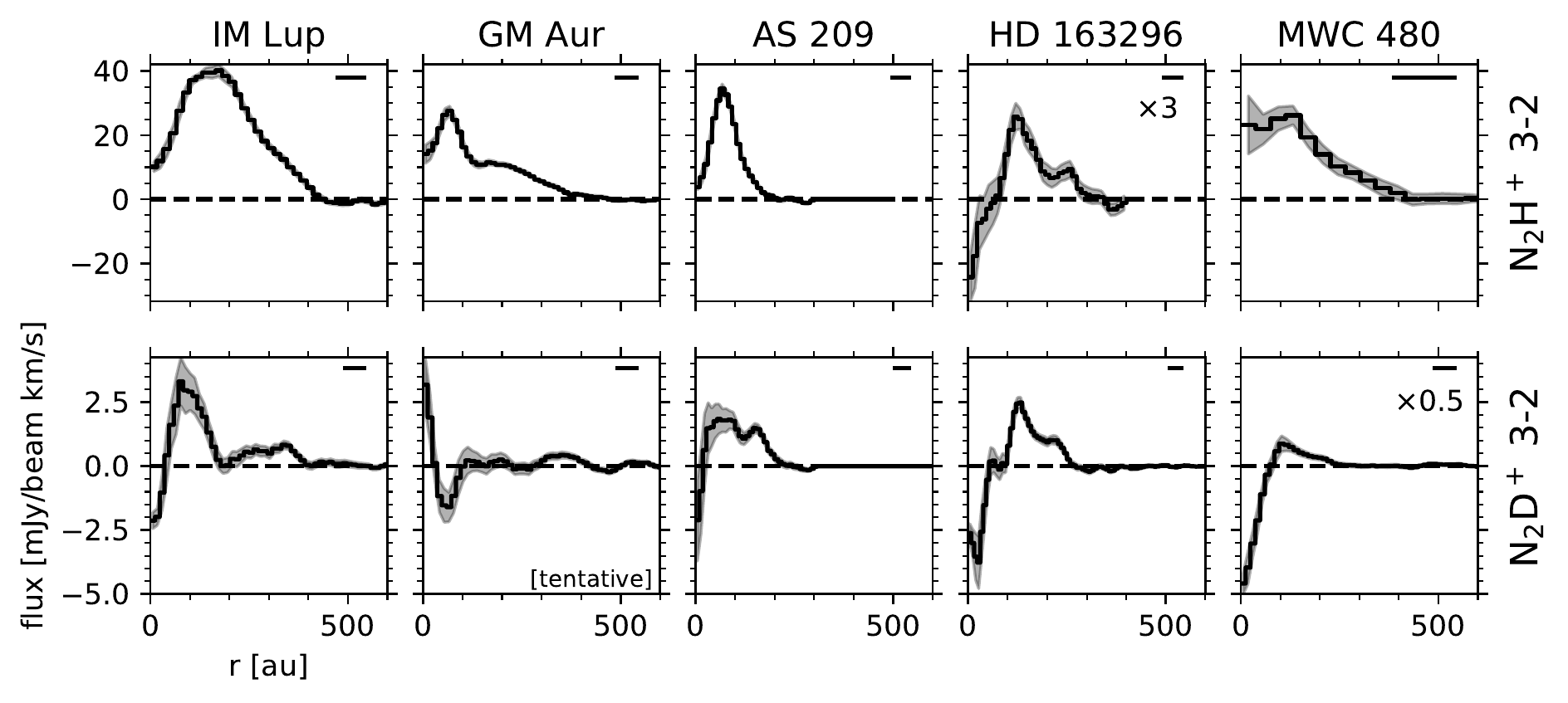}
\caption{Radial emission profiles of \ce{N2H+} 3--2 and \ce{N2D+} 3--2. The horizontal dashed line indicates the zero flux level. The shaded area shows the $\pm1\sigma$ error. If the profile has been scaled by a constant factor, the scaling is indicated in the upper right of the panel. The beam major axis is shown as horizontal black line in the upper right of each panel. \ce{N2D+} 3--2 toward GM~Aur is marked as tentatively detected based on the measurement of the disk-integrated flux and the matched filter analysis. \label{fig:N2H+_radial_profiles_gallery}}
\end{figure*}

Figure \ref{fig:N2H+_radial_profiles_gallery} shows the radial emission profiles of \nthp 3--2 and \ntdp 3--2. These profiles are also produced by azimuthally averaging a zeroth moment map generated with a Keplerian mask only (i.e., without a $\sigma$-clipping mask). The \nthp 3--2 radial emission profiles of IM~Lup, GM~Aur, AS~209, and MWC~480 are consistent with the ones presented in \citet{Qi19} and \citet{Loomis20}.

\nthp and \ntdp show similar ring emission structures, although their detailed structure differs from source to source. \citet{Qi19} proposed that the morphology of the \nthp emission reflects the vertical temperature structure of the disk. A narrow ring with extended tenuous emission (GM~Aur) is expected for a disk that is vertically isothermal up to substantial heights ($z/r\approx0.2$) above the midplane. Conversely, disks with a vertical temperature gradient above the midplane are expected to present broad rings (IM~Lup, AS~209). For HD~163296 and MWC~480, observations with higher SNR and, in the case of MWC~480, higher angular resolution would be useful to firmly distinguish between these two cases.

For the \ntdp emission, the four sources with a clear detection show ring structures. Most interestingly, IM~Lup clearly shows a double ring structure peaking at $\sim$100 au and $\sim$330 au, similar to the rings detected in DCN (Section \ref{sec:HCN_radial_emission_profiles}) and \ce{DCO+} \citep{Oberg15}. \nthp in IM~Lup also shows a subtle shoulder at the location of the outer ring of \ntdp. The relation between the structures seen for DCN, \ntdp and \ce{DCO+} will be discussed more in detail in Section \ref{sec:DCN_N2D+_DCO+_comparison}.

We note the presence of negative \ntdp emission toward the center of the HD~163296 and MWC~480 disks. This is probably due to the difficulty of achieving a precise continuum subtraction for this line, as discussed in Appendix \ref{appendix:negative_flux}.

\subsection{Azimuthally averaged spectra}\label{sec:az_averaged_spectra}
In order to compute radial column density profiles, we will model the azimuthally averaged spectra of equally spaced, deprojected annuli (radial bins). However, at each spatial pixel of a data cube, the spectrum is shifted with respect to the systemic velocity due to the Keplerian rotation of the gas. Therefore, in the azimuthally averaged spectrum, the emission is spread over a broad range of velocities. This results in a suboptimal SNR. Furthermore, the hyperfine structure of HCN and \ce{N2H+} is not resolved in such a broad spectrum and cannot be used to constrain the column density. Therefore, we apply the following procedure prior to azimuthally averaging \citep{Teague16,Yen16,Matra17}: we shift each spectrum by the projected Keplerian velocity, which is given by
\begin{equation}\label{eq:v_Kep}
    v_\mathrm{Kep}(r, \theta) = \sqrt{\frac{GM_\star}{r}}\sin i \cos \theta,
\end{equation}
where $r$ and $\theta$ are the (deprojected) radial and azimuthal coordinates of the disk, $M_\star$ is the (dynamical) stellar mass, and $i$ is the inclination \citep[see][for the adopted values]{Oberg21_MAPS}. Here we are assuming that emission originates from the midplane. The center of the disk is assigned to the proper motion-corrected stellar position. This procedure centers each spectrum at the systemic velocity. When averaging azimuthally, the SNR is increased and the hyperfine structure remains spectrally resolved.

The calculation of the error bars of the averaged spectra is described in Appendix \ref{appendix:az_averaged_spectra_errors}. The size of the radial bins was chosen to be equal to half of the mean of the beam major and minor axes.

The west side of the AS~209 disk is known to be affected by foreground cloud contamination \citep{Oberg11_SMA,Huang16,Guzman18}, which should be most relevant for Band 3 data. Thus, for the 1--0 transitions of HCN and H$^{13}$CN, we follow \citet{Teague18_AS209} and extract all spectra from a $\pm$55\arcdeg wedge that encompasses the uncontaminated eastern side of the disk. The other disks are not affected by cloud contamination and thus are averaged over the full azimuth.

Figures \ref{fig:AS_209_HCN_example_fitted_models} and \ref{fig:IM_Lup_N2H+_example_fitted_models} show examples of extracted spectra. The full gallery of spectra can be found in Appendix \ref{appendix:az_averaged_spectra_gallery}. In Figure \ref{fig:AS_209_HCN_example_fitted_models}, the hyperfine structure of HCN 1--0 and 3--2 is readily visible, except in the innermost region, where the finite spatial resolution causes considerable broadening of the spectra. Hyperfine structure is also seen for \ce{N2H+} in Figure \ref{fig:IM_Lup_N2H+_example_fitted_models}.

\section{Analysis}\label{sec:analysis}

\subsection{Radial column density profiles of HCN and DCN}\label{sec:HCN_DCN_fitting}
To derive the radial column density profiles of HCN and DCN, we use the 3--2 transitions of HCN and DCN covered in Band 6, and the 1--0 transitions of HCN and H$^{13}$CN covered in Band 3. Our modeling includes the hyperfine satellite lines, which help to derive reliable column densities even if the main components are optically thick. Table \ref{tab:atomic_parameters_HCN} provides an overview of the hyperfine lines used in the analysis. Combining the 3--2 and 1--0 transitions allows us to constrain the gas temperature.

For each radial bin, we fit the azimuthally averaged spectra calculated in Section \ref{sec:az_averaged_spectra} of all four lines simultaneously, using the image cubes with a circular 0.3$\arcsec$ beam. We start by assuming local thermodynamic equilibrium (LTE), that is, the excitation temperature $T_\mathrm{ex}$ is the same for all transitions and equals the kinetic gas temperature $T_\mathrm{kin}$. The free parameters are the excitation temperature, the HCN column density, the DCN column density, and additional parameters describing the line width and velocity offsets of the spectra. This results in a total of nine free parameters (see Table \ref{tab:HCN_free_parameters_overview}). We assume that HCN, DCN and H$^{13}$CN have the same temperature, that is, that they are cospatial. Although DCN might be more concentrated toward the midplane compared with HCN, their theoretically expected spatial distributions \citep{Aikawa18} are similar enough to justify this first-order approximation. The general similarity of the radial emission profiles of HCN 3--2 and DCN 3--2 (Fig.\ \ref{fig:HCN_radial_profiles_gallery}) also supports this assumption \citep{Huang17}.

We fix the H$^{13}$CN/HCN ratio to the ISM value of $^{13}$C/$^{12}$C=1/68 \citep{Milam05}. To explore the parameter space, we use the MCMC method implemented in the \texttt{emcee} package \citep{Foreman-Mackey13}. The details of the model and the fitting procedure are described in Appendix \ref{appendix:HCN_DCN_fitting}. A few example fits are shown in Figure \ref{fig:AS_209_HCN_example_fitted_models} for AS~209, and the full gallery of fits is shown in Appendix \ref{appendix:az_averaged_spectra_gallery}.

As can be seen in Fig.\ \ref{fig:AS_209_HCN_example_fitted_models}, in the innermost region (roughly within one beam FWHM from the disk center, i.e.\ two radial bins, corresponding to 30--47\,au depending on the disk), the lines are strongly broadened by the velocity gradient within the beam and the hyperfine structure is not resolved (see also Figures \ref{fig:IM_Lup_HCN_data_spectra} to \ref{fig:MWC_480_HCN_data_spectra} in Appendix \ref{appendix:az_averaged_spectra_gallery}). We caution that the strong broadening in the inner two radial bins might introduce additional uncertainties for the inferred column densities that are not reflected by our error bars.

\begin{figure*}
\plotone{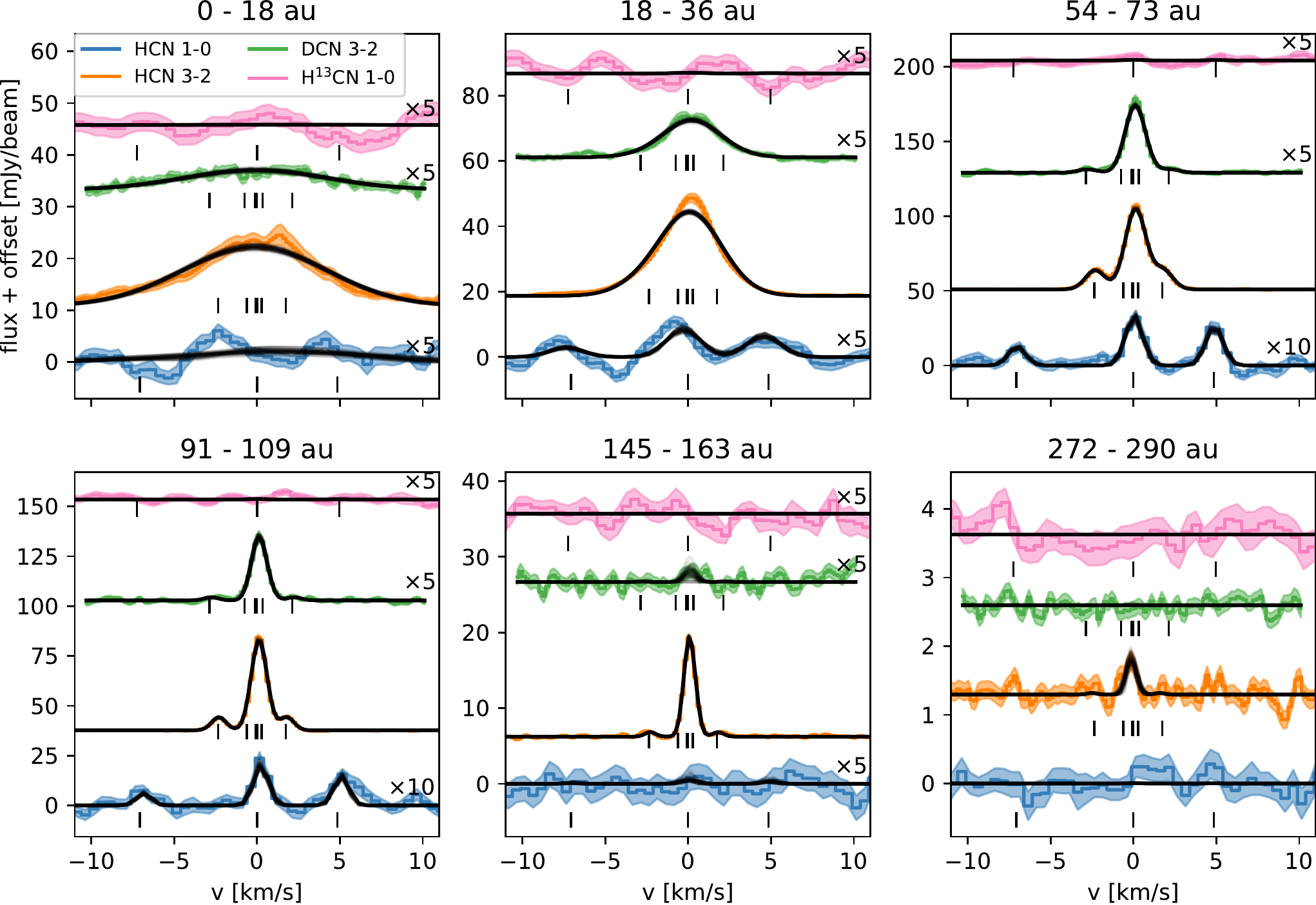}
\caption{Examples of HCN, DCN, and H$^{13}$CN model spectra fit to azimuthally averaged spectra of AS~209 for a few radial bins. Spectra are centered on the systemic velocity. The colored lines show the data, with the shaded regions corresponding to the 1$\sigma$ uncertainty. The black curves show 50 randomly selected models drawn from the Markov Chain Monte Carlo (MCMC), with the selection probability proportional to the posterior probability of the model. The small black vertical lines mark the hyperfine components. Spectra that have been scaled show the corresponding scaling factor on their right. Spectra are vertically offset for clarity. \label{fig:AS_209_HCN_example_fitted_models}}
\end{figure*}

\begin{figure*}
\plotone{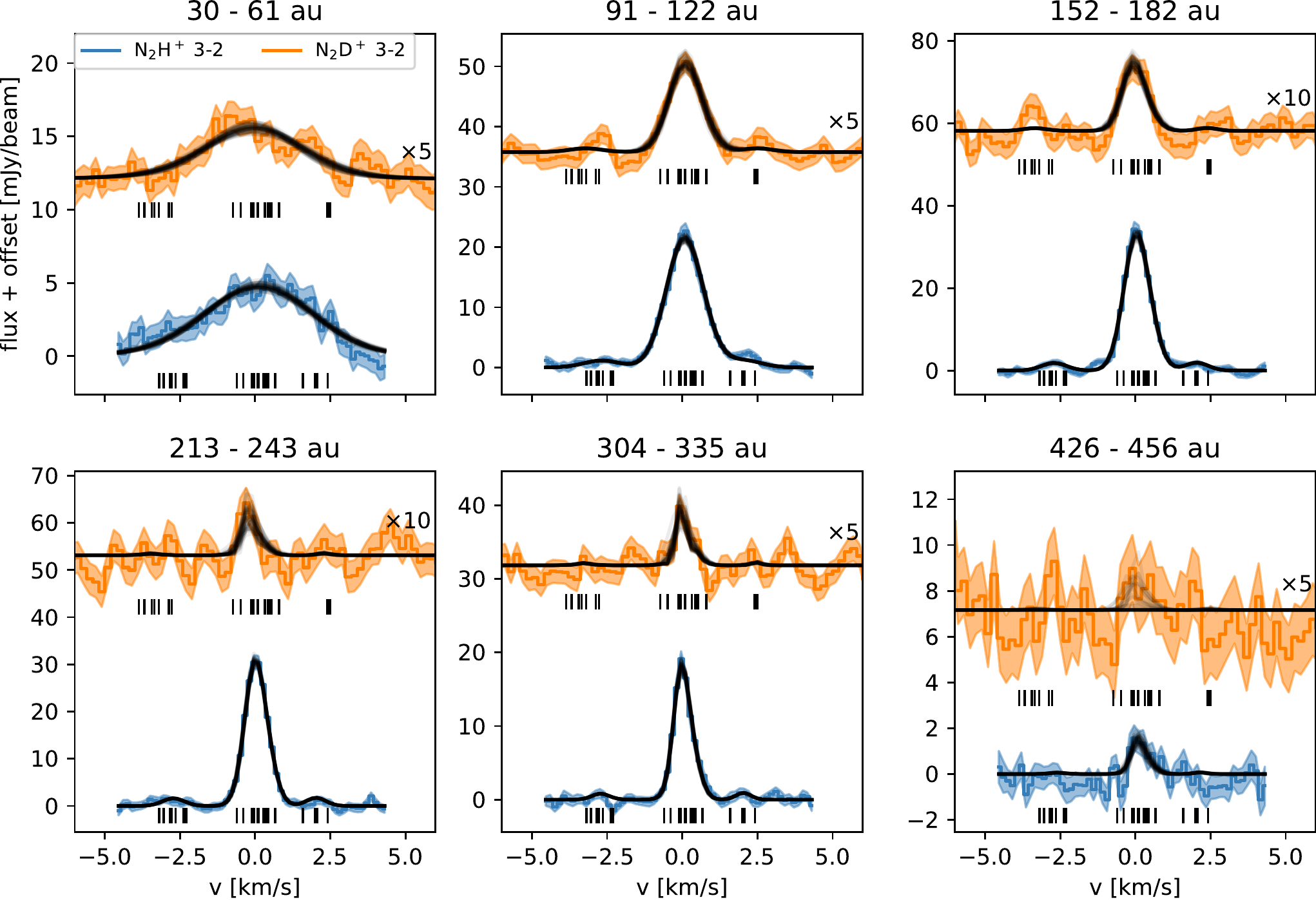}
\caption{Examples of \nthp and \ntdp model spectra fit to azimuthally averaged spectra of IM~Lup for a few radial bins. Spectra are centerd on the systemic velocity. The colored lines show the data, with the shaded regions corresponding to the 1$\sigma$ uncertainty. The black curves show 50 randomly selected models drawn from the MCMC with the selection probability proportional to the posterior probability of the model. The small black vertical lines mark the hyperfine components. The \ntdp spectra have been scaled and show the corresponding scaling factor on their right. Spectra are vertically offset for clarity. \label{fig:IM_Lup_N2H+_example_fitted_models}}
\end{figure*}

\begin{figure*}
\plotone{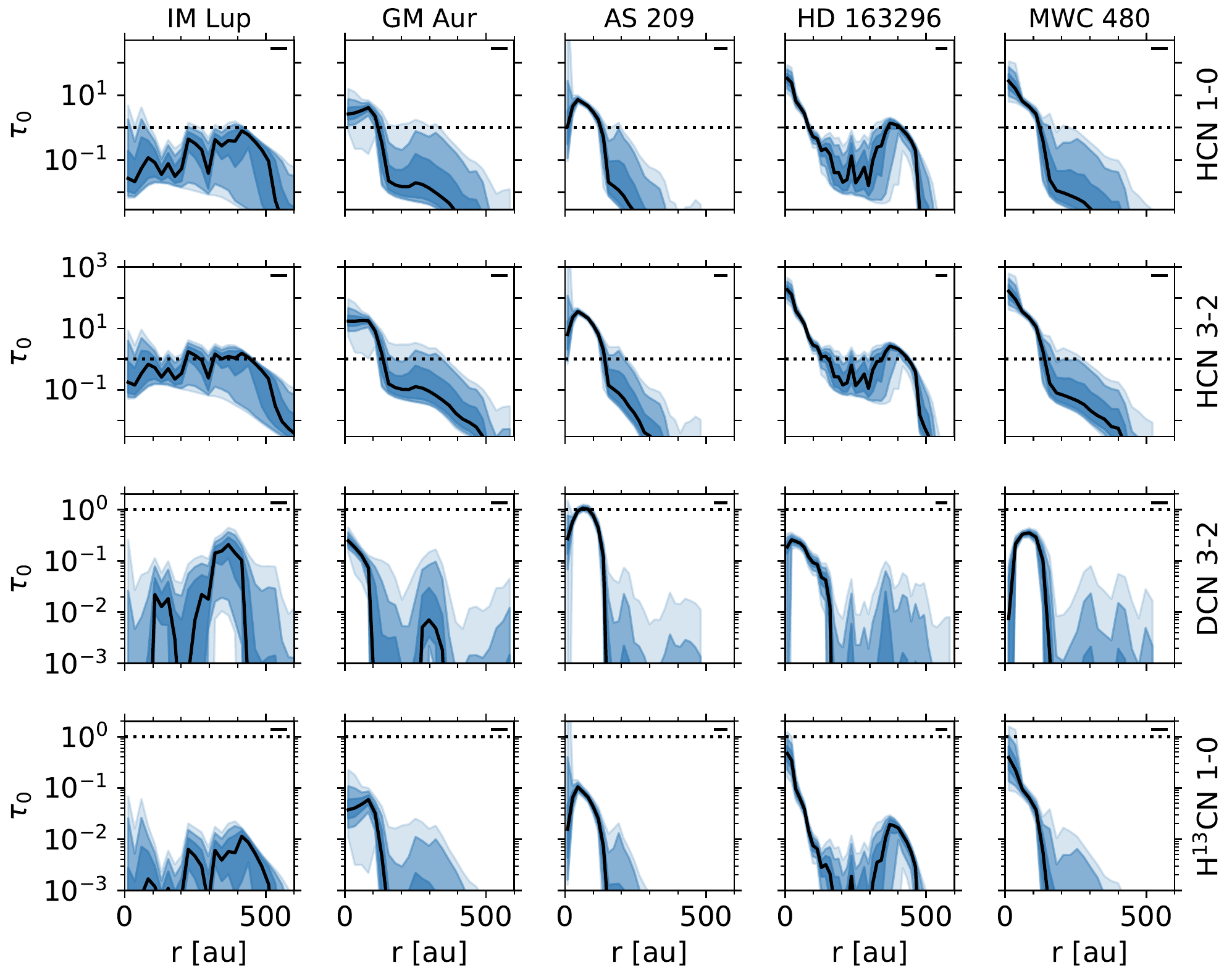}
\caption{Optical depths from the MCMC calculation assuming LTE (i.e.\ $T_\mathrm{ex}=T_\mathrm{kin}$, see Appendix \ref{appendix:HCN_DCN_fitting_LTE}) for the HCN 1--0, HCN 3--2, DCN 3--2 and H$^{13}$CN 1--0 lines. The black solid line shows the median. The blue shaded regions encompass the 16th to 84th, 2.3th to 97.7th, and 0.15th to 99.85th percentile regions. The horizontal dotted line marks an optical depth of 1. The beam major axis is shown as a horizontal black line in the top right of each panel.\label{fig:HCN_DCN_tau}}
\end{figure*}

\begin{figure*}
\plotone{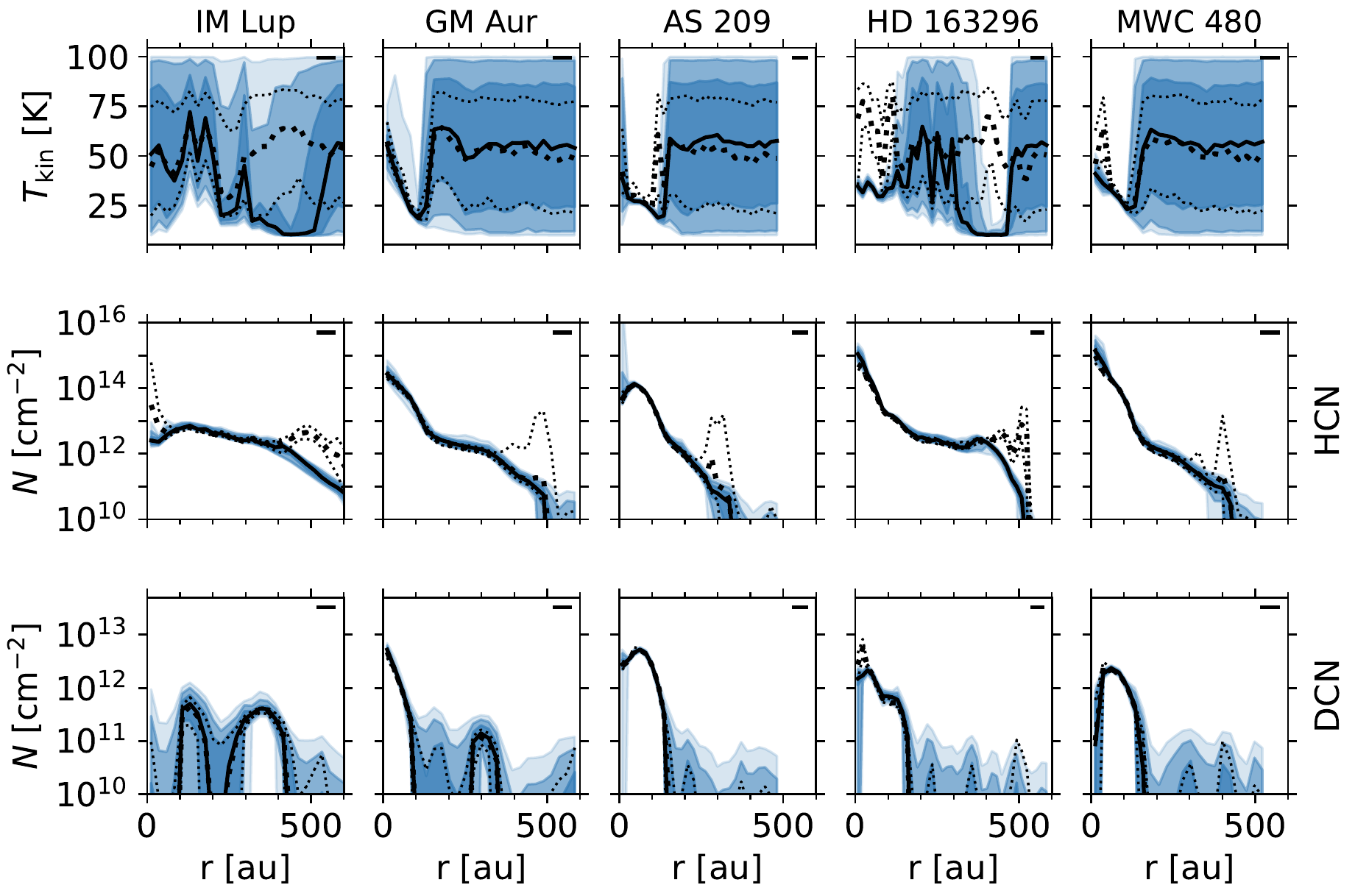}
\caption{Temperatures and HCN and DCN column densities derived from MCMC fits of azimuthally averaged spectra assuming LTE (i.e.\ $T_\mathrm{ex}=T_\mathrm{kin}$, see Appendix \ref{appendix:HCN_DCN_fitting_LTE}). The solid black line shows the median. The blue shaded regions encompass the 16th to 84th, 2.3th to 97.7th, and 0.15th to 99.85th percentile regions. For comparison, the median values (thick dotted lines) and 16th and 84th percentile (thin dotted lines) of the kinetic temperature and column densities derived from non-LTE fits are shown. The beam major axis is shown as a horizontal black line in the upper right of each panel.\label{fig:HCN_DCN_columndensities_T_LTE}}
\end{figure*}

Figures \ref{fig:HCN_DCN_tau} and \ref{fig:HCN_DCN_columndensities_T_LTE} show the derived optical depths, temperatures, and column densities for all five sources. The HCN 3--2 transition is optically thick ($\tau_0>1$) in the inner $\sim$100\,au for all sources except IM~Lup. 

\begin{figure*}
\plotone{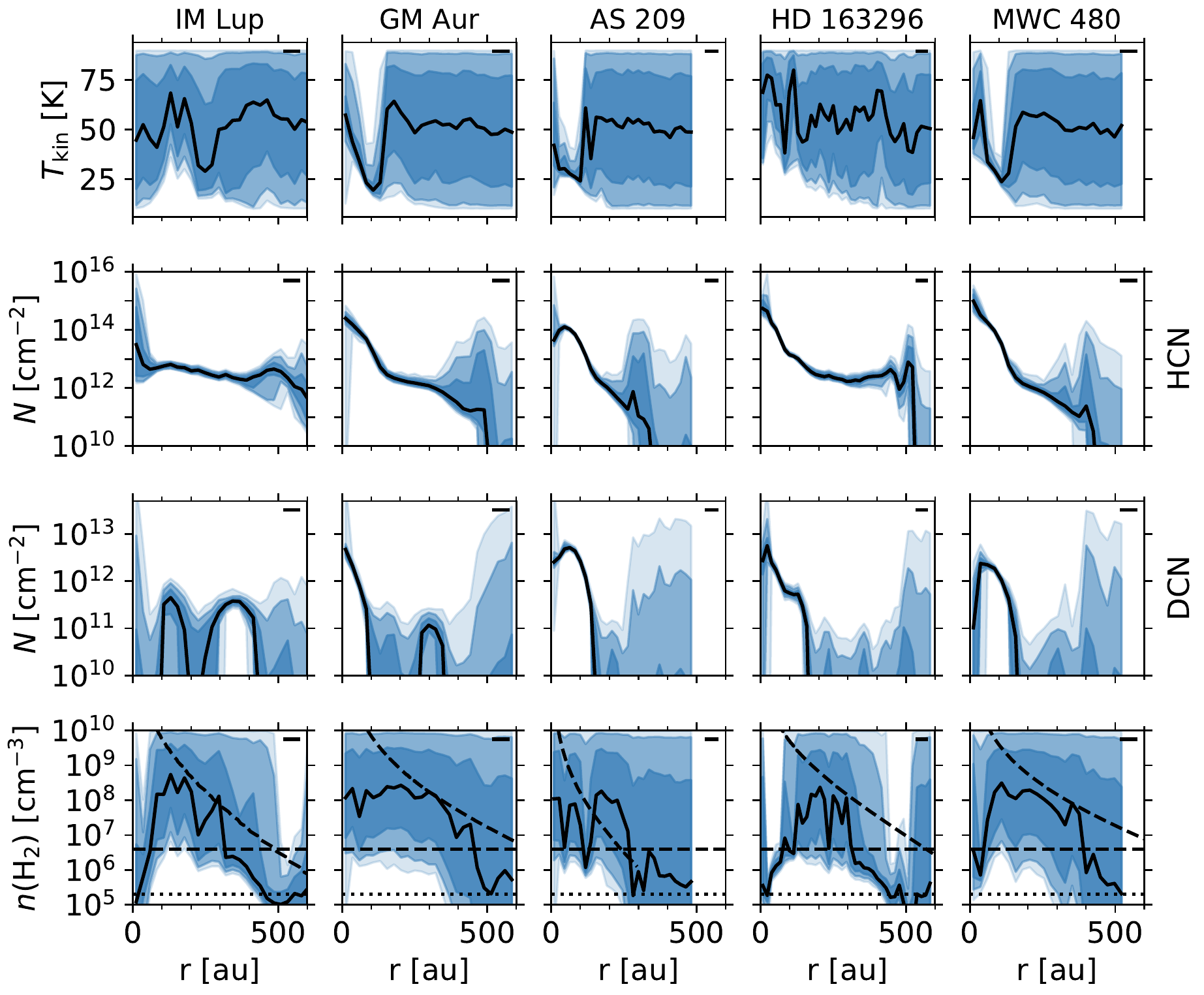}
\caption{Kinetic gas temperature, HCN and DCN column densities, and \ce{H2} number density derived from non-LTE MCMC fits of azimuthally averaged spectra. The solid black line shows the median. The blue shaded regions encompass the 16th to 84th, 2.3th to 97.7th, and 0.15th to 99.85th percentile regions. The beam major axis is shown as a horizontal black line in the upper right of each panel. The black dashed curves in the last row show the \ce{H2} number density in the midplane of the MAPS reference disk models by \citet{Zhang21_MAPS}, while the horizontal dotted and dashed lines show the critical density of HCN 1--0 and 3---2, respectively.\label{fig:HCN_DCN_columndensities_T_nH2_nonLTE}}
\end{figure*}

We next performed a fit without assuming LTE, that is, the excitation temperatures are not necessarily equal to the kinetic gas temperature. In this case, we fit for the kinetic gas temperature $T_\mathrm{kin}$ and the \ce{H2} number density. Details are again given in Appendix \ref{appendix:HCN_DCN_fitting}. The results are shown in Figure \ref{fig:HCN_DCN_columndensities_T_nH2_nonLTE}. The \ce{H2} number density is constrained to $\gtrsim10^6$\,cm$^{-3}$ for all disks for $r\lesssim$200--400\,au. For comparison, the critical density is $2\times10^5$\,cm$^{-3}$ and $4\times10^6$\,cm$^{-3}$ for HCN 1--0 and 3--2, respectively \citep{Dutrey97}. The fitted \ce{H2} number density is mostly consistent with the midplane \ce{H2} number density from the MAPS reference disk models \citep{Zhang21_MAPS}, shown with black dotted lines. The column densities are not significantly different from the LTE case, as can be seen in Figure \ref{fig:HCN_DCN_columndensities_T_LTE}, where the non-LTE results are overplotted for comparison. This suggests that the emission is indeed in LTE. The exception is the region around 500\,au of the disks around IM~Lup and HD~163296, where non-LTE conditions might prevail (see Section \ref{sec:LTE_nonLTE_comparison}).

\subsection{Radial column density profiles for \texorpdfstring{\ce{N2H+}}{N2H+} and \texorpdfstring{\ce{N2D+}}{N2D+}}\label{sec:N2H+_N2D+_fitting}
We derive column densities of \nthp and \ntdp with the same procedure as in Section \ref{sec:HCN_DCN_fitting}, assuming LTE. However, since we observed only one transition for each molecule, we did not fit the excitation temperature. Theoretical models predict that the \nthp abundance peaks around the threshold temperature of CO freeze-out \citep[e.g.][]{Aikawa15,Aikawa18}, which is constrained by observations to $\sim$20\,K \citep{Qi11,Schwarz16,Pinte18}. Thus, for simplicity we fixed the excitation temperature of both molecules to 20\,K for our fiducial fits. This results in a total of six free parameters that are fitted for each radial bin: the column densities, the velocity offsets, and the parameter describing line broadening for each of the two lines. These parameters and their priors are listed in Table \ref{tab:N2H+_free_parameters_overview}. The hyperfine structure is taken into account in the same way as for HCN and DCN. Tables \ref{tab:atomic_parameters_N2H+} and \ref{tab:atomic_parameters_N2D+} list the 28 and 25 hyperfine components considered for the fitting of \nthp and \ntdp, respectively. Figure \ref{fig:IM_Lup_N2H+_example_fitted_models} shows example fits for IM~Lup for a few selected radial bins. The complete gallery of fits is shown in Figures \ref{fig:IM_Lup_N2H+_data_spectra} to \ref{fig:MWC_480_N2H+_data_spectra}. Similar to the HCN and DCN column densities, we caution that the column densities of the two innermost radial bins might be less reliable than those in the outer regions because of strong line broadening (see, e.g., Figures \ref{fig:IM_Lup_N2H+_data_spectra} and \ref{fig:GM_Aur_N2H+_data_spectra} for extreme examples of broadening of \nthp 3--2).

\begin{figure*}
\plotone{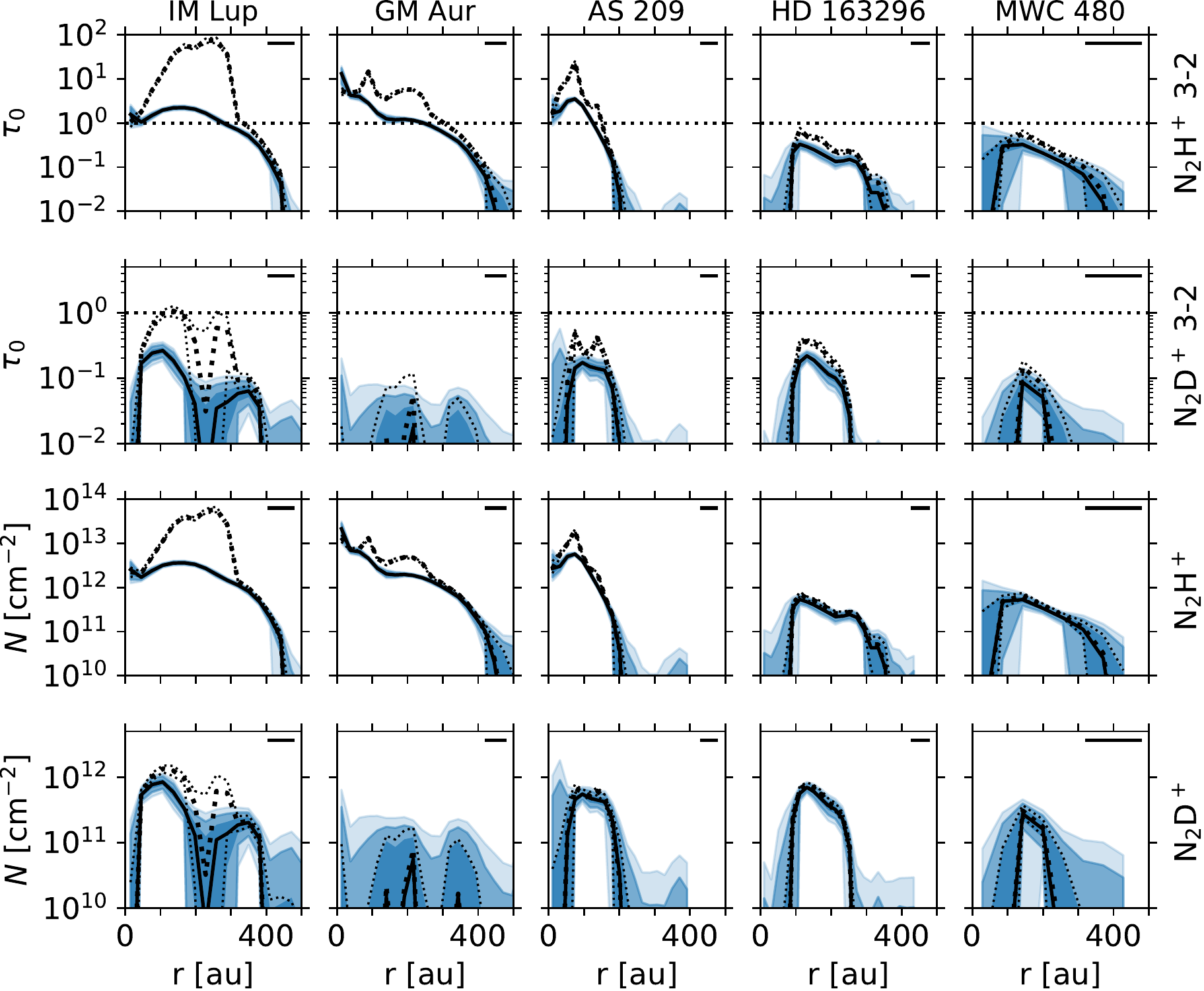}
\caption{Optical depths and column density profiles of \nthp and \ntdp. The results from the fiducial fit ($T_\mathrm{ex} = 20$\,K) are shown with the solid black line (median) and blue shaded regions encompassing the 16th to 84th, 2.3th to 97.7th, and 0.15th to 99.85th percentile regions. For comparison, the median (thick dotted lines) and 16th and 84th percentile (thin dotted lines) from the alternative fit ($T_\mathrm{ex} = T_\mathrm{mid}$) are shown. In the first two rows, the horizontal dotted line marks an optical depth of 1. The beam size is shown as a horizontal black line in the upper right of each panel.\label{fig:N2H+_N_tau}}
\end{figure*}

Figure \ref{fig:N2H+_N_tau} shows the derived optical depth and column density profiles for all five sources. The \nthp 3--2 transition is optically thick around the emission peak for IM~Lup, GM~Aur, and AS~209. The column densities of \nthp are close to $10^{13}\,\mathrm{cm^{-2}}$ for these sources, while they only reach $\lesssim$$10^{12}\,\mathrm{cm^{-2}}$ for the two warmest disks in the sample around the Herbig stars HD~163296 and MWC~480. For all sources, the \ntdp 3--2 transition is optically thin throughout the whole disk. The \ntdp peak column densities exceed $10^{11}\,\mathrm{cm^{-2}}$ except for GM~Aur.

In order to check the dependence of the column density on the assumed excitation temperature, we also run fits assuming that the excitation temperature equals the midplane gas temperature $T_\mathrm{mid}$ shown in Figure \ref{fig:radial_temp_model} of Appendix \ref{appendix:HCN_DCN_fitting_LTE}. These temperature profiles are extracted from the MAPS reference models \citep{Zhang21_MAPS}. As can be seen in Figure \ref{fig:N2H+_N_tau}, there are some areas where the two assumptions about the excitation temperature result in different column densities. The strongest difference is seen for IM~Lup, where the \nthp column density increases by more than an order of magnitude if assuming $T_\mathrm{ex}=T_\mathrm{mid}$. This is because for IM~Lup, the model midplane temperature is as low as $\sim$7\,K. However, even with such a high column density, the fit actually underpredicts the flux of the main \nthp component by a factor of $\sim$2, while at the same time overpredicting the flux of the hyperfine components (see Fig.\ \ref{fig:IM_Lup_N2H+_data_spectra}). This strongly suggests that the \nthp excitation temperature is actually $\gtrsim$10\,K. Beyond $\sim$300\,au where the model midplane temperature rises to $\sim$15\,K, the column densities derived for the two excitation temperature choices agree again. Compared to \nthp, the \ntdp optical depth and column density of IM~Lup show a less strong increase if we assume $T_\mathrm{ex}=T_\mathrm{mid}$.

Similarly, when assuming $T_\mathrm{ex}=T_\mathrm{mid}$, the \nthp column density toward AS~209 is higher by a factor $\sim$4 around $\sim$75\,au (where $T_\mathrm{mid}\approx11$\,K, see Fig.\ \ref{fig:radial_temp_model}), but again the \nthp data are not fit well by this model (see Fig.\ \ref{fig:AS_209_N2H+_data_spectra}), in contrast to the fit where $T_\mathrm{ex}=20$\,K. Finally, Fig.\ \ref{fig:N2H+_N_tau} shows that, when assuming $T_\mathrm{ex}=T_\mathrm{mid}$, the \nthp column density toward GM~Aur is increased by up to a factor of $\sim$3 in the region between $\sim$50\,au and $\sim$250\,au. In this region, $T_\mathrm{mid}\approx12$\,K. However, in contrast to IM~Lup and AS~209, the model assuming $T_\mathrm{ex}=T_\mathrm{mid}$ fits the \nthp data as good as the model assuming $T_\mathrm{ex}=20$\,K (see Fig.\ \ref{fig:GM_Aur_N2H+_data_spectra}). Thus, if one believes that the temperature in the emitting region is indeed as low as 12\,K, the \nthp column density toward GM~Aur would be a factor of $\sim$3 larger. From a theory point of view, this low temperature is not unreasonable; the models by \citet[][see their Fig.\ 10]{Aikawa15} show that, while the \nthp abundance should peak around the CO freeze-out temperature ($\sim$20\,K), it can remain substantial even below the \ce{N2} freeze-out temperature of $\sim$17\,K \citep[see also][]{Aikawa21_MAPS}.

\subsection{Radial profiles of the deuteration fraction} \label{sec:deuteration_profiles}

\subsubsection{DCN/HCN}\label{sec:DCN2HCN}
\begin{figure*}
\plotone{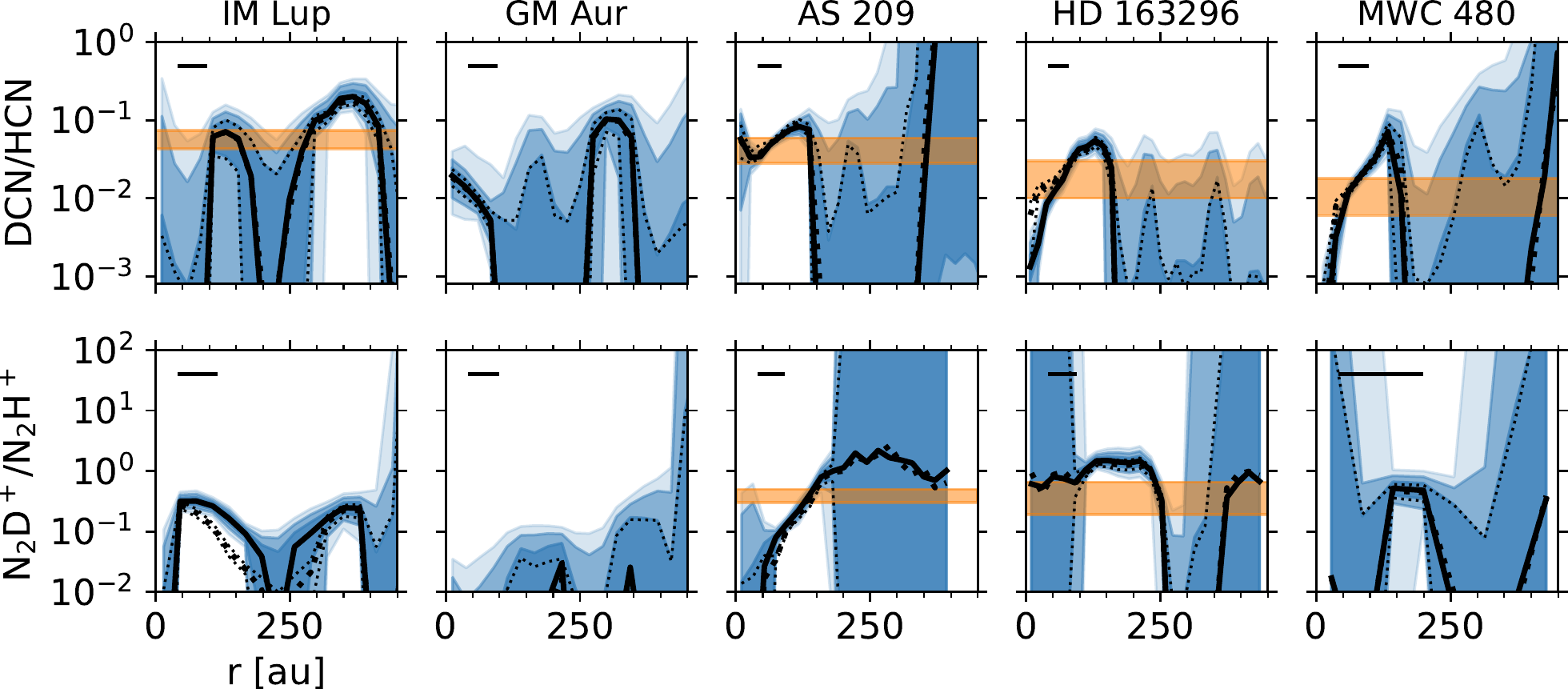}
\caption{Radial profiles of the HCN/DCN (top) and \ce{N2D+}/\ce{N2H+} column density ratios. The solid black lines show the median of the LTE fits for DCN/HCN and of the fits with $T_\mathrm{ex}$=20\,K for \ntdp/\nthp. The blue shaded regions encompass the 16th to 84th, 2.3th to 97.7th, and 0.15th to 99.85th percentile regions. Medians (thick dotted line) and the 16th and 84h percentiles (thin dotted lines) of the alternative fits (DCN/HCN: non-LTE; \ntdp/\nthp: $T_\mathrm{ex}=T_\mathrm{mid}$) are shown for comparison. The horizontal orange shaded areas show the disk-integrated HCN/DCN ratios inferred by \citet[][IM~Lup, AS~209, HD~163296, MWC~480]{Huang17} and \citet[][HD~163296]{Salinas17} and disk-integrated \ce{N2D+}/\ce{N2H+} ratios by \citet[][AS~209]{Huang15} and \citet[][HD~163296]{Salinas17}. The beam major axis is shown with the horizontal black line in the upper left.\label{fig:deuteration_profiles}}
\end{figure*}

Using the column density profiles of HCN and DCN calculated in Section \ref{sec:HCN_DCN_fitting}, we now derive the radial profile of the deuteration fraction of HCN. Figure \ref{fig:deuteration_profiles} (upper panel) shows the radial profiles of the DCN/HCN ratio. The results from the LTE and non-LTE fits agree well. The inferred DCN/HCN ratios range from $\sim$$10^{-3}$ to $\sim$$10^{-1}$. For IM~Lup, the outer DCN ring at $\sim$350\,au is more strongly deuterated than the inner ring at $\sim$100\,au. The data also suggest that the deuteration further decreases inward of the inner ring, although not at high significance. For the disk of GM Aur, the outer ring at $\sim$300\,au is more strongly deuterated than the inner $\sim$100\,au of the disk. AS~209 shows a weak decreasing trend of the deuteration toward the inner parts of the disk. Finally, for the two disks around the Herbig stars HD~163296 and MWC~480, the deuteration in the innermost region is reduced by almost two orders of magnitude compared to the deuteration at $\sim$150\,au. Thus, while there is some diversity in the deuteration profiles toward our five targets, generally the outer disk regions are more strongly deuterated in HCN compared with the inner disk. In the inner 100\,au of the disk around GM~Aur, the deuteration seems to be increasing toward the star instead, although higher SNR and angular resolution data would be needed for confirmation. These trends are further discussed in Section \ref{sec:radial_DCN2HCN_discussion}.

\subsubsection{\texorpdfstring{\ce{N2D+}/\ce{N2H+}}{N2D+/N2H+}}
The lower panels of Figure \ref{fig:deuteration_profiles} show radial profiles of the \ntdp/\nthp column density ratio. We find a \ntdp/\nthp ratio typically between a few times $10^{-2}$ and 1. The profiles do not show a strong dependence on the assumed excitation temperature. The radial distribution of the \ntdp/\nthp ratio varies among sources. For IM~Lup, the ratio has a minimum located at $\sim$220\,au; this structure reflects the gap in the column density profile of \ntdp. For GM~Aur, \ntdp remains undetected while strong emission of \nthp is seen, resulting in upper limits on the ratio. For AS~209, the ratio monotonically increases with radius from $\sim$50\,au to 170\,au. For the disk around HD~163296, the profile is rather flat. For MWC~480, the \ntdp/\nthp ratios is a few times 10$^{-1}$, but the low angular resolution (beam major axis of 0.94$\arcsec$) of the \nthp data precludes an accurate determination of the \ntdp/\nthp profile.

\section{Discussion}\label{sec:discussion}

\subsection{HCN and DCN}

\subsubsection{Comparing LTE and non-LTE}\label{sec:LTE_nonLTE_comparison}
It is interesting to compare the HCN gas temperatures derived from the LTE and non-LTE fits (Fig.\ \ref{fig:HCN_DCN_columndensities_T_LTE} top row). In general, the temperature is not well constrained, although there are some disk regions where information on the temperature can be extracted. For example, $T\gtrsim30$\,K around 150\,au in the disk around IM~Lup, and $T\approx25$\,K in the inner $\sim$100\,au of the disk around AS~209. Despite the large uncertainties, we find significant differences between LTE and non-LTE temperature estimates in some disk regions: from 300 to 550\,au for IM~Lup, at $\sim$110\,au for AS~290, inward of 100\,au and between 300 and 500\,au for HD~163296, and inward of 70\,au for MWC~480. We find that in these regions, non-LTE generally provides a better fit to the data. In particular, the LTE fits tend to underpredict the HCN 1--0 emission. As an example, this can clearly be seen in Fig.\ \ref{fig:IM_Lup_HCN_data_spectra} around 500\,au for IM~Lup and in Fig.\ \ref{fig:HD_163296_HCN_data_spectra} around 450\,au for HD~163296. For these two outer disk regions, our fits thus suggest that the gas density is low enough for non-LTE conditions to prevail. This is further supported by Fig.\ \ref{fig:HCN_DCN_columndensities_T_nH2_nonLTE} that shows that the \ce{H2} density might be below the critical density of HCN 3--2. The HCN column density derived from the non-LTE fit is higher in those two regions compared with the LTE case.

However, non-LTE seems more unlikely an explanation for the disk regions $\lesssim100$\,au mentioned above, where the gas density is expected to be higher. Instead, for those regions, the difference between the LTE and non-LTE temperatures might indicate that our assumption of a single temperature being able to describe the emission of all lines is invalid. For example, since the HCN emission is optically thick in those regions (Fig.\ \ref{fig:HCN_DCN_tau}), it could be that HCN 3--2 is emitting from a different vertical layer with a different temperature than HCN 1--0 and DCN 3--2. Similarly, it could be that the hyperfine components trace a different temperature compared with the main component due to optical depth. Another possibility is effects due to dust optical depth that might affect the 3--2 and 1--0 transitions differently. Source-specific radiative transfer modeling would be necessary to investigate these ideas. Fortunately, the column densities do not differ strongly between the LTE and non-LTE fits in those inner $\sim$100\,au regions.

\subsubsection{Comparison with previous observations}
\citet{Bergner19,Bergner20_erratum} derived HCN column density profiles toward AS~209, HD~163296, and MWC~480 by using observations of the 3--2 transition of HCN and H$^{13}$CN. Their data were of significantly lower resolution: $\sim$0.5$\arcsec$ for AS~209 and HD~163296, and $\sim$2.7$\arcsec$ for MWC~480. Our HCN column density profiles show a much greater dynamical range compared with \citet{Bergner19,Bergner20_erratum}, but they show order of magnitude agreement in terms of mean column densities. \citet{Bergner19,Bergner20_erratum} derive excitation temperatures mostly below 20\,K, that is, lower than the temperatures we derive. To test whether these differences arise because of the higher spatial resolution of our data (i.e.\ less beam dilution), we convolved our data to match the resolution of the \citet{Bergner19} data and repeated the fitting. We find that even in that case, significant differences in column density and excitation temperature are present. This suggests that the lower resolution of the \citet{Bergner19,Bergner20_erratum} data is not the primary reason for the observed differences but rather that their analysis did not include the HCN 1--0 transition and that they were unable to use the hyperfine structure of HCN 3--2.

\citet{Guzman21_MAPS} and \citet{Bergner21_MAPS} also fitted spectra extracted from the MAPS data to derive HCN column density profiles. \citet{Bergner21_MAPS} only used the HCN 3--2 data, while \citet{Guzman21_MAPS} used HCN 3--2 and 1--0. Neither of them used H$^{13}$CN 1--0. In general, the profiles show good agreement with ours. Some differences are observed in the inner 50--100\,au, where \citet{Guzman21_MAPS} and \citet{Bergner21_MAPS} derive up to an order of magnitude higher HCN column densities for GM~Aur, AS~209, HD~163296, and MWC~480. Adopting those values would further strengthen the trend of low DCN/HCN toward the disk center. We refer to Appendix~C of \citet{Guzman21_MAPS} for a comparison of the derived HCN column densities.

The disk-integrated DCN/HCN ratios derived by \citet{Huang17} and \citet{Salinas17} (orange shaded areas in Figure \ref{fig:deuteration_profiles}) are in good agreement with our results. Concerning the radial variation of DCN/HCN, our results are in line with trends of enhanced deuteration in the outer disks seen in previous work. \citet{Huang17} discuss the possibility of enhanced deuteration in the outer disks of V4046~Sgr, LkCa~15, and HD~163296 by comparing the emission morphologies of H$^{13}$CN and DCN. More recently, \citet{Facchini21} presented a radial profile of the DCN/HCN column density ratio for the PDS~70 disk at 0.42$\arcsec$ (47\,au) resolution. They also identify an increasing HCN deuteration with increasing radius. Similarly, \citet{Oberg21_TWHya} found that the DCN emission in the TW~Hya disk drops toward the center, while the HCN emission is centrally peaked \citep{Hily-Blant19}. While a determination of the column density DCN/HCN ratio is pending, TW~Hya may well show the same trend of an increasing DCN/HCN ratio with radius.

\subsubsection{Comparison of HCN and DCN column densities to model predictions}
In this section, we compare the derived HCN and DCN column densities to some models presented in the literature \citep{Willacy07,Walsh12,Aikawa18,Cleeves18}. Among these models, the \citet{Cleeves18} model is the only one tailored to a specific MAPS source: IM~Lup. The others are generic models of disks around T~Tauri stars. This should especially be kept in mind when comparing these models to the disks around the Herbig Ae stars HD~163296 and MWC~480.

We first consider the HCN column density. The models by \citet{Willacy07}, \citet{Walsh12}, and \citet{Aikawa18} predict a flat HCN profile (except for the innermost few au). In contrast, IM~Lup shows an HCN profile slowly rising inward (except for the innermost 100\,au), while the disks around AS~209, GM~Aur, HD~163296, and MWC~480 all show HCN column densities that are steeply rising toward the star inward of 100--200\,au (Figure \ref{fig:HCN_DCN_columndensities_T_LTE}). \citet{Cleeves18} modeled the HCN column density for a range of elemental C/O ratios and cosmic-ray (CR) ionisation rates. Their models show increasing HCN column densities toward the star. Therefore, the \citet{Cleeves18} models actually better reproduce the morphology of the HCN column density profiles for all disks (depending on the assumed C/O), although they were tailored to IM~Lup. The \citet{Cleeves18} models also reproduce reasonably well the order of magnitude values of the HCN column densities we derive. One possible reason might be that \citet{Cleeves18} assumed an elevated interstellar radiation field strength ($G_{\rm 0}=4$). Also, for the UV radiation from the central star, they scaled the TW~Hya UV spectrum to be consistent with the flux density of IM~Lup. These points might matter, since UV photons can dissociate HCN, and the UV radiation field is thus an important parameter influencing the HCN column density profile \citep[e.g.,][]{Bergin03,Chapillon12,Bergner21_MAPS}. The \citet{Cleeves18} study also demonstrated that the HCN column density depends strongly on the C/O ratio. Thus, future work might be able to constrain the C/O ratio using observed HCN column density profiles. This could then be compared to the C/O ratio inferred from, for example, the study of \ce{C2H} \citep[e.g.][]{Bergin16,Alarcon21_MAPS,Bosman21_MAPS_CtoO} or sulfur-bearing molecules \citep{LeGal21_MAPS}.

In the innermost $\sim$5\,au, the models by \citet{Willacy09}, \citet{Walsh12}, and \citet{Aikawa18} all predict an up to six orders of magnitude increase of the HCN column density due to thermal desorption. The angular resolution of our data is insufficient to test this prediction directly. A detailed analysis of the line kinematics may be able to probe this innermost region \citep[e.g.,][]{Bosman21_MAPS_inner20au}, but this is beyond the scope of the present paper.

Finally, we compare our DCN column density profiles to the models by \citet{Willacy07} and \citet{Aikawa18}. They typically predict DCN column densities of $\sim$10$^{12}$\,cm$^{-2}$, that is, the order of magnitude agrees with the column densities we derive (see Figure \ref{fig:HCN_DCN_columndensities_T_LTE}).

In summary, the HCN and DCN column density profiles derived here provide useful benchmarks for disk chemistry models. In particular, we expect that parameters such as the UV field or the C/O ratio can be constrained in future modeling work.

\subsubsection{The radial variation of the DCN/HCN ratio}\label{sec:radial_DCN2HCN_discussion}
A major result of our work is the variation of the DCN/HCN ratio within the disks: generally, DCN/HCN decreases when moving from the outer disk regions ($r\gtrsim100$\,au) toward the disk center (Figure \ref{fig:deuteration_profiles}). For IM~Lup, this decrease is seen when comparing the outer ring to the inner ring, as well as when comparing the inner ring to the disk center. For GM~Aur, the decrease is seen by comparing the outer ring at $\sim$300\,au to the disk's central $\sim$100\,au, although DCN/HCN might be increasing from 100\,au toward the disk center. For AS~209, the decrease is quite weak from $\sim$150\,au toward the disk center. For all these T~Tauri disks, the decrease of DCN/HCN is not more than an order of magnitude. On the other hand, for the disks around the Herbig stars HD~163296 and MWC~480, the ratio decreases by almost two orders of magnitude, from $10^{-1}$ to almost $10^{-3}$, when moving inward from $\sim$150\,au.

Generally speaking, we interpret the decrease of the DCN/HCN ratio toward the disk center as a confirmation that HCN fractionation is due to in situ exothermic exchange reactions such as those shown in Equations \ref{eq:HD_exchange_reaction_lowT}--\ref{eq:HD_exchange_reaction_highT2}. Indeed, the temperature is expected to increase toward the disk center, which in turn increases the efficiency of the reverse reactions, thus lowering the deuteration. To get further insight, we also compare our DCN/HCN profiles to models from the literature. Interestingly, the disk chemistry model (assuming a T~Tauri host star) by \citet{Favre15} predicts an increasing DCN/HCN ratio from 60\,au inward. Unfortunately, the DCN/HCN inward of 60\,au is not constrained well for our disks, and the model does not go to larger radii. A comparison is therefore difficult, but we can at least say that the qualitative trend of the DCN/HCN model profile is different from the data, with the notable exception of the inner $\sim$100\,au of the disk around GM~Aur. On the other hand, the disk model (assuming a T~Tauri host star) by \citet{Aikawa18} predicts a decreasing DCN/HCN ratio when moving inward from 300 to 10\,au (see their Figure 5d). One of the updates of the \citet{Aikawa18} model compared with \citet{Favre15} is the inclusion of the ortho/para ratio of H$_2$. An elevated ortho/para ratio of \ce{H2} in the inner disk compared with the outer regions can reduce the efficiency of the deuteration.

For a more detailed interpretation of the DCN/HCN radial profiles, source-specific modeling will be needed. In any case, if the gas we observe is the main source of HCN for forming icy bodies in the disk, we expect the formation location to be imprinted in the DCN/HCN ratio of the body, that is, the higher the DCN/HCN ratio, the further out in the disk the body formed. However, there remains a caveat that the gas we observe may not be the primary HCN source of forming comets (see Section \ref{sec:comparison_to_comets}).

\subsubsection{DCN/HCN as a function of temperature}
\begin{figure*}
\plotone{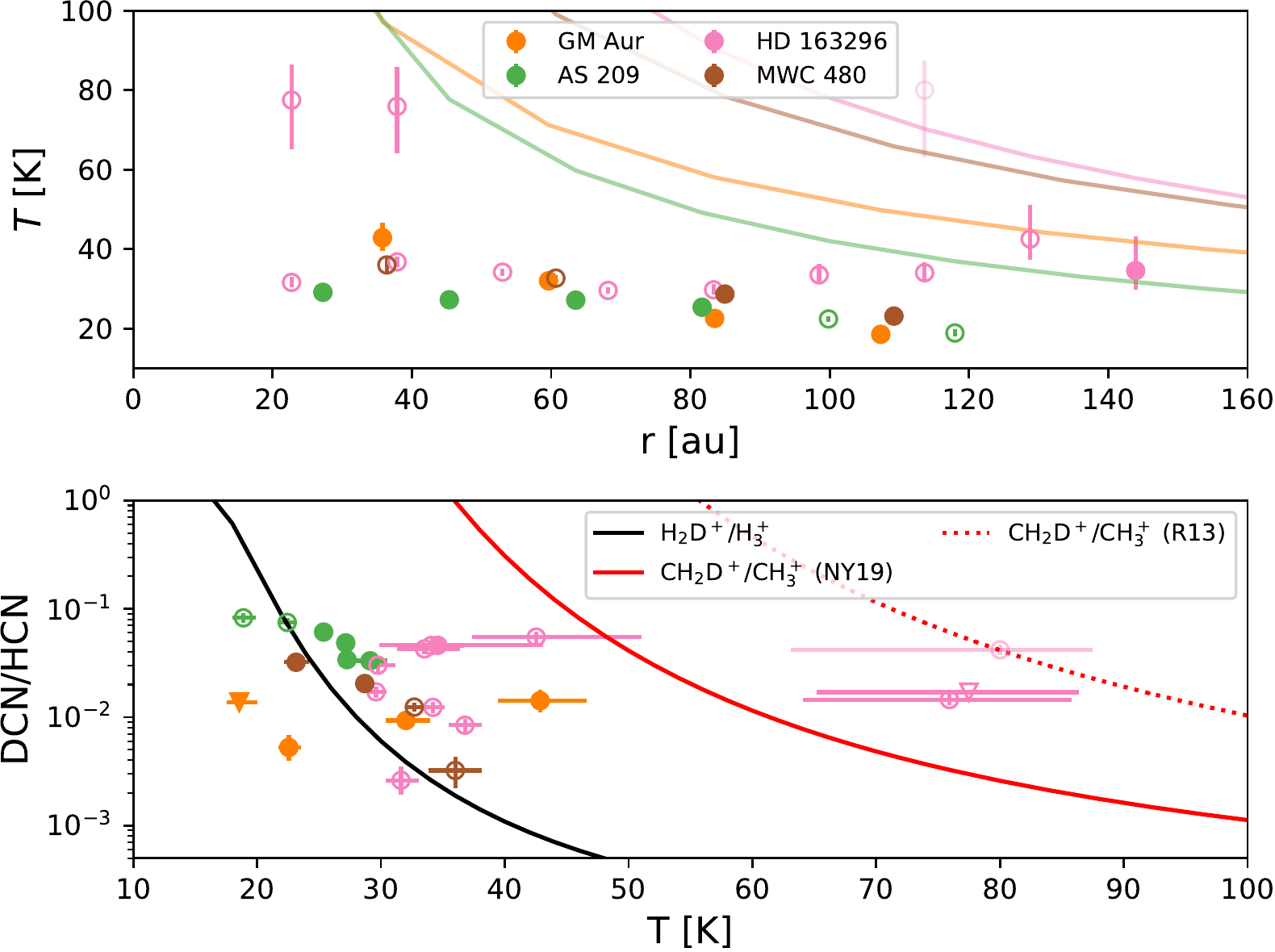}
\caption{\textit{Top}: the gas temperature as a function of radius. The solid lines show power-law fits to the $^{12}$CO 2--1 brightness temperature derived by \citet{Law21_MAPS_surfaces_vertical_distributions}. \textit{Bottom}: The DCN/HCN column density ratio as a function of gas temperature. Upper and lower triangles mark upper and lower limits corresponding to the 99th and 1st percentiles, respectively. The black and red lines are upper limits on the HCN deuteration from the low- and high-temperature pathways, respectively. They show the \ce{H2D+}/H$_3^+$ and \ce{CH2D+}/CH$_3^+$ ratios calculated by balancing Equations \ref{eq:HD_exchange_reaction_lowT} and \ref{eq:HD_exchange_reaction_highT1}, respectively. For \ce{CH2D+}/CH$_3^+$, we show ratios calculated using the exothermicity of \citet[][solid]{Nyman19} and \citet[][dotted]{Roueff13}. In both panels, only data points with a relative error on the temperature smaller than 20\% are included. See the main text for more details on the data point selection. Data points from disk regions where the LTE and non-LTE temperatures disagree are plotted with open symbols to indicate the possibility of systematic uncertainty. Data points exceeding the CO temperature are plotted fainter.\label{fig:DCN2HCN_vs_T}}
\end{figure*}

In order to study the contributions of the low- and high-temperature pathways to HCN deuteration, we now look at the deuteration fraction versus the temperature estimated from the multiline analysis. In Figure \ref{fig:DCN2HCN_vs_T}, we show the temperature as a function of radius (top) and the DCN/HCN ratio as a function of temperature (bottom). We only include points with a relative error on the temperature smaller than 20\%. This is a somewhat arbitrary choice, but we found that a stricter cutoff results in only a few points available for analysis, while a more generous cutoff results in a large number of data points with poorly constrained temperatures.

If, for a given radius, the LTE and non-LTE fits give similar results (relative difference smaller than 15\% for both the fitted temperature and DCN/HCN ratio), we only consider the LTE fit. Data points within half a beam FWHM from the disk center are excluded, because the additional line broadening makes the inference of column densities and temperatures more uncertain. We further define that the upper or lower bound of the fitted DCN/HCN is unconstrained whenever the 99th percentile is larger than 10 or the 1st percentile is smaller than $10^{-4}$, respectively. We exclude data points where both bounds are unconstrained by this definition, and plot upper and lower limits whenever only the lower or upper bound is unconstrained, respectively.

In section \ref{sec:LTE_nonLTE_comparison}, we noticed that the gas temperatures derived from the LTE and non-LTE fits differ in certain regions of the disks. For the regions beyond 300\,au in IM~Lup and HD~163296, these differences can be explained by non-LTE conditions, that is, the excitation temperature does not equal the gas temperature. As a consequence, we only consider the gas temperature derived from the non-LTE fits for $r>300$\,au for IM~Lup and HD~163296, but it turns out that no non-LTE data points from those regions pass the temperature error bar cutoff. For the inner regions of AS~209, HD~163296, and MWC~480 where the LTE and non-LTE temperatures disagree as well, non-LTE seems unlikely, making the inferred temperatures uncertain. Thus, we plot data points from those regions (90 to 140\,au for AS~209, 0 to 130\,au for HD~163296, and 0 to 70\,au for MWC~480) with open symbols to indicate the possibility of systematic uncertainty. We also plot in Fig.\ \ref{fig:DCN2HCN_vs_T} power-law fits to the $^{12}$CO 2--1 brightness temperature \citep{Law21_MAPS_surfaces_vertical_distributions} for guidance, since the HCN temperature is unlikely to exceed the CO temperature.

The bottom panel of Fig.\ \ref{fig:DCN2HCN_vs_T} also shows the \ce{H2D+}/H$_3^+$ and \ce{CH2D+}/CH$_3^+$ ratios calculated by balancing Equations \ref{eq:HD_exchange_reaction_lowT} (low-temperature pathway) and \ref{eq:HD_exchange_reaction_highT1} (high-temperature pathway), respectively. We assumed a thermalized ortho-to-para ratio of H$_2$, which is theoretically expected for protoplanetary disks \citep{Aikawa18,Furuya19}. For the \ce{CH2D+}/CH$_3^+$ ratio, we show two curves, corresponding to the exothermicities by \citet{Roueff13} and \citet{Nyman19}. The \ce{H2D+}/H$_3^+$ and \ce{CH2D+}/CH$_3^+$ ratios are upper limits on the HCN deuteration by the low- and high-temperature pathways, respectively, since, for example, the destruction of \ce{H2D+} by CO or the destruction of \ce{CH2D+} by electrons is neglected\footnote{The effect of the destruction of \ce{H2D+} by CO is small (a factor of two at most), because at temperatures above 20\,K and with a thermalized \ce{H2} ortho/para ratio, the main destruction partner of \ce{H2D+} is \ce{H2} as long as CO/\ce{H2} is lower than $10^{-4}$. Quantifying the \ce{CH2D+} destruction by electrons is not straightforward, because we do not know the electron abundances in disks.}. We see that for a considerable number of points, the low-temperature pathway alone cannot explain the observed deuteration fraction. Thus, a contribution from the high-temperature pathway is needed.

Naively, we might expect that the deuteration should decrease with increasing temperature. There are suggestions of such a trend with temperature in the bottom panel of Fig.\ \ref{fig:DCN2HCN_vs_T} for the data points from AS~209 and MWC~480. However, the presence of two deuteration pathways as well as possible systematic uncertainties in the derived temperatures complicates the interpretation. More data points with a well-constrained temperature and DCN/HCN ratio would be necessary to study the temperature dependence of the deuteration in more detail.

\subsubsection{Is the drop of DCN/HCN in the inner disks an artefact caused by high dust optical depth?}
Here we consider the possibility that the observed decrease of the DCN/HCN ratio toward the disk centers is merely an artifact caused by high dust optical depth. Optically thick dust can block line photons from escaping \citep[e.g.][]{Cleeves16}. Furthermore, continuum subtraction can lead to underestimated line flux \citep{Weaver18}. Naively, we might expect that taking the ratio of DCN/HCN at least partially cancels out such effects, but since DCN and HCN are not necessarily perfectly cospatial, there is a possibility that DCN is more strongly affected.

Both of the effects described above would occur in regions of high dust optical depth. Thus, we consider the dust optical depth at ALMA Band 6 wavelengths ($\sim$1.3\,mm), which should be higher than in Band 3 ($\sim$3\,mm). Previous studies suggested that the dust is mostly optically thin in ALMA Band 6, with the exception of the inner 10\,au and 50\,au for IM~Lup and MWC~480, respectively \citep{Huang18,Liu19,Huang20}. However, these studies did not consider the effects of scattering. \citet{zhu19} found that dust scattering can considerably reduce the emission from an optically thick region, which leads to underestimation of the dust optical depth. Therefore, we consider radial profiles of the vertical\footnote{All MAPS disk have fairly low inclination, ranging from 35$\arcdeg$ for AS~209 to 53$\arcdeg$ for GM~Aur \citep{Oberg21_MAPS}.} dust optical depth derived from models including scattering by \citet{Sierra21_MAPS}. These models show that the dust emission is optically thin for $r\gtrsim$60\,au. The exception is the disk around IM~Lup that could be optically thick even for $r>150$\,au. Generally, we find that the DCN/HCN ratio starts to decrease at radii well beyond 60\,au, even when considering the finite beam size. In addition, there is no clear correlation between the DCN/HCN profiles and the dust optical depth profiles by \citet{Sierra21_MAPS}. This suggests that dust optical depth is not the dominant cause for the decrease of DCN/HCN toward the disk centers.

\subsubsection{Comparison with the DCN/HCN ratio in comets}\label{sec:comparison_to_comets}
In this section, we attempt to connect our results to the formation history of cometary bodies. Unfortunately, the number of comets with a measured DCN/HCN ratio is limited \citep[e.g.,][]{Bockelee-Morvan15}. For comet Hale-Bopp, $\mathrm{DCN/HCN}= (2.3 \pm 0.4)\times 10^{-3}$ \citep{Meier98,Crovisier04}. Upper limits of DCN/HCN$<10^{-2}$ are reported for C/1996 B2 \citep[Hyakutake,][]{Bockelee-Morvan98} and 103P/Hartley 2 \citep{Gicquel14}. The DCN/HCN ratios measured in our sample of protoplanetary disks are typically $\gtrsim10^{-2}$, with the exception of the inner $\sim$50\,au of the HD~163296 and MWC~480 disks (see Fig.\ \ref{fig:deuteration_profiles}). The higher DCN/HCN ratios we derive compared with Hale-Bopp suggest that comets do not directly form from the bulk of the gas reservoir we observe. The simplest explanation might be that comets form from a gas reservoir located closer to the star than we probe here \citep{Ceccarelli14}. Indeed, in the framework of the Nice model, both the Oort cloud (the reservoir of long-period comets) and the scattered disk (the reservoir of short-period, that is, Jupiter-family comets) originate from dynamical scattering of objects originally located in the Uranus-Neptune zone (at $\sim$30\,au) by the migrating giant planets \citep{Brasser13}. Although the resolution of our data is insufficient to put strong constraints on the DCN/HCN in these inner regions of the disks, the observed trends of decreasing DCN/HCN toward the disk centers of IM~Lup, AS~209, HD~163296, and MWC~480 are consistent with this picture. On the other hand, in the inner $\sim$100\,au of the disk around GM~Aur, DCN/HCN is increasing toward the star, making such a scenario less likely for this target.

An alternative interpretation would be that comets do actually form in the outer disk regions but not from the gas we are observing here. Instead, they might form from ISM ices that were incorporated into the disk. The origin of cometary material (direct inheritance of ISM ice versus condensation in the protoplanetary disk, or both) is still actively debated in the literature \citep[e.g.][and references therein]{Bockelee-Morvan15,Willacy15,Rubin20}. HCN and DCN are chemically stable and have relatively high desorption energies \citep{Noble13}. Thus, a significant amount of HCN and DCN ice formed in molecular clouds could survive in the disk and be incorporated into comets.

We expect that radially resolved deuteration profiles as presented in this work will be useful to constrain models of comet formation in the solar nebula such as those presented by \citet{Mousis00}.

\subsubsection{Comparison with the DCN/HCN ratio in earlier evolutionary stages}
The DCN/HCN column density ratios inferred for class~I young stellar objects by \citet{LeGal20} and \citet{Bergner20_volatile_evolution} are within the range of values found for our sources. The DCN/HCN ratios inferred for even earlier evolutionary stages of star formation are also comparable to the results from our work: infrared dark clouds and hot molecular cores \citep{Gerner15} as well as high-mass star-forming clumps \citep{Feng19} show DCN/HCN ratios roughly of the order of $10^{-2}$. Considering the relatively high desorption energy of HCN \citep[3370 K,][]{Noble13}, these single-dish observations should mostly probe HCN and DCN formed in the gas phase rather than sublimated from ice (the sublimation zone is small compared with the beam). In summary, this suggests that the gas-phase deuteration chemistry is similar between a wide range of ISM environments and protoplanetary disks.

\subsection{\texorpdfstring{\ce{N2H+}}{N2H+} and \texorpdfstring{\ce{N2D+}}{N2D+}}

\subsubsection{Comparison with previous observation of \texorpdfstring{\ce{N2D+}}{N2D+} and theoretical predictions}
\ntdp has previously been detected toward only two protoplanetary disks \citep{Huang15,Salinas17}. \citet{Huang15} observed \ntdp 3--2 toward the disk around AS~209 with ALMA and compared it with a Submillimeter Array observation of \nthp by \citet{Oberg11_SMA}. Because the resolution ($\sim$1$\arcsec$) and the sensitivity were not high enough to resolve the disk structure, they derived a disk-averaged column densities for two fixed excitation temperatures: 10 and 25\,K. They find column densities of \ntdp and \nthp of $(1.4-2.1) \times 10^{11}$\,cm$^{-2}$ and $(3.1-6.3)\times 10^{11}$\,cm$^{-2}$, respectively. The \ntdp/\nthp column density ratio is then estimated as 0.3--0.5. While we derive higher column densities, our \nthp deuteration ratio agrees with their disk-averaged value (Fig.\ \ref{fig:deuteration_profiles}).

\citet{Salinas17} observed the \ntdp 3--2 transition toward the HD~163296 disk with ALMA and derived a disk-averaged \ntdp column density of $(2.5\pm 0.3)\times 10^{11}$\,cm$^{-2}$ for the assumed excitation temperature of 10\,K (for 25\,K, they derive $(1.6\pm 0.2) \times 10^{11}$\,cm$^{-2}$). They also estimated the \nthp column density from the \nthp 3--2 observation by \citet{Qi15}, and derived an \ntdp/\nthp column density ratio of $0.34\pm 0.15$ and $0.45\pm 0.21$ for the assumed excitation temperatures of 10 and 25\,K, respectively. We derive a slightly higher \ntdp column density and \ntdp/\nthp ratio (Figures \ref{fig:N2H+_N_tau} and \ref{fig:deuteration_profiles}). One reason could be that their values are averaged over the whole disk, including the inner disk, which shows negative emission, while our higher-resolution observations allow us to ignore that inner region. Indeed, their integrated \ntdp 3--2 flux (from which they estimate the column density) is significantly lower compared with our value ($61.6\pm7.5$\,mJy\,km\,s$^{-1}$, compared with our $135\pm15$\,mJy\,km\,s$^{-1}$, where the latter is derived by excluding the inner 50\,au, see Table \ref{tab:disk_integrated_fluxes}).

The radial emission profile of \ntdp 3--2 in \citet{Salinas17} also shows a morphology similar to our data (see Figure \ref{fig:N2H+_radial_profiles_gallery}): a ring extending from $\sim$100 to 300\,au. The line flux is negative in the inner disk region in both the \citet{Salinas17} and our profile. This negative emission is probably due to the difficulty of achieving an accurate continuum subtraction, possibly caused by an atmospheric absorption feature. We confirmed, however, that our \ntdp profile is not affected by this issue outside of 50\,au. See Appendix \ref{appendix:negative_flux} for a more detailed discussion.

Based on the chemical disk modeling by \citet{Aikawa18}, \ntdp is expected to be present mainly in the cold outer regions ($\gtrsim$ 50\,au) of the disk. We find that for all of our sources where \ntdp is detected, it is indeed distributed in the outer regions, peaking beyond $\sim$ 50\,au. This is consistent with the theoretical expectation that the deuteration of \nthp proceeds via the low-temperature channel. In addition to the radial profile, the N$_2$D$^+$/N$_2$H$^+$ ratio derived from the observations is in reasonable order of magnitude agreement with theoretical models. In \citet{Aikawa18}, the ratio reaches almost unity in the outer disk, while it declines to $\lesssim0.1$ inside a radius of $\sim$100\,au.

In summary, we find that both previous observations and predictions by chemical disk models are broadly consistent with our analysis of the \nthp and \ntdp data. The overall picture of efficient \nthp deuteration in the cold midplane of the outer disk is confirmed.

\subsubsection{\texorpdfstring{\ce{N2D+}}{N2D+} as CO snow line tracer}
\begin{figure*}
\epsscale{1.1}
\plotone{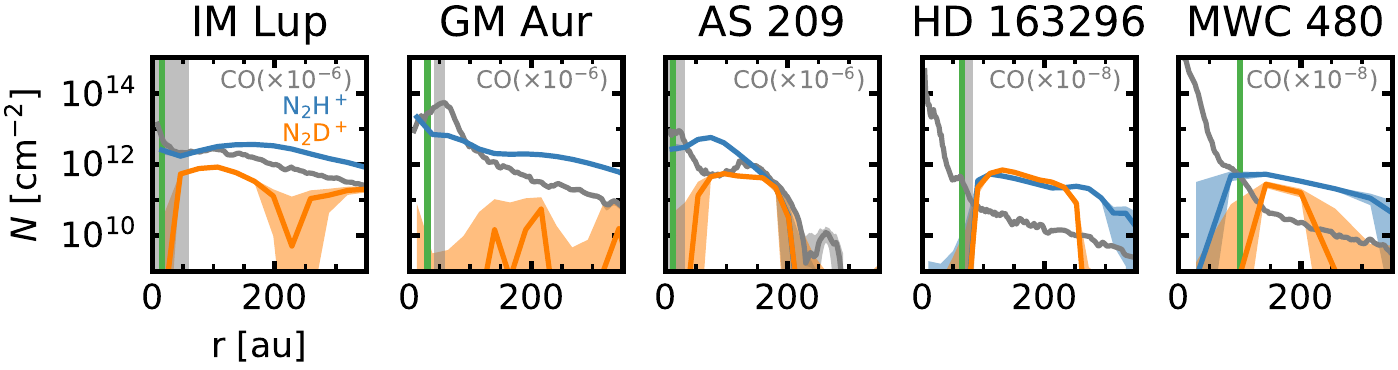}
\caption{The column density profiles of \nthp (blue), \ntdp (orange), and CO (gray, scaled by $10^{-8}$ for HD~163296 and MWC~480 and $10^{-6}$ for the other targets). The CO column densities are from \citet{Zhang21_MAPS}. The color-shaded regions extend from the 16th to 84th percentile of the MCMC fits for \nthp and \ntdp, and over the $1\sigma$ uncertainty for CO. The vertical green bars indicate the CO snow line locations inferred from the MAPS reference models \citep{Zhang21_MAPS}. The vertical gray-shaded regions indicate the CO snow line locations estimated from \nthp observations by \citet{Qi15, Qi19}.\label{fig:CO_and_N2H+}}
\end{figure*}

In this section, we discuss the prospects of using \ntdp as a CO snow line tracer. Determining the CO snow line from observations of CO emission lines is generally challenging, because even outside of the CO snow line, there is a warm molecular layer with CO gas. This can be seen in Figure \ref{fig:CO_and_N2H+}, which shows the CO column density profiles for the MAPS disks determined from \ce{C^{18}O} observations \citep{Zhang21_MAPS}. The vertical green bars indicate the model estimates for the CO snow line, that is, the radii where the abundances of CO gas and CO ice become the same in the MAPS reference models \citep{Zhang21_MAPS}. In the two disks around the Herbig stars, HD~163296 and MWC~480, the CO column density profiles rapidly increase inward across the CO snow lines, tracing CO evaporation \citep{Zhang21_MAPS}. However, no clear trend is seen for the other disks, that is, the snow line cannot be determined from the CO column density profile.

Instead, the snow line is usually determined from indirect tracers. In particular, \nthp has been used to trace the CO snow line \citep{Qi13,Qi15,Qi19}. Since \nthp is destroyed by CO (\nthp + CO $\rightarrow$ HCO$^+$ + N$_2$), \nthp is expected to be abundant outside of the CO snow line. Therefore, the inner edge of the \nthp emission can be used as an estimate of the CO snow line. However, models by \citet{vantHoff17} and \citet{Aikawa18} predict that \nthp can, in addition to the midplane region, also form in a surface layer of the disk. This surface layer can make the inference of the CO snow line location considerably more uncertain \citep{vantHoff17}. Thus, \ntdp might be a better snow line tracer, since no surface layer is expected to form \citep{Aikawa18}. Here we test how well \ntdp traces the CO snow line.

In Figure \ref{fig:CO_and_N2H+}, we plot the column density profiles of \nthp and \ntdp (remember that the column densities might carry additional systematic uncertainties for $r\lesssim 50$\,au due to severe broadening; see Section \ref{sec:N2H+_N2D+_fitting}), and the CO snow line estimate derived from \nthp observations \citep[gray-shaded area,][]{Qi15,Qi19} and from the MAPS reference models \citep[green bar,][]{Zhang21_MAPS}. Except for GM~Aur, where the SNR of \ntdp is low, the estimates for the CO snow line approximately correspond to the inner edges of the \ntdp column density profiles. Unfortunately, an accurate measurement of the inner edge of \ntdp can be challenging since it is difficult to achieve an accurate continuum subtraction for the \ntdp 3--2 line, which can result in unphysical negative emission in the inner disk region (see Appendix \ref{appendix:negative_flux}). Furthermore, \ntdp emission is generally weaker than \nthp. Still, our results suggest that \ntdp can in principle be used as an alternative or complement to \nthp for constraining the CO snow line.

\subsection{Outer emission rings of DCN, \texorpdfstring{\ce{N2D+}}{N2D+} and \texorpdfstring{\ce{DCO+}}{DCO+}}\label{sec:DCN_N2D+_DCO+_comparison}
\begin{figure*}
\epsscale{1.1}
\plotone{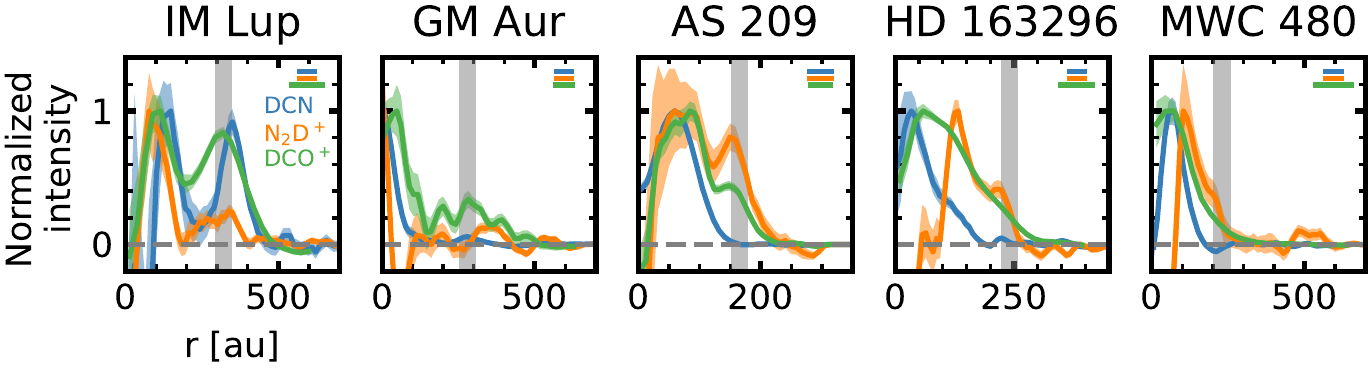}
\caption{Comparison of the radial emission profiles of DCN 3--2 (blue), \ce{N2D+} 3--2 (orange), and \ce{DCO+} (green). For the latter, we show the 4--3 transition in GM~Aur and the 3--2 transition in the other disks. The vertical gray lines indicate the outer edge of the millimeter dust continuum emission determined by \citet{Law21_MAPS_radial_profiles}. The size of the beam major axis for each emission line is indicated by a colored horizontal line in the upper right corner of each panel.\label{fig:DCN_N2D+_DCO+}}
\end{figure*}

In this section, we discuss ring structures seen in the radial emission profiles of the deuterated molecules DCN, \ntdp and \ce{DCO+}. In Figure \ref{fig:DCN_N2D+_DCO+}, we show DCN 3--2 and \ntdp 3--2 from this work (0.3$\arcsec$ resolution) together with \ce{DCO+} 4--3 for GM~Aur (PI C.\ Qi, project code 2015.1.00678.S, 0.35$\arcsec$ resolution) and \ce{DCO+} 3--2 for the other disks \citep[][0.26$\arcsec$--0.7$\arcsec$ resolution]{Huang17,Favre19}. The \ce{DCO+} emission profiles were generated from the archival data in the same way as for the MAPS data, that is, by azimuthal averaging of a zeroth moment map generated with a Keplerian mask \citep{Law21_MAPS_radial_profiles}.

The disks around the T~Tauri targets IM~Lup, GM~Aur and AS~209 all show outer rings (i.e., a ring beyond the most central emission component) in at least two of the three deuterated molecules. IM~Lup shows a double-ring structure in all three deuterated molecules. For GM~Aur, both DCN and \ce{DCO+} have an outer ring at $\sim$300\,au. For AS~209, \ce{N2D+} and \ce{DCO+} show an outer ring at $\sim$150\,au. In addition, by employing a deconvolution technique, \citet{Flaherty17} revealed an outer \ce{DCO+} ring at 215\,au\footnote{We scaled the radial distance of the ring to account for the different distance of HD~163296 assumed in our study (101\,pc) compared to \citet[][122\,pc]{Flaherty17}.} in HD~163296, in good agreement with the shoulder of the \ntdp emission profile \citep[see also][]{Salinas18}. Interestingly, all of these outer rings coincide with the edge of the millimeter dust continuum, which is indicated by the vertical gray line in Figure \ref{fig:DCN_N2D+_DCO+}. There is also a shoulder in the \ntdp profile of MWC~480 just inside the millimeter dust edge.

The MAPS disks are not the only sources where rings of deuterated molecules are observed. A DCN ring at the millimeter dust edge is also seen in the disk around LkCa~15 \citep{Huang17}. Furthermore, non-deuterated molecules can show similar associations to the dust edge: \citet{Law21_MAPS_radial_profiles} find such behavior among some of the MAPS sources for HCN and \ce{H2CO}. \citet{Carney17}, \citet{Pegues20} and \citet{Guzman21_MAPS} also report examples of enhanced \ce{H2CO} emission at the dust edge.

\citet{Oberg15} suggested that the outer \ce{DCO+} ring in the IM~Lup disk results from nonthermal desorption of CO ice by UV photons at the edge of the dust disk where the UV penetration depth increases. The newly supplied CO gas then reacts with \ce{H2D+}, which is abundant due to the low temperature in these outer regions, to form \ce{DCO+}. In addition, CO might return to the gas phase at these large radii as a result of a thermal inversion: generic disk models by  \citet{Cleeves16_COThermalInversion} and \citet{Facchini17} predict that the removal of large grains due to radial drift leads to an increase in dust temperature at the dust disk edge.

Similar to \ce{DCO+} forming from photodesorbed CO, an \ce{N2D+} ring might be formed from photodesorbed \ce{N2} \citep{Oberg09} that reacts with \ce{H2D+}. To get a ring in both \ntdp and \ce{DCO+}, the CO gas phase abundance should be neither too high (otherwise \ntdp gets destroyed) nor too low \citep[otherwise \ce{DCO+} will not form efficiently, e.g.,][]{Pagani11}. As for DCN, in cold gas it forms from \ce{DCNH+ + e- -> DCN + H}. \ce{DCNH+} is produced by the reaction of HNC with either \ce{DCO+} or \ce{D3+} \citep{Willacy07}. Like the CO required for \ce{DCO+} formation, the HNC might originate from UV photodesorption from icy grains or alternatively be formed from other species (e.g.\ HCN) that are photodesorbed. No direct laboratory measurements of HNC or HCN photodesorption are available, but naively we might expect that the photodesorption yield is roughly similar to other molecules such as CO, \ce{H2O}, or \ce{N2} for which those measurements exist \citep{Cuppen17}. A thermal inversion could also prompt the desorption of \ce{N2} and other precursor molecules of the deuterated species discussed here.

Not all deuterated molecules show rings at the dust disk edge in all five disks. For example, there is no DCN ring at the millimeter dust edges of AS~209 and MWC~480. This suggests that the mechanism(s) for forming deuterated molecules at the dust disk edge depend on disk properties.

\section{Summary}\label{sec:summary}
In this work, we study the distribution of the deuterated molecules DCN and \ce{N2D+} in five protoplanetary disks observed as part of the ALMA MAPS Large Program. We combine the observations of DCN 3--2 and \ntdp 3--2 with additional observations of HCN 1--0 and 3--2 and H$^{13}$CN 1--0 from MAPS, and \ce{N2H+} 3--2 from the archive. We model azimuthally averaged spectra (including hyperfine structure) to derive radial column density profiles, from which we determine the profiles of the DCN/HCN and \ce{N2D+}/\ce{N2H+} column density ratios. Our results are summarized as follows:
\begin{enumerate}
    \item DCN 3--2 is detected in all five sources, with the detection toward GM~Aur reported for the first time. \ce{N2D+} 3--2 is detected in four sources (IM~Lup, AS~209, HD~163296, and MWC~480), two of which (IM~Lup and MWC~480) are newly reported detections. This brings the total number of protoplanetary disks with \ce{N2D+} detected to four. In addition, we tentatively detect \ce{N2D+} 3--2 in GM~Aur.

    \item We find a variety of DCN emission profile morphologies. This hints at a complex dependence of HCN deuteration on individual disk properties such as the temperature or ionization structure.

    \item \ce{N2D+} is expected to be abundant in the cold midplane beyond the CO snow line. This is reflected in the observed radial emission profiles that all show ring-like structures. For IM~Lup, we discover a double ring that corresponds well to double rings seen for DCN and DCO$^+$. This is the first time the radial distribution of the N$_2$D$^+$ column density is derived in protoplanetary disks.

    \item The DCN/HCN ratio shows a wide range of values across our sample, extending from a few times $10^{-3}$ in the inner regions of HD~163296 and MWC~480 to $\sim$0.2 in the outer ring of IM~Lup. We observe a general trend of decreasing DCN/HCN when moving from the outer disk regions toward the disk center. The strongest decrease of almost two orders of magnitude is seen for the disks around the Herbig stars HD~163296 and MWC~480. We interpret this trend as a confirmation that deuteration proceeds via in situ exchange reactions, because the backward endothermic reactions should become more efficient at the higher temperatures expected in the inner disk.

    \item Even for temperatures $\gtrsim$30\,K, we find DCN/HCN column density ratios above $10^{-2}$, which cannot be explained by the low-temperature deuteration channel alone. We conclude that both the low- and high-temperature deuteration channels are active.

    \item The derived DCN/HCN column density ratios are mostly larger than observed for comet Hale-Bopp \citep{Meier98}. This suggests that either Hale-Bopp originated from smaller radii than probed here ($\lesssim30$\,au), or inherited its HCN from the ice of the prestellar cloud.

    \item We find \ce{N2D+}/\ce{N2H+} column density ratios typically between a few times $10^{-2}$ and 1, consistent with model predictions of efficient \nthp deuteration in the cold midplane of the outer disk. The disks around IM Lup and HC~163296 show a relatively flat \ce{N2D+}/\ce{N2H+} profile, while the ratio strongly increases outward between $\sim$50 and $\sim$170\,au for AS~209.

    \item The inner edge of the observed \ce{N2D+} rings is consistent with the snow line locations inferred from \nthp observations \citep{Qi15,Qi19} and predicted by the MAPS reference models \citep{Zhang21_MAPS}. This suggests that \ntdp might be a useful tracer of the CO snow line. However, a careful continuum subtraction is essential to derive the inner edge of \ntdp 3--2 (see appendix \ref{appendix:negative_flux}).

    \item All disks show rings or shoulders in the emission profiles of at least one of the deuterated molecules DCN, \ntdp and \ce{DCO+} at the edge of the millimeter dust disks. These rings are probably linked to an increased UV flux due to the decreasing dust optical depth, with UV photodesorption releasing CO, \ce{N2} and other precursor molecules. In addition, a thermal inversion at the dust disk edge might prompt thermal desorption of ices that results in the observed rings.

\end{enumerate}

\acknowledgments
We would like to thank the anonymous referee for an exceptionally careful review of our manuscript that helped us to clarify many aspects of our paper. We are grateful to Satoshi Yamamoto and Yoko Oya for helpful discussions on spectroscopic properties of the observed lines. We thank C\'{e}cile Favre for sharing the \ce{DCO+} data of AS~209. We also thank Marie-Lise Dubernet and Yaye Awa Ba for help with \texttt{SPECTCOL}, and Holger S.\ P.\ M\"uller for help with the CDMS. This paper makes use of the following ALMA data: ADS/JAO.ALMA\#2018.1.01055.L,\\ADS/JAO.ALMA\#2015.1.00678.S,\\ ADS/JAO.ALMA\#2012.1.00681.S,\\ADS/JAO.ALMA\#2015.1.00657.S,\\ ADS/JAO.ALMA\#2013.1.00226.S,\\ADS/JAO.ALMA\#2015.1.00486.S. ALMA is a partnership of ESO (representing its member states), NSF (USA) and NINS (Japan), together with NRC (Canada), MOST and ASIAA (Taiwan), and KASI (Republic of Korea), in cooperation with the Republic of Chile. The Joint ALMA Observatory is operated by ESO, AUI/NRAO and NAOJ. This research has made use of NASA’s Astrophysics Data System and the SIMBAD database, operated at CDS, Strasbourg, France.

G.C. is supported by the NAOJ ALMA Scientific Research grant code 2019-13B. Y.Y. is supported by IGPEES, WINGS Program, the University of Tokyo. Y.A. acknowledges support by NAOJ ALMA Scientific Research grant code 2019-13B and Grant-in-Aid for Scientific Research (S) 18H05222, and Grant-in-Aid for Transformative Research Areas (A) 20H05844 and 20H05847. J.B.B., J.H., I.C., K.R.S., and K.Z. acknowledge the support of NASA through Hubble Fellowship grants HST-HF2-51429.001-A, HST-HF2-51460.001-A, HST-HF2-51405.001-A, HST-HF2-51419.001, and HST-HF2-51401.001, awarded by the Space Telescope Science Institute, which is operated by the Association of Universities for Research in Astronomy, Inc., for NASA, under contract NAS5-26555. V.V.G. acknowledges support from FONDECYT Iniciaci\'on 11180904 and ANID project Basal AFB-170002. J.H. and S.M.A. acknowledge funding support from the National Aeronautics and Space Administration under grant No.\ 17-XRP17 2-0012 issued through the Exoplanets Research Program. E.A.B. and A.D.B. acknowledge support from NSF AAG grant No.\ 1907653. A.S.B. acknowledges the studentship funded by the Science and Technology Facilities Council of the United Kingdom (STFC). L.I.C. gratefully acknowledges support from the David and Lucille Packard Foundation and Johnson \& Johnson's WiSTEM2D Program. J.D.I. acknowledges support from the Science and Technology Facilities Council of the United Kingdom (STFC) under ST/T000287/1. C.J.L. acknowledges funding from the National Science Foundation Graduate Research Fellowship under grant No.\ DGE1745303. R.L.G. acknowledges support from a CNES fellowship grant. Y.L. acknowledges the financial support by the Natural Science Foundation of China (grant No.\ 11973090). F.L. and R.T. acknowledge support from the Smithsonian Institution as a Submillimeter Array (SMA) Fellow. F.M. acknowledges support from ANR of France under contract ANR-16-CE31-0013 (Planet-Forming-Disks) and ANR-15-IDEX-02 (through CDP ``Origins of Life"). H.N. acknowledges support by NAOJ ALMA Scientific Research grant code 2018-10B and Grant-in-Aid for Scientific Research 18H05441. K.I.\"O. acknowledges support from the Simons Foundation (SCOL \#321183) and an NSF AAG Grant (\#1907653). T.T. is supported by JSPS KAKENHI grant Nos.\ JP17K14244 and JP20K04017. C.W. acknowledges financial support from the University of Leeds, STFC and UKRI (grant Nos.\ ST/R000549/1, ST/T000287/1, MR/T040726/1). K.Z. acknowledges the support of the Office of the Vice Chancellor for Research and Graduate Education at the University of Wisconsin–Madison with funding from the Wisconsin Alumni Research Foundation.

\vspace{5mm}
\facilities{ALMA}

\software{\texttt{astropy} \citep{astropy13,astropy18}, \texttt{bettermoments} \citep{Teague18_bettermoments}, \texttt{CASA} \citep{McMullin07}, \texttt{CBcolors.py} (\url{https://gist.github.com/thriveth/8560036}), \texttt{corner.py} \citep{Foreman-Mackey16}, \texttt{emcee} \citep{Foreman-Mackey13}, \texttt{gofish} \citep{Teague19_gofish}, \texttt{matplotlib} \citep{Hunter07}, \texttt{NumPy} \citep{vanderWalt11}, \texttt{pythonradex} (\url{https://github.com/gica3618/pythonradex}), \texttt{SciPy} \citep{Virtanen20}, \texttt{SPECTCOL} (VAMDC Consortium, \url{http://www.vamdc.org}), \texttt{VISIBLE} \citep{Loomis18}.}

\appendix

\section{Brief summary of the JvM correction}\label{appendix:JvM_correction_summary}
We first briefly summarize the CLEAN algorithm in order to set the stage for the JvM correction. CLEAN first initializes a ``residual image" equal to the dirty image (i.e.,\ the Fourier transform of the calibrated visibilities), and a ``CLEAN model" equal to a blank image. Then a CLEAN component is placed into the CLEAN model at the position of the current peak in the residual map. At the same time, the newly placed CLEAN component is deconvolved from the residual map by subtracting its convolution with the dirty beam. This procedure is repeated until a stopping criterion is reached, which in the case of MAPS is a threshold of the peak SNR in the residual map. Then the CLEAN model is convolved with the CLEAN beam, which is a Gaussian fit to the dirty beam. Lastly, the residual image is added to the convolved CLEAN model to obtain the final image cube. This last step is necessary because the residual image still contains astrophyical flux below the noise threshold. A more detailed discussion as well as visualization of the CLEAN algorithm is presented by \citet{Czekala21_MAPS}.

A problem can arise in the final step of the workflow described above: the CLEAN model has units of flux/(CLEAN~beam), while the residual map has units of flux/(dirty~beam). Adding together these two maps produces a final image with inconsistent flux units. This is not a concern as long as the difference between the dirty and CLEAN beams is small. However, the combination of visibilities from short and long baseline antenna configurations (such as for MAPS) can lead to a significantly non-Gaussian dirty beam and therefore an inconsistent flux scale. Thus, MAPS employed a correction first discussed by \citet[][hereafter JvM correction]{Jorsater95} and also used by, for example, \citet{Walter99}, \citet{Walter08}, \citet{Koda19} and \citet{Pinte20}. MAPS implemented the correction by changing the units of the residual map to flux/(CLEAN beam) prior of combining it with the CLEAN model \citep{Czekala21_MAPS}. This is achieved by scaling the residual map by the ratio of the beam areas,
\begin{equation}
\epsilon = \frac{\Omega_\mathrm{CLEAN}}{\Omega_\mathrm{dirty}}
\end{equation}
The more $\epsilon$ deviates from 1, the more severe the mismatch between the CLEAN and dirty beam is, and the more important the JvM correction becomes. We list the $\epsilon$ factors for the image cubes used in this study in Tables \ref{tab:HCN_image_params} and \ref{tab:N2Hp_image_params}.

The noise level of the final image cube is set by the residual map. Without the JvM correction, the noise level in the final cube is wrong by a factor $\epsilon$. Furthermore, for weak lines (such as those from DCN or \ntdp), a significant fraction of the total flux is below the noise threshold and thus resides in the residual map. For those cases, the JvM correction is crucial to determine the correct line flux. As an illustration, \citet{Czekala21_MAPS} consider a channel of DCN 3--2 toward AS~209. About half of the total flux resides in the residual map. With an $\epsilon=0.36$, this implies a $\sim$50\% difference between the total fluxes inferred from the corrected and uncorrected images, respectively.

\section{Negative emission of \texorpdfstring{\ce{N2D+}}{N2D+} 3--2}\label{appendix:negative_flux}
The radial profiles of the \ntdp 3--2 emission are strongly negative inside of $\sim$50\,au for HD~163296 and MWC~480. The negative emission is much deeper than the statistical error bars (see Figure \ref{fig:N2H+_radial_profiles_gallery}), implying that this feature is highly unlikely to be just random noise. Furthermore, the \ntdp 3--2 profile for HD~163296 derived by \citet{Salinas17} from an independent ALMA data set shows similarly negative emission, further suggesting that this is a systematic effect. The fact that both sources show strong continuum emission within $\sim$50\,au \citep{Huang18,Liu19,Sierra21_MAPS} suggests that the effect is related to the continuum subtraction \citep{Salinas17}. The difficulties with the continuum subtraction of \ntdp 3--2 are probably due to the fact that the line is located at the edge of a strong atmospheric feature, which may affect the bandpass calibration.

\begin{figure}
\plotone{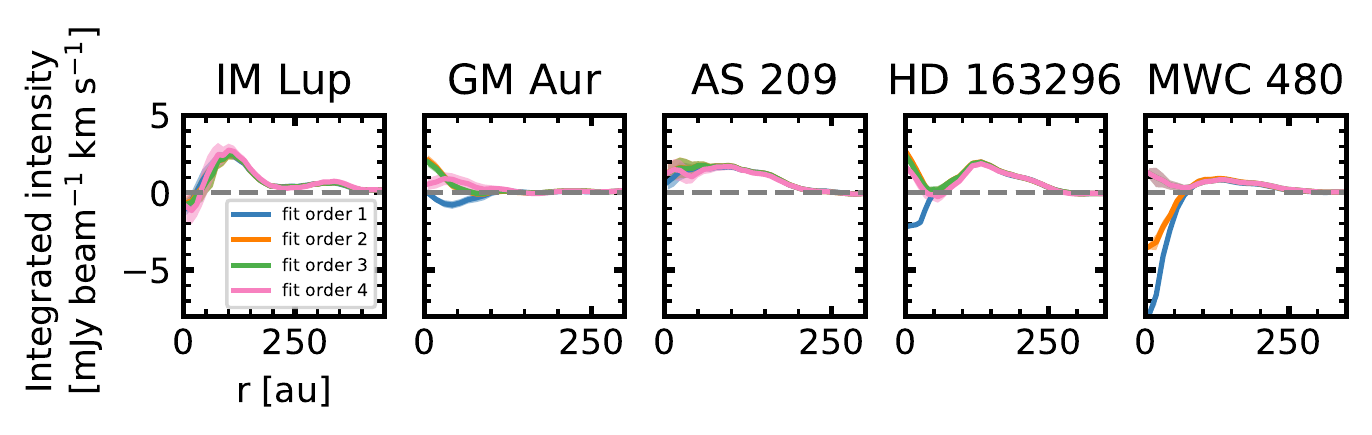}
\caption{Radial emission profiles of \ntdp 3-2 for different orders of the polynomials used to subtract the continuum in the $uv$ plane. In this example, the polynomials were only fitted over wavelengths without line emission. Strong variation is seen in the inner region of GM~Aur, HD~163296, and MWC~480, depending on the adopted fit order. Similar variations are seen when the polynomial fit is performed using the whole spectral window. \label{fig:N2D+_continuum_subtraction_tests}}
\end{figure}

We inspected \ntdp spectra extracted from the central disk region from a data cube where the continuum has not been subtracted. We found that the continuum indeed shows a wave-like structure, making accurate continuum subtraction difficult. The calibration procedure adapted by the MAPS team uses the \texttt{CASA} task \textit{uvcontsub} to subtract the continuum in the $uv$ plane by fitting first-order polynomials to those wavelengths without line emission \citep{Oberg21_MAPS}. For the \ntdp 3--2 data, this results in an overprediction of the continuum level at the wavelength of the line. In order to investigate the effect of the continuum subtraction on the radial emission profiles, we subtracted the continuum using \texttt{uvcontsub} with polynomials of order 1 to 4. We tested two cases: (1) fitting the polynomial to wavelengths outside of the line and (2) fitting the polynomial over all wavelengths of the spectral window, including the channels were line emission is expected. The latter approach is motivated by the fact that the \ntdp 3--2 emission is weak and might not influence the fit significantly. This might allow a more precise continuum subtraction at the wavelengths of the line emission. For both cases, we find that in the inner $\sim$50\,au of MWC~480 and HD~163296 (and also GM~Aur), the radial profiles strongly vary depending on the adopted fit order and the wavelength range chosen for the fit (Figure \ref{fig:N2D+_continuum_subtraction_tests}). On the other hand, in the outer region, the profiles are independent of the adopted fit order. This supports the hypothesis that the negative emission is due to oversubtraction of the continuum. Importantly, it also tells us that the radial profile is not affected by this problem beyond $\sim$50\,au.

\begin{figure}
\plotone{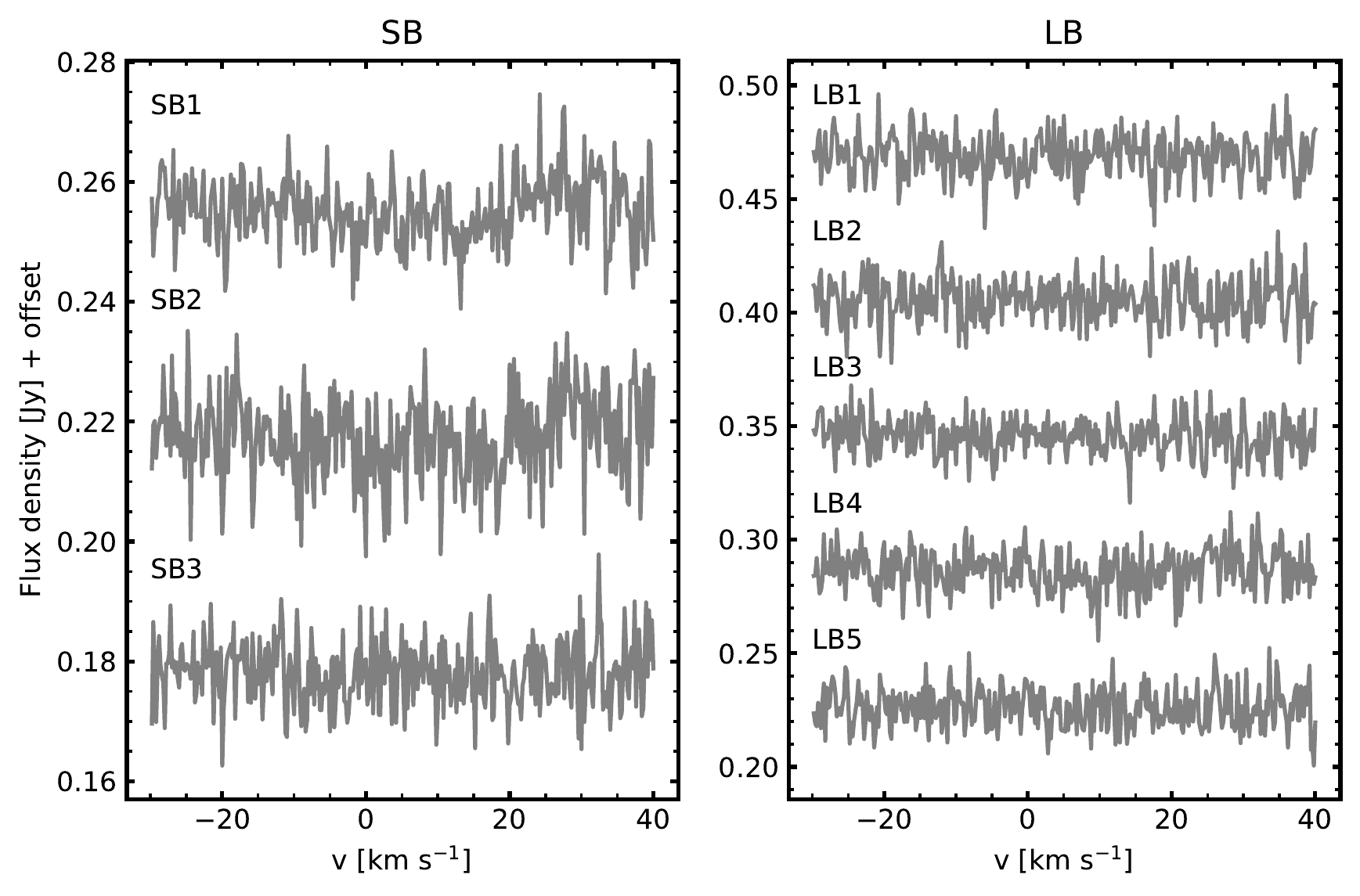}
\caption{Spectra of \ntdp 3--2 without continuum subtraction toward MWC~480 for different EBs, obtained by integrating over the central 0.3$\arcsec$ region. The observations consist of a total of eight EBs, three with short baselines (SB, i.e., a compact array configuration) and five with long baselines \citep[LB, i.e., an extended array configuration,][]{Oberg21_MAPS}. Spectra are vertically offset for clarity. \label{fig:N2D+_line_continuum_spectra}}
\end{figure}

For MWC~480, we also investigated whether the problem comes from a single execution block (EB), for example due to poor weather. We imaged the continuum-subtracted visibilities for each EB separately and derived the radial emission profiles. We did not find a correlation between the strength of the negative emission and the precipitable water vapor for individual EBs, suggesting that variable weather conditions are not the primary cause of the issue. However, we find that the short baseline (SB) executions show the strongest negative emission. We then imaged the visibilities without the continuum subtracted for each EB separately. Figure \ref{fig:N2D+_line_continuum_spectra} shows spectra of MWC~480 extracted from the innermost circular region ($\sim$ beam size) of these image cubes for each EB. The spectra from the SB executions indeed show the most obvious wavy structure of the continuum. We experimented with continuum subtractions tailored to individual EBs (e.g., excluding the high-velocity region of SB1 and SB2 for the continuum fit; see Figure \ref{fig:N2D+_line_continuum_spectra}). However, the inner region of the emission profile still showed strong variations depending on the adopted fit order and fitted wavelength range.

In summary, our tests suggest that nonoptimal continuum subtraction makes the \ce{N2D+} 3--2 radial profiles of HD~163296 and MWC~480 unreliable within the inner $\sim$50\,au. The disk beyond $\sim$50\,au is not affected by this problem, and therefore our main results still hold. However, the negative emission means that the inner edge of the \ntdp 3--2 emission is difficult to determine, which could reduce its usefulness as a CO snow line tracer.

\section{Matched filter analysis in the \texorpdfstring{$uv$}{uv} plane}\label{appendix:matched_filter}
\begin{figure}
\plotone{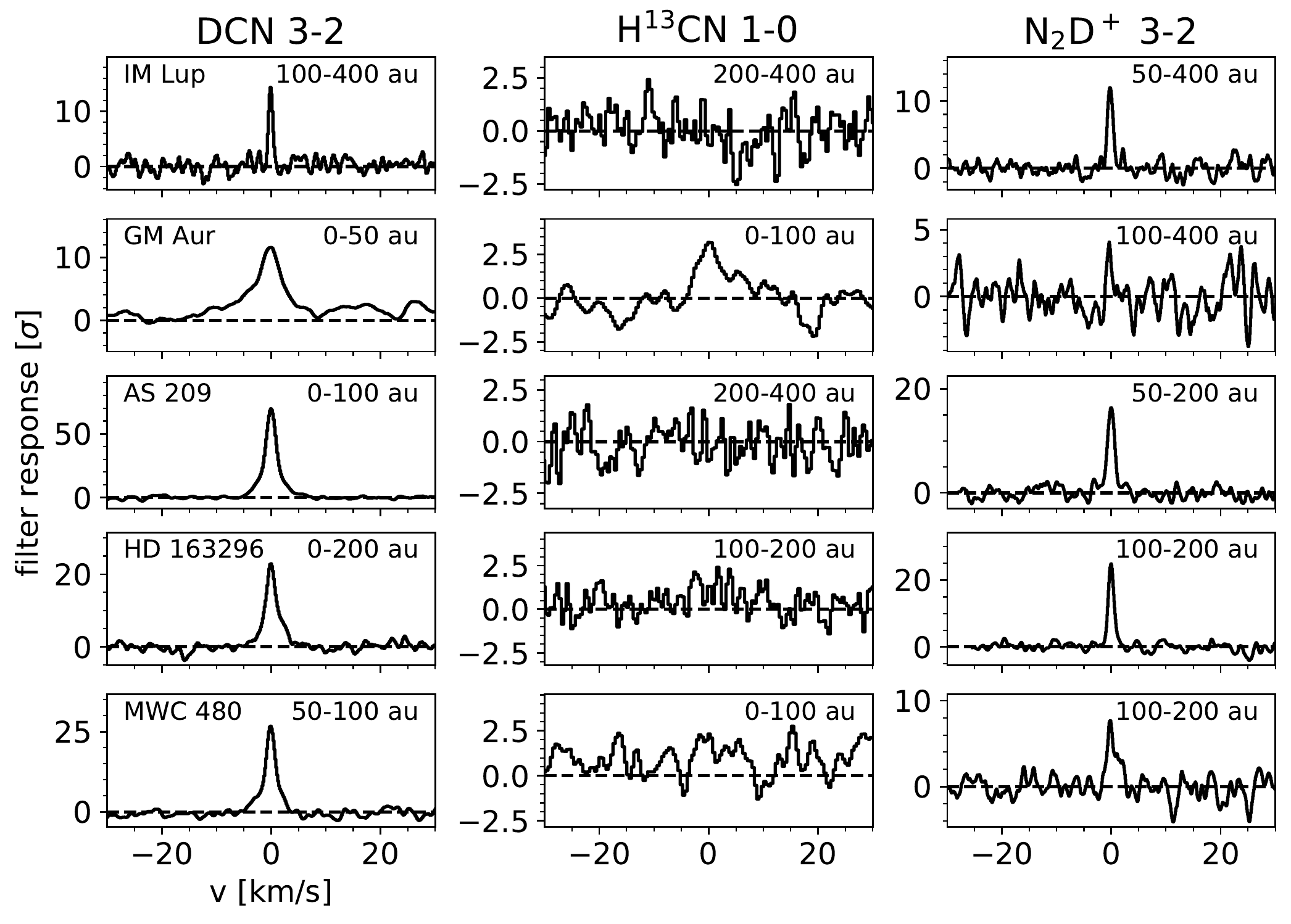}
\caption{Matched filter response for DCN 3--2, \ce{H^{13}CN} 1--0, and \ce{N2D+} 3--2. For each disk and emission line, the filter with the Keplerian mask that maximizes the SNR was chosen, as indicated in the upper right of each panel. The x-axis is centered on the systemic velocity. The y-axis is in units of the noise in the filter response.\label{fig:matched_filter}}
\end{figure}

\citet{Loomis18} proposed the application of a matched filter in the $uv$ plane to detect weak lines. With this method, visibilities are calculated from a Keplerian disk model or from a model based on another, strong emission line that is believed to have a spatial distribution similar to the weak line. The model visibilities are then cross-correlated with the observed visibilities, producing a response for each channel of the data. We applied this procedure for a series of Keplerian disk models with varying radial extents using the \texttt{VISIBLE} software \citep{Loomis18}. In Figure \ref{fig:matched_filter} we present the matched filter responses for DCN 3--2, \ce{H^{13}CN} 1--0, and \ce{N2D+} 3--2. For each disk and emission line, we selected the Keplerian mask that maximizes the filter response at the systemic velocity. DCN is detected above 10$\sigma$ in all sources. \ce{N2D+} is detected above 5$\sigma$ in all sources except GM~Aur. This is consistent with our analysis in the image plane. There is a marginal \ce{N2D+} detection in GM~Aur at 4$\sigma$. However, several peaks of similar magnitude in the spectrum cast some doubt on its significance, so we consider it only tentatively detected. \ce{H^{13}CN} is tentatively detected at 3$\sigma$ for GM~Aur and not detected in any of the other disks.

\section{Azimuthally averaged spectra}\label{appendix:az_averaged_spectra}

\subsection{Calculation of errors}\label{appendix:az_averaged_spectra_errors}
We recall that the first step for calculating azimuthally averaged spectra is to shift the spectrum at each spatial position of the data cube according to the projected Keplerian velocity (see Section \ref{sec:az_averaged_spectra}). For each velocity $v_i$, the averaged flux is then computed as an azimuthal average over a deprojected ring within channel $i$ of the (shifted) cube. The error $\epsilon_i$ on the averaged flux is given by
\begin{equation}\label{eq:error_mean_spectra}
    \epsilon_i = \frac{\sigma_i}{\sqrt{N_\mathrm{indep}^i}}
\end{equation}
where $\sigma_i$ is the standard deviation of the pixels that were averaged and $N_\mathrm{indep}^i$ is the number of independent samples represented by those pixels. It is calculated as
\begin{equation}
    N_\mathrm{indep}^i = \sum_k\frac{1}{n^{i,k}_\mathrm{corr}}
\end{equation}
where the index $k$ runs over all pixels included in the average and $n^{i,k}_\mathrm{corr}$ is the number of pixels correlated with pixel $k$ (where only pixels included in the average are considered). For example, consider the case where all pixels within a beam are correlated. Then $n^{i,k}_\mathrm{corr}=N_\mathrm{beam}$ where $N_\mathrm{beam}$ is the number of pixels per beam. In that case, $N_\mathrm{indep}^i=N/N_\mathrm{beam}$ with $N$ being the total number of pixels included in the average. In other words, $N_\mathrm{indep}^i$ equals the number of beams included in the average. However, in general $n^{i,k}_\mathrm{corr}<N_\mathrm{beam}$ because (1) the deprojected ring over which the average runs can be narrower than the beam, so that for a given pixel, some neighboring pixels are not included in the average, even though they are within a beam distance, and (2) the spectral shifting procedure applied to the data cube prior to azimuthally averaging means that pixels within a beam can be uncorrelated if their relative spectral shift exceeds the spectral resolution \citep[see also Section 2.2 in][]{Yen16}. Therefore, for a given pixel $k$, we calculate the number of correlated pixels as the number of pixels that satisfy all of the following conditions: they are in the same channel of the shifted data cube, are within the deprojected ring, are within a beam centered onto pixel $k$, and have a spectral shift relative to pixel $k$ smaller than the spectral resolution.

As an alternative, we also computed the standard deviation of the azimuthally averaged spectrum measured at velocities without line emission, which we shall denote $\delta$. We find that generally, $\delta\approx\epsilon_i$. Note that $\epsilon_i$ depends on velocity, while $\delta$ does not. Most of the time, $\epsilon_i$ is slightly larger than $\delta$. The exception is the innermost disk region, where $\epsilon_i$ at times is significantly smaller than $\delta$. This points to an important limitation of estimating the error bar with $\epsilon_i$: when most averaged pixels are correlated, $N_\mathrm{indep}^i\approx1$, and thus $\epsilon_i \approx \sigma_i$. But $\sigma_i$ is small if most pixels included in the average are correlated, leading to an underestimation of the error bar. This problem only occurs in the innermost disk where all pixels within the deprojected ring under consideration are part of the same beam. To be conservative, we adopt $\max(\epsilon_i,\delta)$ as our final error bar.

\subsection{Gallery of azimuthally averaged spectra}\label{appendix:az_averaged_spectra_gallery}
Figures \ref{fig:IM_Lup_HCN_data_spectra} to \ref{fig:MWC_480_HCN_data_spectra} show the azimuthally averaged spectra of HCN 1--0, HCN 3--2, DCN 3--2, and H$^{13}$CN 1--0 as described in Section \ref{sec:az_averaged_spectra}, together with LTE and non-LTE model spectra. Figures \ref{fig:IM_Lup_N2H+_data_spectra} to \ref{fig:MWC_480_N2H+_data_spectra} show the corresponding spectra for \ce{N2H+} and \ce{N2D+}, together with model spectra where $T_\mathrm{ex}=20$\,K and $T_\mathrm{ex}=T_\mathrm{mid}$. In Figure \ref{fig:disk_integrated_spec_N2D+_3-2}, we show the azimuthally averaged spectra over the whole disk for \ntdp 3--2. The line is clearly detected toward IM~Lup, AS~209, HD~163296, and MWC~480 and tentatively detected toward GM~Aur. This is consistent with the disk-integrated flux (Section \ref{sec:disk_integrated_flux_mom0}) and the results of the matched filter analysis (Appendix \ref{appendix:matched_filter}).

\begin{figure}
\plotone{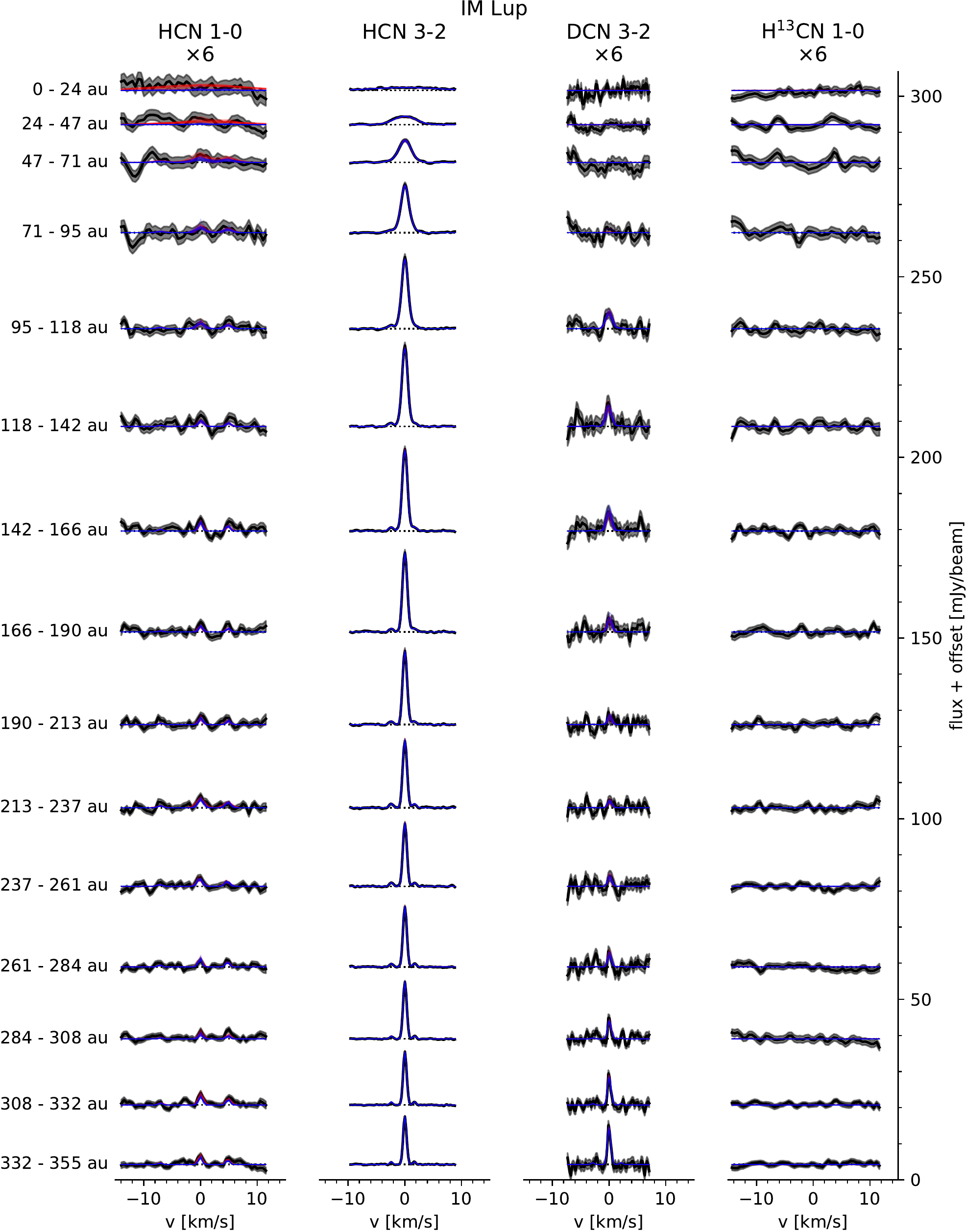}
\caption{Azimuthally averaged spectra of HCN 1--0, HCN 3--2, DCN 3--2, and H$^{13}$CN 1--0 toward IM~Lup. Spectra are vertically offset for clarity and centered on the systemic velocity. If a scaling is applied to the spectra of a transition, it is indicated at the top of the column. The shaded region marks the 1$\sigma$ error. The horizontal dotted line marks the zero flux level. For each spectrum, 20 randomly selected models from the MCMC chain (with the selection probability proportional to the model's posterior probability) are shown by the blue lines (LTE) and the red lines (non-LTE), respectively. For most spectra, the models overlap closely so that only the blue lines are visible. \label{fig:IM_Lup_HCN_data_spectra}}
\end{figure}

\begin{figure}
\plotone{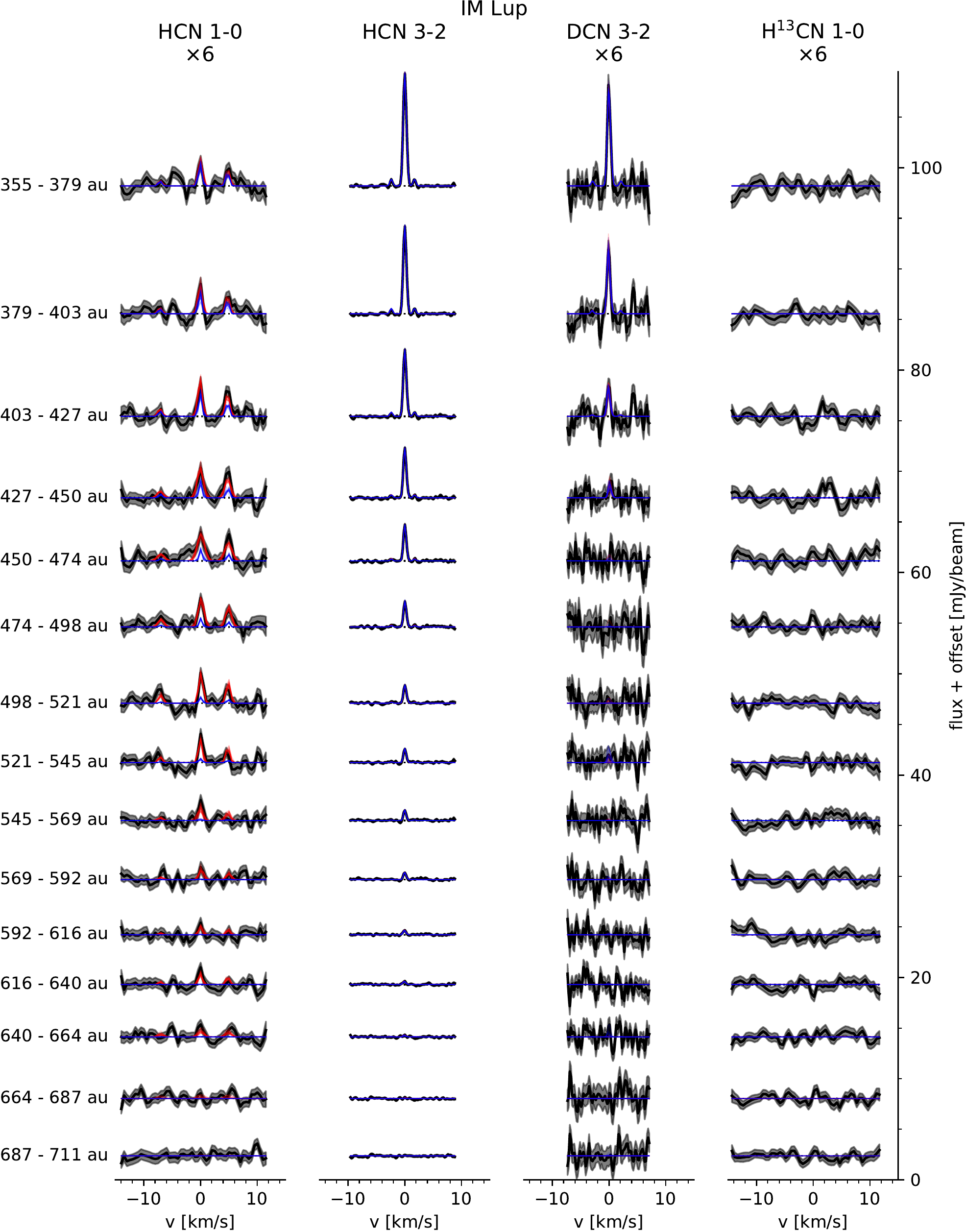}
\figurenum{\ref*{fig:IM_Lup_HCN_data_spectra}}
\caption{(Continued.)}
\end{figure}

\begin{figure}
\plotone{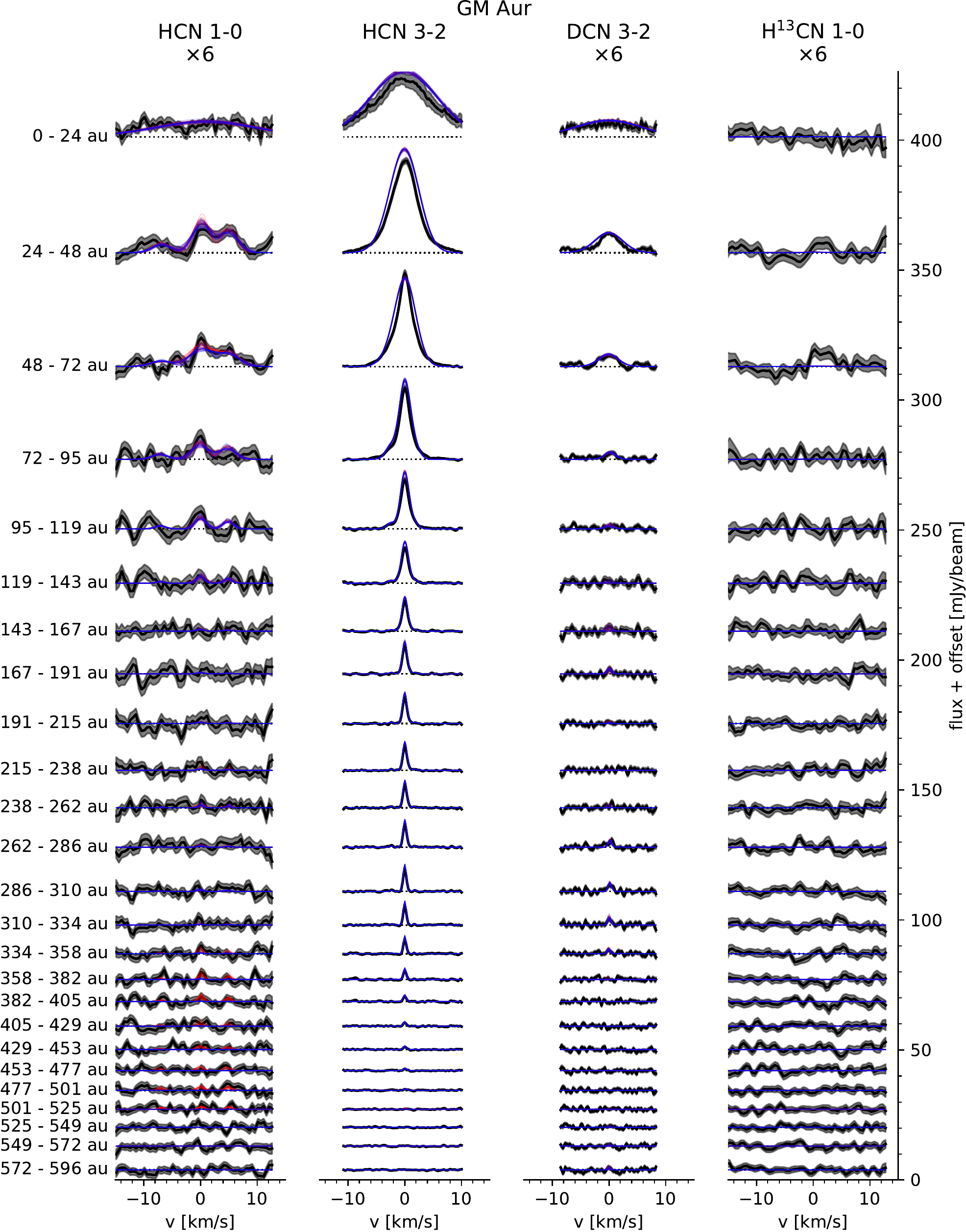}
\caption{Same as Figure \ref{fig:IM_Lup_HCN_data_spectra}, but for GM~Aur. \label{fig:GM_Aur_HCN_data_spectra}}
\end{figure}

\begin{figure}
\plotone{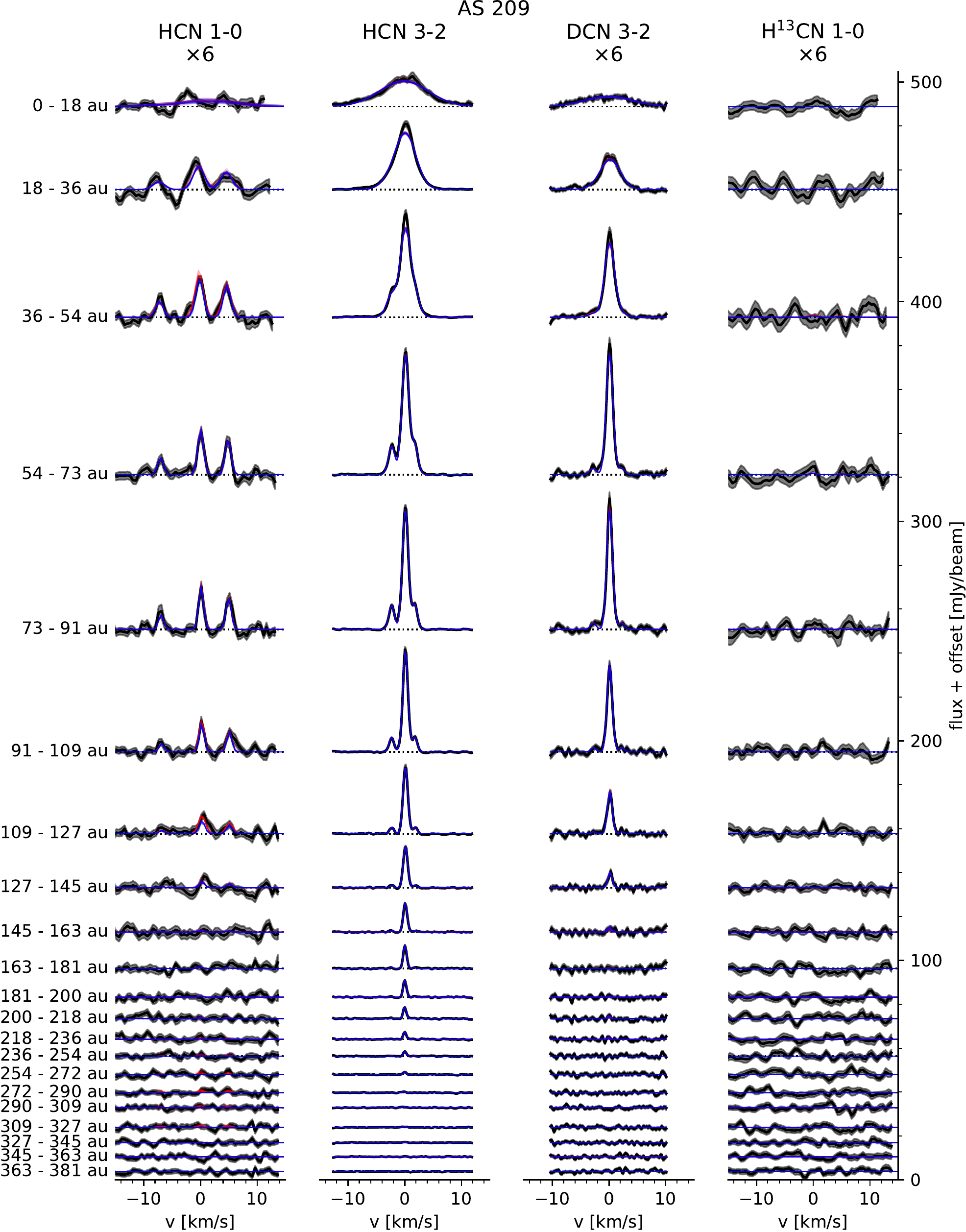}
\caption{Same as Figure \ref{fig:IM_Lup_HCN_data_spectra}, but for AS~209. Note the hyperfine structure of HCN 1--0 and 3--2. \label{fig:AS_209_HCN_data_spectra}}
\end{figure}

\begin{figure}
\plotone{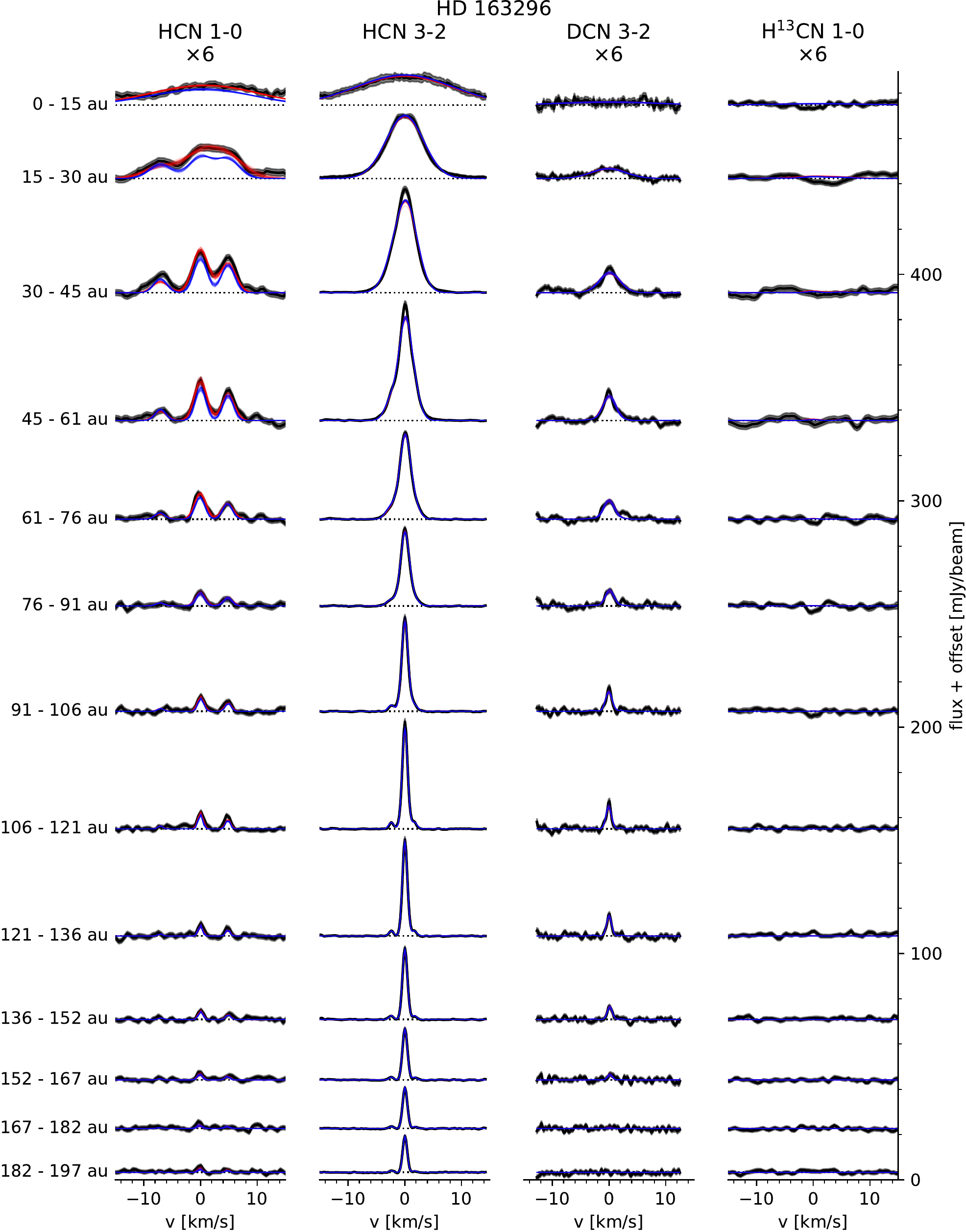}
\caption{Same as Figure \ref{fig:IM_Lup_HCN_data_spectra}, but for HD~163296. \label{fig:HD_163296_HCN_data_spectra}}
\end{figure}

\begin{figure}
\plotone{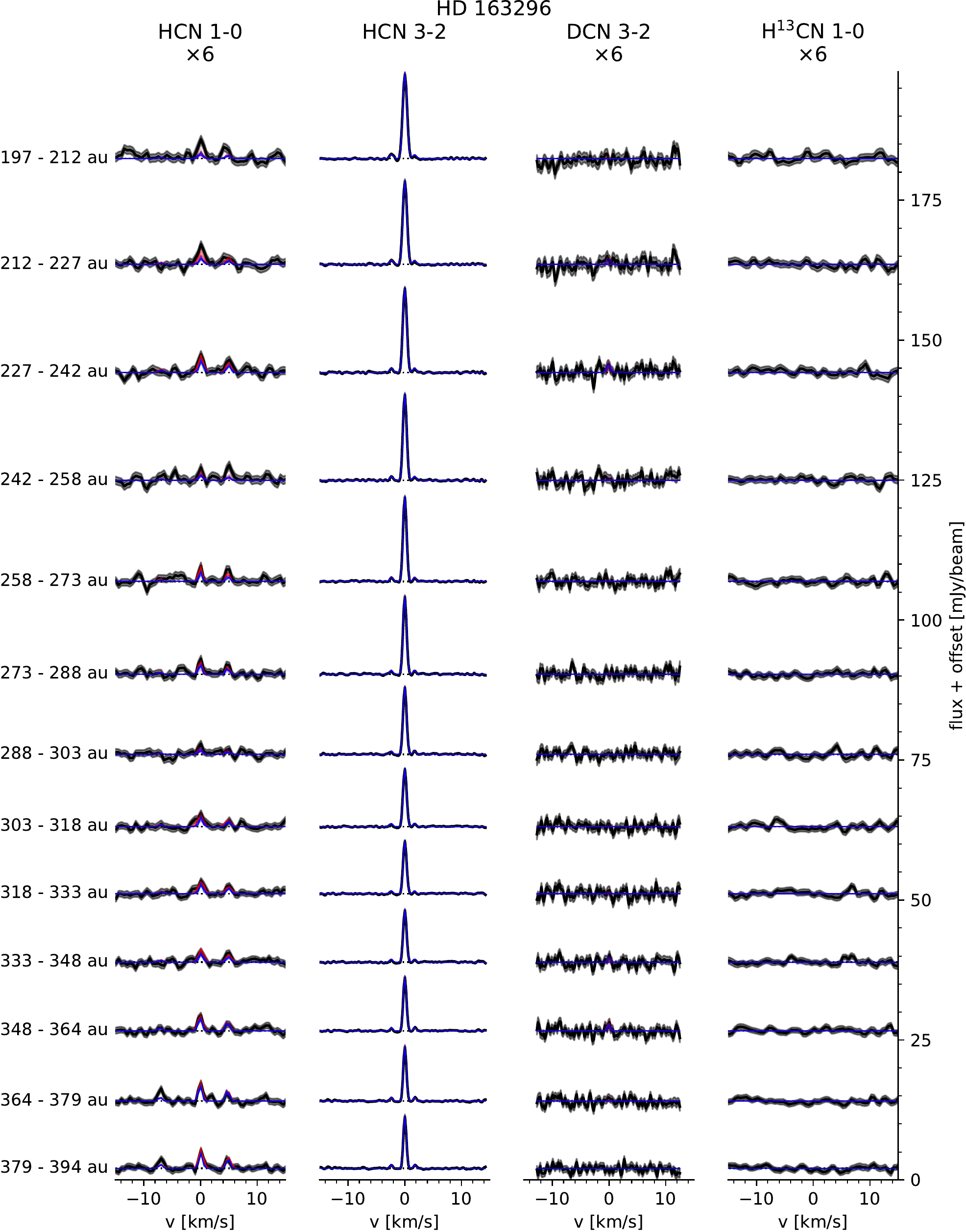}
\figurenum{\ref*{fig:HD_163296_HCN_data_spectra}}
\caption{(Continued.)}
\end{figure}

\begin{figure}
\plotone{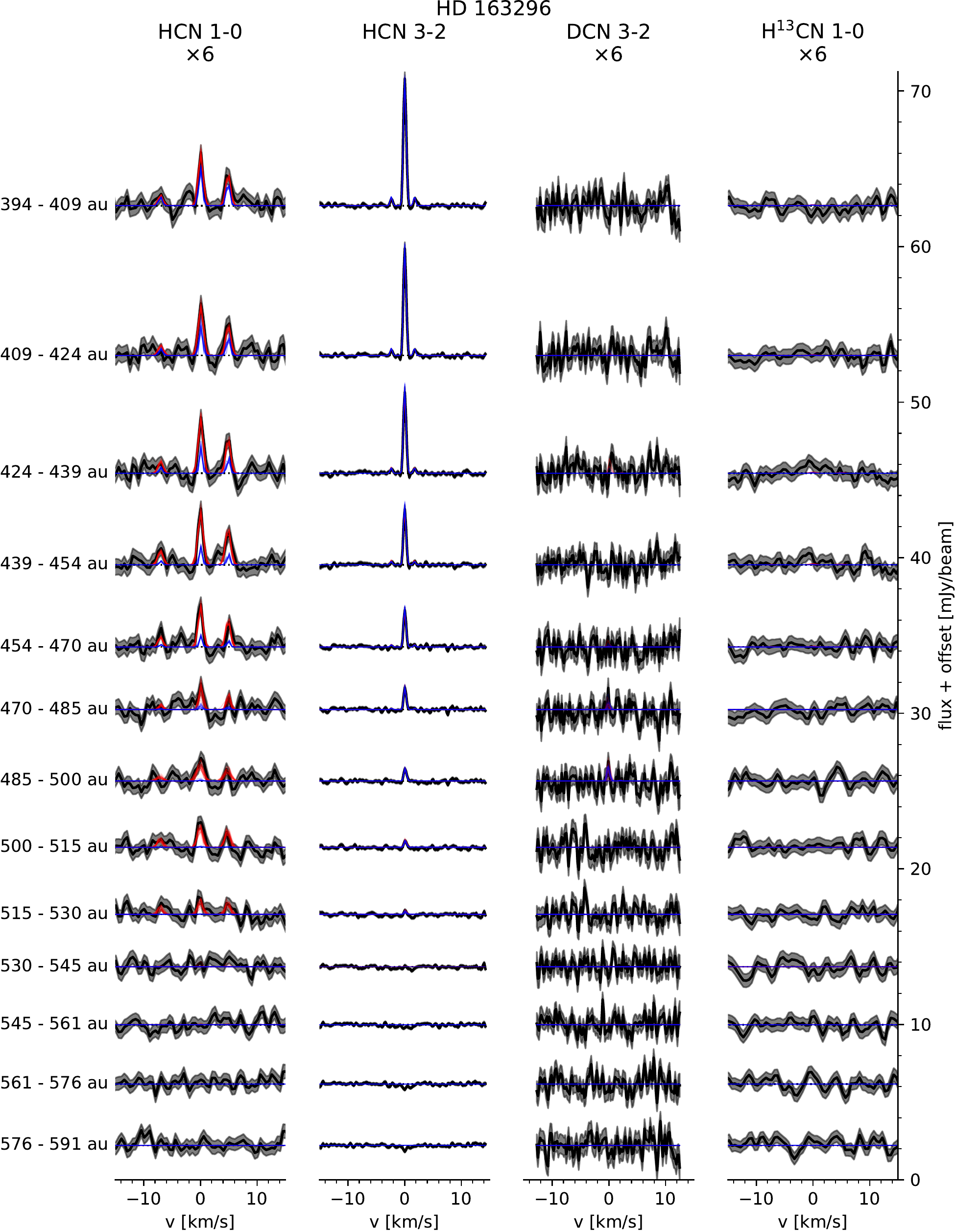}
\figurenum{\ref*{fig:HD_163296_HCN_data_spectra}}
\caption{(Continued.)}
\end{figure}

\begin{figure}
\plotone{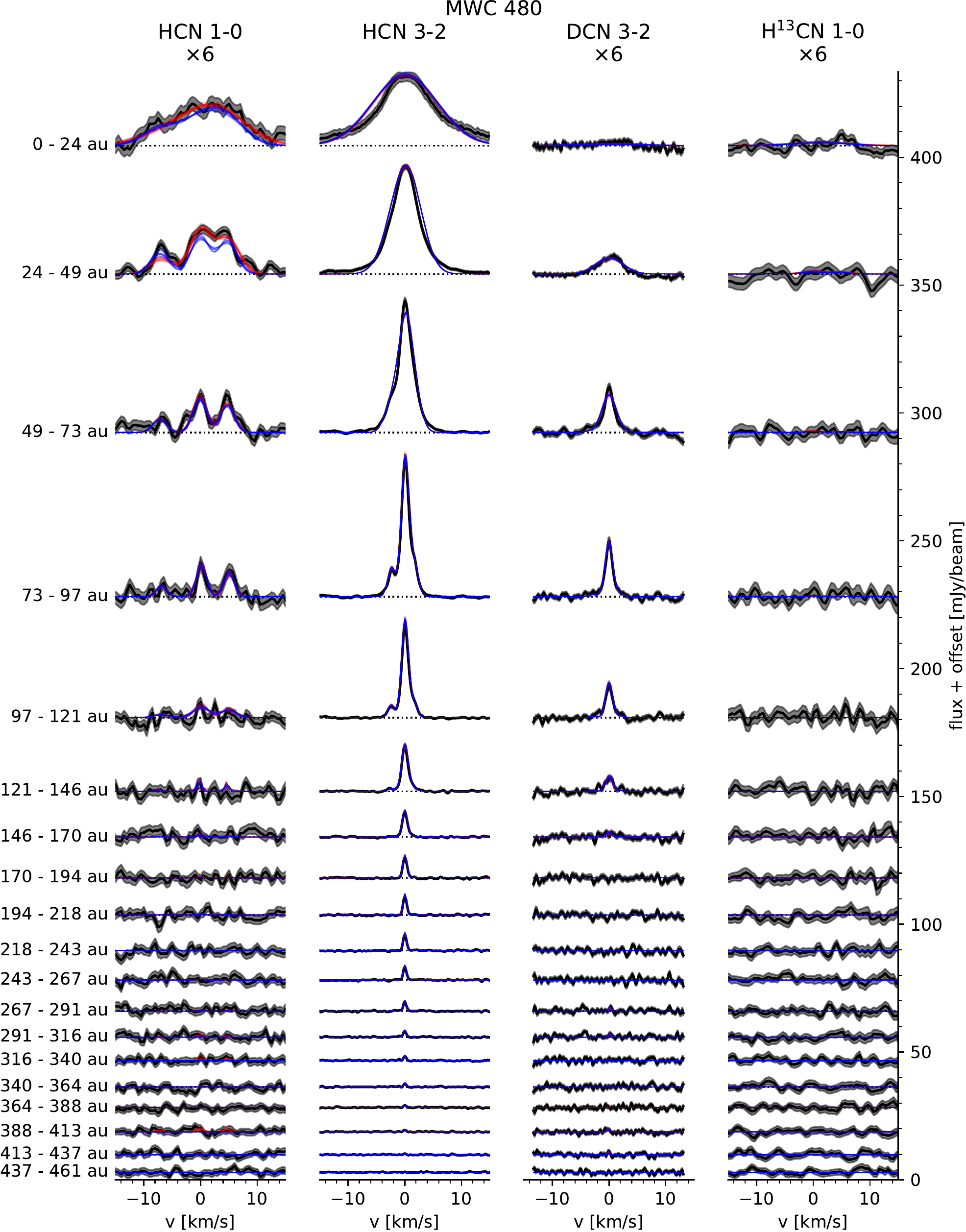}
\caption{Same as Figure \ref{fig:IM_Lup_HCN_data_spectra}, but for MWC~480. \label{fig:MWC_480_HCN_data_spectra}}
\end{figure}

\begin{figure}
\plotone{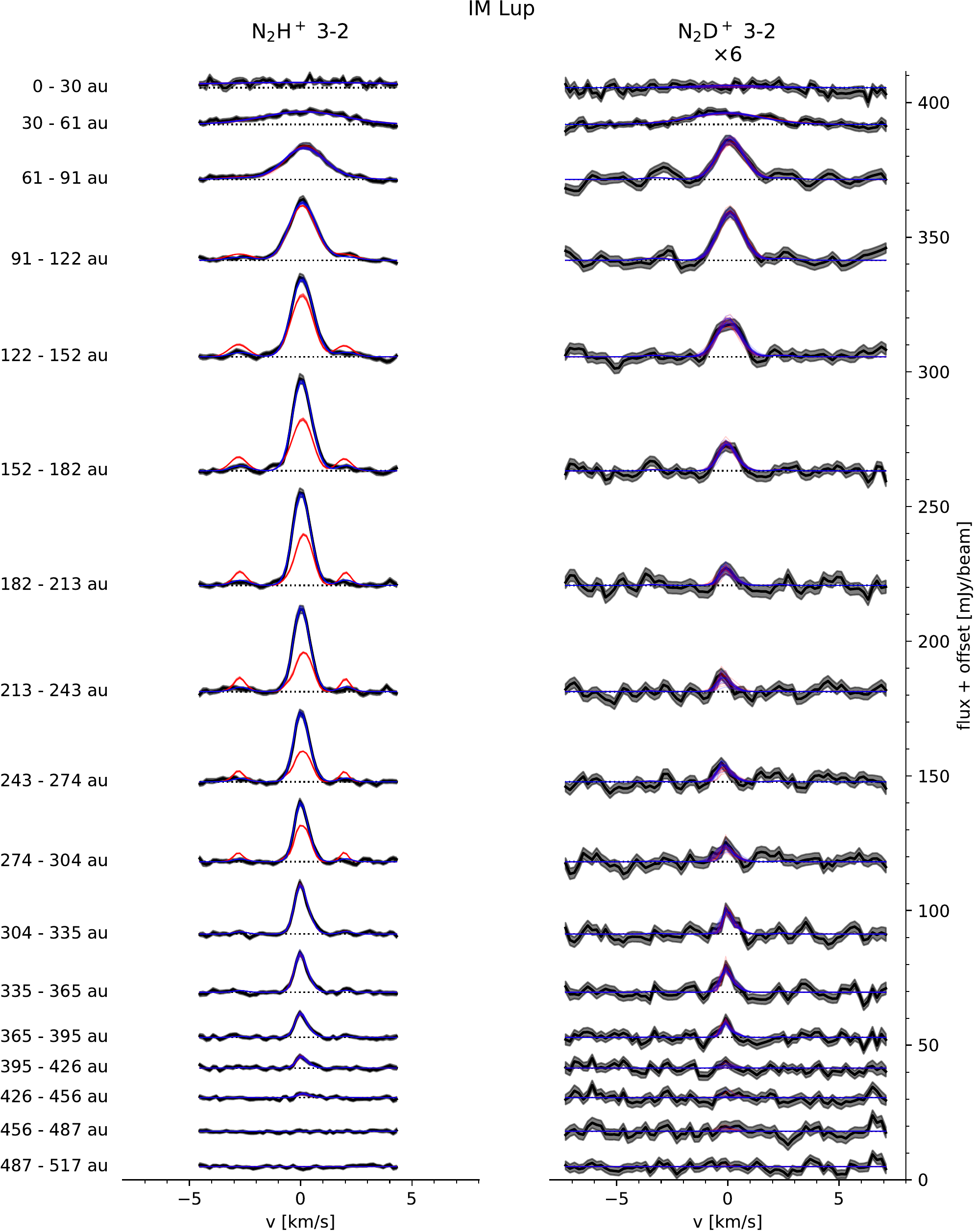}
\caption{Azimuthally averaged spectra for IM~Lup of the 3--2 transitions of \ce{N2H+} and \ce{N2D+}. Spectra are vertically offset for clarity and centered on the systemic velocity. If a scaling is applied to the spectra of a transition, it is indicated at the top of the column. The shaded region marks the 1$\sigma$ error.  The horizontal dotted line marks the zero flux level. For each spectrum, 20 randomly selected models from the MCMC chain (with the selection probability proportional to the model's posterior probability) are shown by the blue lines ($T_\mathrm{ex}=20$\,K) and red lines ($T_\mathrm{ex}=T_\mathrm{mid}$), respectively. For most spectra, the models overlap closely so that only the blue lines are visible.\label{fig:IM_Lup_N2H+_data_spectra}}
\end{figure}

\begin{figure}
\plotone{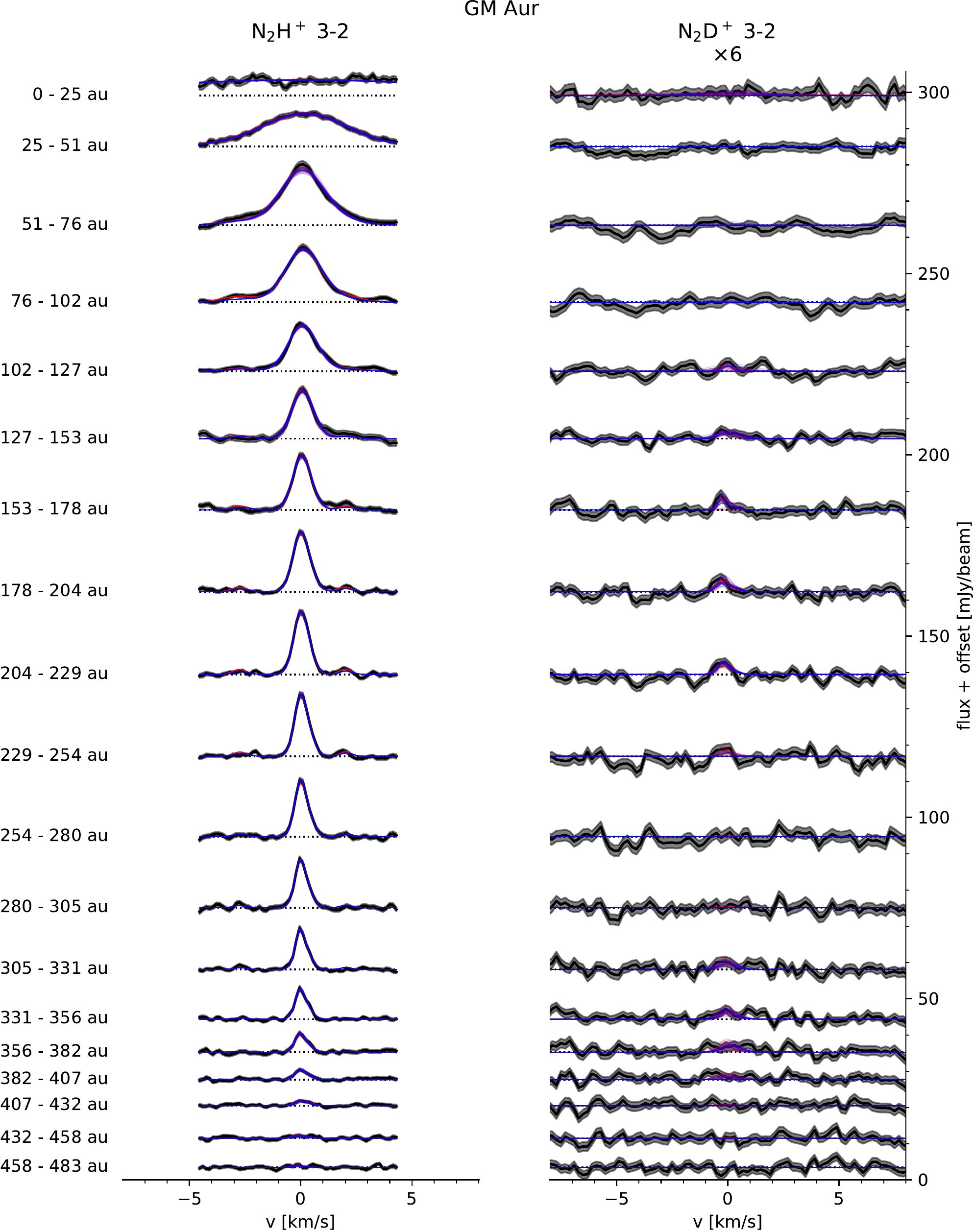}
\caption{Same as Figure \ref{fig:IM_Lup_N2H+_data_spectra}, but for GM~Aur. \label{fig:GM_Aur_N2H+_data_spectra}}
\end{figure}

\begin{figure}
\plotone{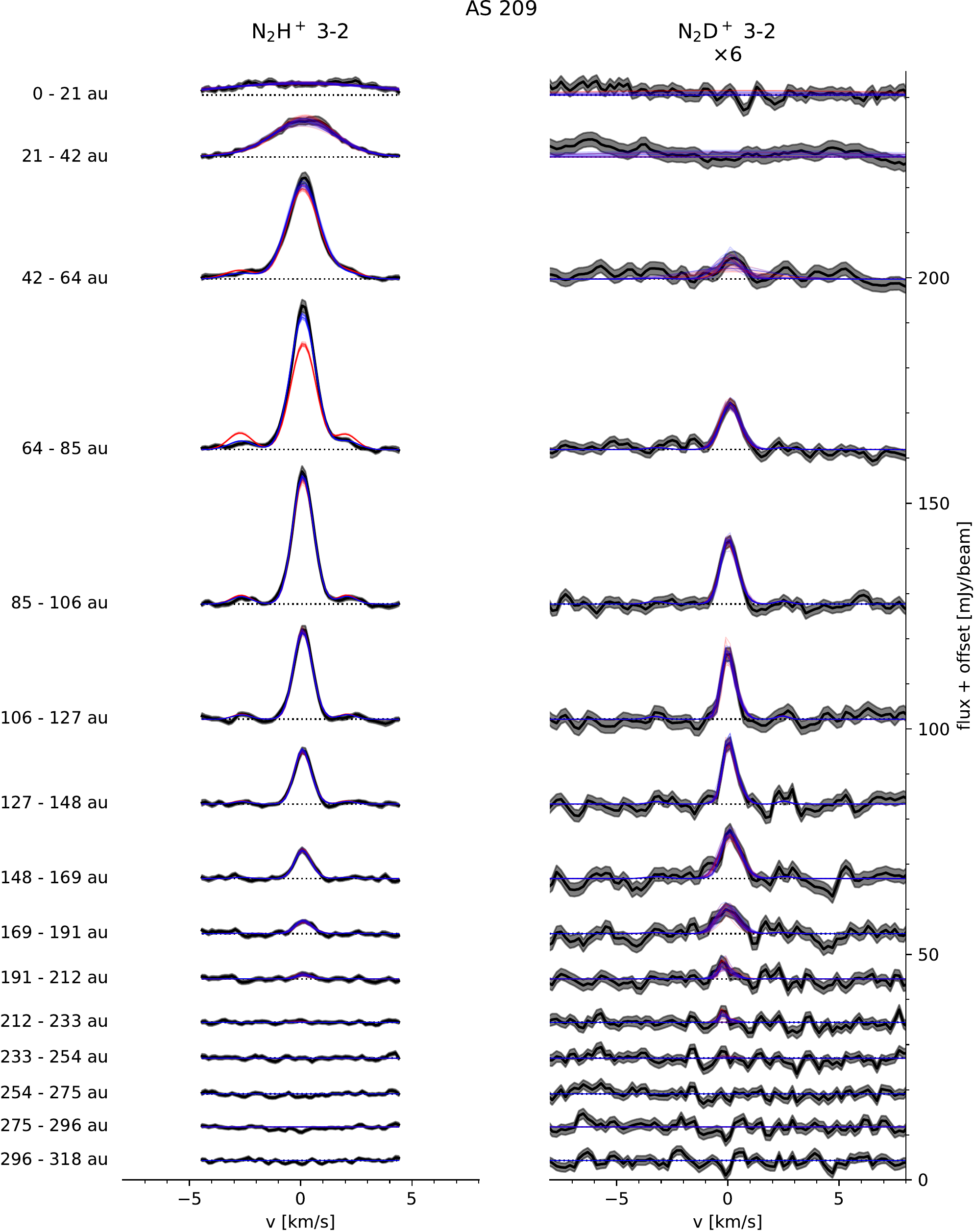}
\caption{Same as Figure \ref{fig:IM_Lup_N2H+_data_spectra}, but for AS~209. \label{fig:AS_209_N2H+_data_spectra}}
\end{figure}

\begin{figure}
\plotone{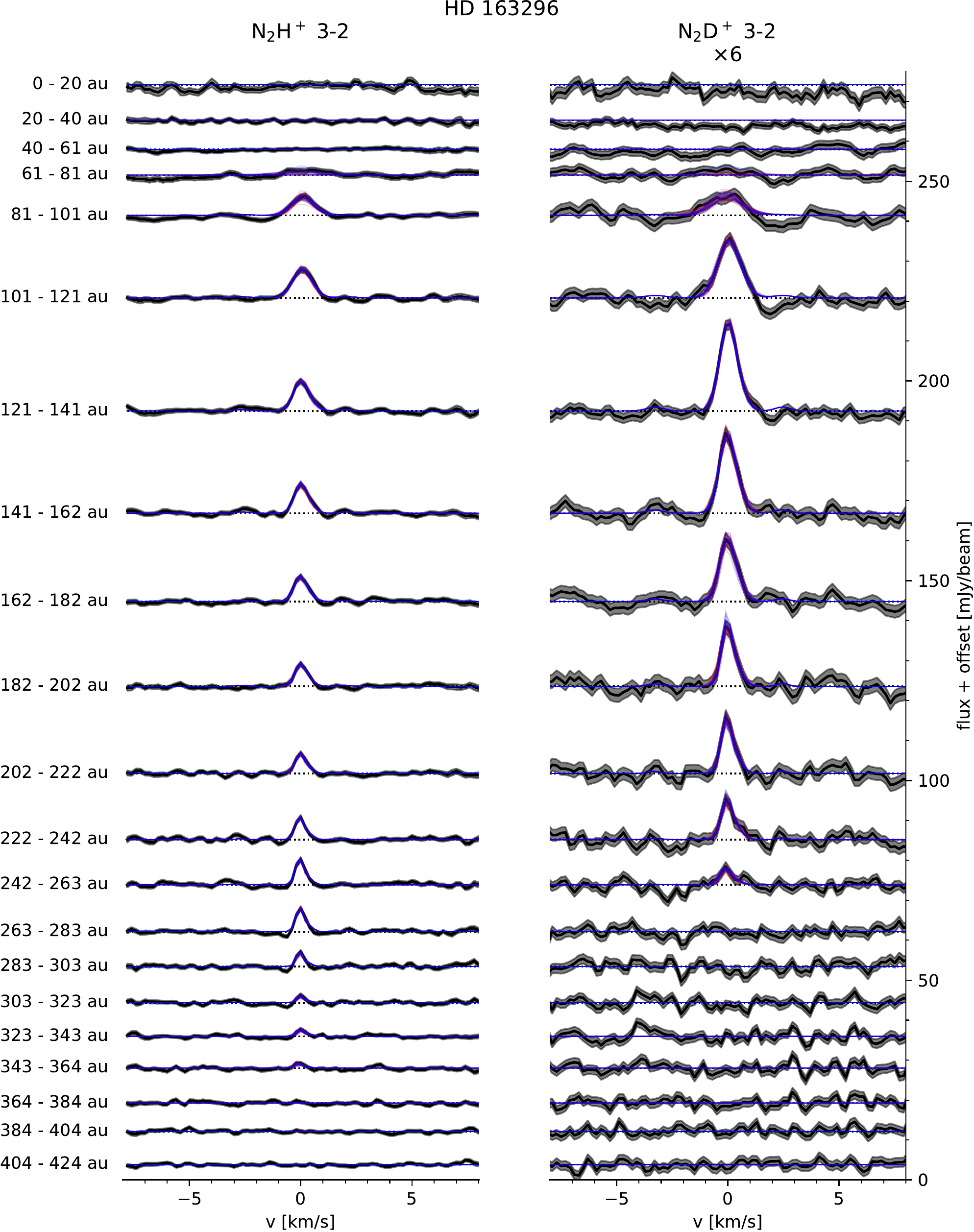}
\caption{Same as Figure \ref{fig:IM_Lup_N2H+_data_spectra}, but for HD~163296. \label{fig:HD_163296_N2H+_data_spectra}}
\end{figure}

\begin{figure}
\plotone{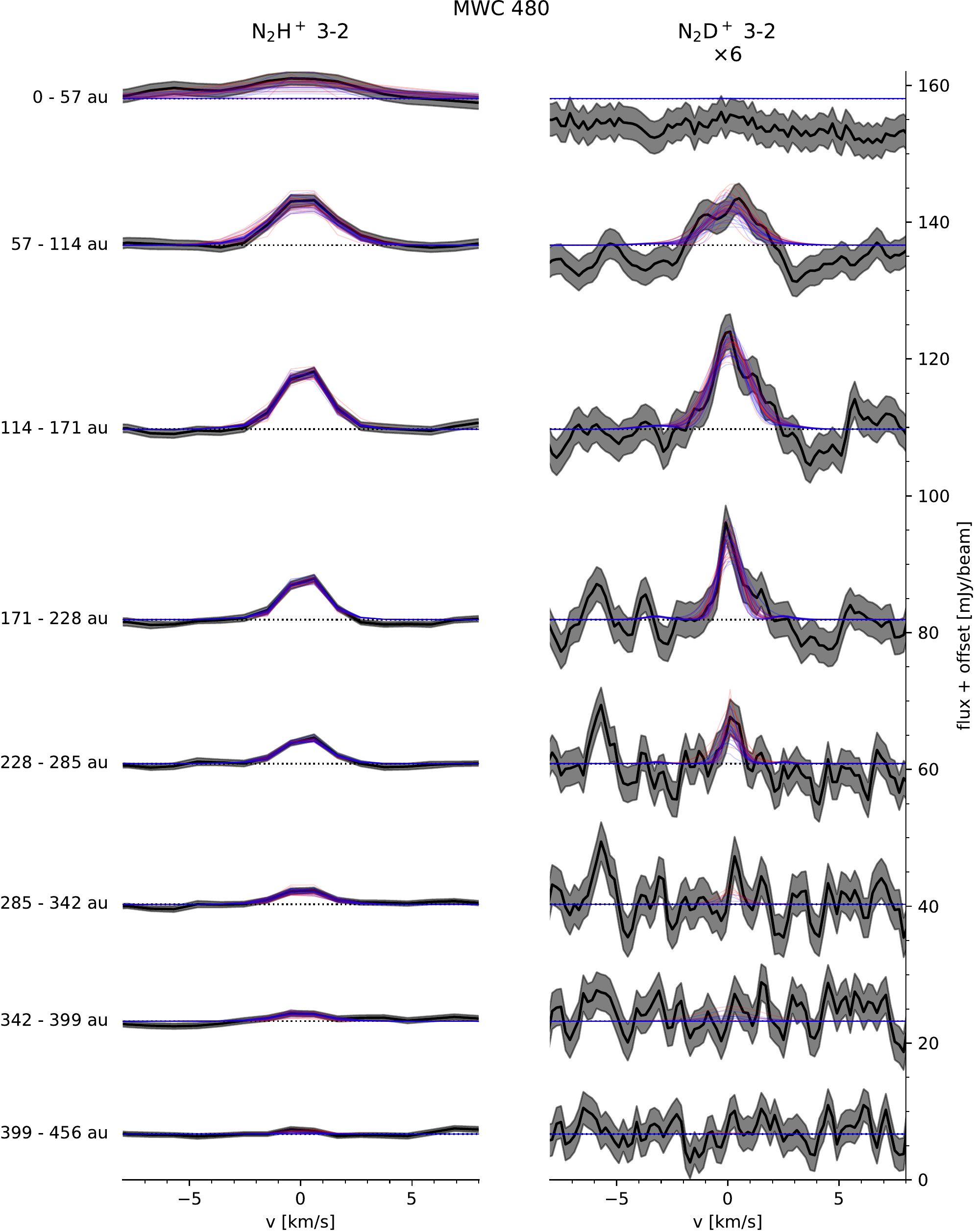}
\caption{Same as Figure \ref{fig:IM_Lup_N2H+_data_spectra}, but for MWC~480. \label{fig:MWC_480_N2H+_data_spectra}}
\end{figure}

\begin{figure}
    \centering
    \plotone{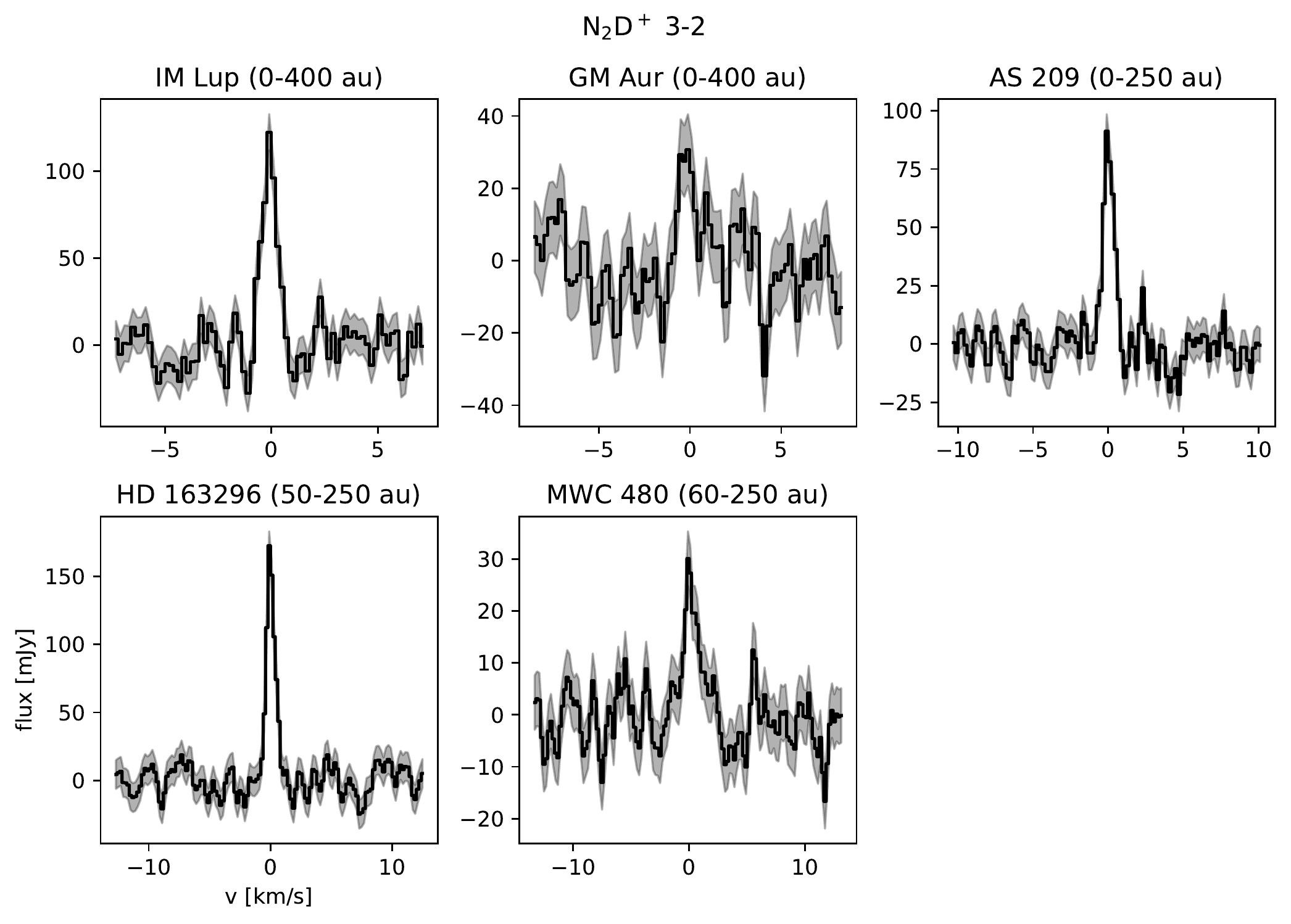}
    \caption{Disk-integrated spectra of \ntdp 3--2 centered on the systemic velocity. The radial integration range is indicated in the title of each panel. The shaded region denotes the 1$\sigma$ uncertainty.}
    \label{fig:disk_integrated_spec_N2D+_3-2}
\end{figure}

\section{Derivation of HCN and DCN column densities}\label{appendix:HCN_DCN_fitting}
In this appendix, we describe how the HCN and DCN column densities were derived by simultaneously fitting the shifted and azimuthally averaged spectra of HCN 1--0 and 3--2, H$^{13}$CN 1--0, and DCN 3--2. An analogous approach was used to derive the column densities of \ce{N2H+} and \ce{N2D+}.

\subsection{LTE}\label{appendix:HCN_DCN_fitting_LTE}
Here we consider the column density calculation under the LTE assumption where the excitation temperature $T_\mathrm{ex}$ is the same for all transitions of a molecule and equals the gas kinetic temperature $T_\mathrm{kin}$. We further assume that the different molecular species are sufficiently cospatial to share the same temperature. Our model has the following free parameters:
\begin{enumerate}
    \item The excitation temperature $T_\mathrm{ex}$.
    \item The logarithms of the column densities: $\log_{10} N_\mathrm{HCN}$ and $\log_{10} N_\mathrm{DCN}$. The column density of H$^{13}$CN is fixed to the ISM ratio of $^{13}$C/$^{12}$C=1/68 of the HCN column density.
    \item The FWHM of the Gaussian convolution kernel used to model instrumental broadening of the lines (see below): $\mathrm{FWHM}_\mathrm{B3}$ (for Band 3 data, i.e.\ the 1--0 transitions) and $\mathrm{FWHM}_\mathrm{B6}$ (for Band 6 data, i.e.\ the 3--2 transitions).
    \item For each line, a velocity offset with respect to the systemic velocity: $\Delta v_\mathrm{HCN1-0}$, $\Delta v_\mathrm{HCN3-2}$, $\Delta v_\mathrm{DCN3-2}$ and $\Delta v_\mathrm{H13CN1-0}$.
\end{enumerate}
This sums up to nine free parameters. For given values of the parameters, model spectra (including hyperfine structure) are calculated by following an approach described in appendix A of \citet{Teague20}. These model spectra can then be compared to the data. The intrinsic emission spectrum is given by
\begin{equation}\label{eq:model_spectrum}
    I_\nu(v) = (B_\nu(T_\mathrm{ex})-B_\nu(T_\mathrm{CMB}))\cdot(1-e^{-\tau_\nu})
\end{equation}
with $B_\nu$ the Planck function and $T_\mathrm{CMB}=2.73$\,K the cosmic microwave background (CMB) temperature (i.e.\ the background radiation is assumed to be the CMB). The frequency-dependent optical depth $\tau_\nu$ is given by
\begin{equation}\label{eq:model_tau_nu}
    \tau_\nu = \tau_0\sum_ir_i\exp\left(-\frac{(v-v_\mathrm{sys}-\delta v_i-\Delta v_X)^2}{2\sigma_\mathrm{v}^2} \right)
\end{equation}
with $v_\mathrm{sys}$ the systemic velocity \citep[taken from][]{Oberg21_MAPS} and $\Delta v_X$ the free parameter describing the velocity offset of the observed emission line $X$. The sum goes over all hyperfine components, with $r_i$ the strength and $\delta v_i$ the velocity offset of each component. The strength $r_i$ is given by
\begin{equation}\label{eq:model_strength_hfs}
    r_i=\frac{g_{u,i}A_{ul,i}}{\sum_j g_{u,j}A_{ul,j}}
\end{equation}
with $g_u$ the weight of the upper level and $A_{ul}$ the Einstein coefficient for spontaneous emission of the hyperfine line. Tables \ref{tab:atomic_parameters_HCN}, \ref{tab:atomic_parameters_N2H+} and \ref{tab:atomic_parameters_N2D+} list the values of the molecular parameters used in this work. $\tau_0$ is the optical depth of the hyperfine-unresolved transition at the line center, given by
\begin{equation}
    \tau_0 = \frac{hc}{4\pi}(x_lB_{lu}-x_uB_{ul})N\frac{1}{\sqrt{2\pi}\sigma_\mathrm{v}}
\end{equation}
with $N$ the column density. Here the Einstein $B$ coefficients and the fraction of molecules in the upper or lower level ($x_u$ and $x_l$) refer to the hyperfine-unresolved transition. The latter are given by
\begin{align}
x_u &= g_u\frac{e^{-E_u/(kT_\mathrm{ex})}}{Z(T_\mathrm{ex})}\label{eq:x_u}\\
x_l &= g_l\frac{e^{-E_l/(kT_\mathrm{ex})}}{Z(T_\mathrm{ex})}\label{eq:x_l}
\end{align}
where the partition function $Z$ is taken from the CDMS database\footnote{\url{https://cdms.astro.uni-koeln.de/cdms/portal/}}.

In contrast to \citet{Teague20}, the width of the intrinsic spectrum is fixed to the thermal width, that is,
\begin{equation}
    \sigma_v=\sqrt{\frac{kT_\mathrm{kin}}{m}}
\end{equation}
where the kinetic temperature $T_\mathrm{kin}=T_\mathrm{ex}$ in LTE, and $m$ is the mass of the molecule. We ignore turbulent broadening given recent results suggesting low turbulence \citep[e.g.,][]{Teague18_CS}. Finally, the spectrum given by equation \ref{eq:model_spectrum} is convolved with a Gaussian kernel with FWHM given by the free parameter $\mathrm{FWHM}_\mathrm{Bx}$. This models any broadening of the line by effects such as beam smearing or imprecise Keplerian shifting. This broadening is highest in the inner region of the disk and gradually becomes smaller at larger radii (see Figures \ref{fig:IM_Lup_HCN_data_spectra} to \ref{fig:MWC_480_N2H+_data_spectra}).

Table \ref{tab:HCN_free_parameters_overview} shows an overview of the free parameters. We explore the parameter space using the MCMC method implemented in the \texttt{emcee} package \citep{Foreman-Mackey13} for each radial bin separately, starting with the innermost bin. We run 200 walkers, each with a total of 5000 steps. The walkers of radial bin $n+1$ are initialized using the result from radial bin $n$. The first 2500 steps are discarded prior to analyzing the results. We choose flat priors over the ranges detailed in Table \ref{tab:HCN_free_parameters_overview}.

To take into account the correlation between neighboring wavelengths in the azimuthally averaged spectra, we rescale the error bars in the spectra by multiplying by the square root of the correlation length \citep{Booth17}. We define the latter as $\max(1,D_\mathrm{corr})$, where $D_\mathrm{corr}$ is the FWHM (in units of spectral pixels) of the autocorrelation of the spectrum. Before calculating the autocorrelation, we remove the region of the spectrum with line emission and then subtract the mean value such that the line-free spectrum has a mean value of zero. The correlation length is typically $\lesssim3$ spectral pixels, meaning that the rescaling of the error bars is marginal. However, we find that the correlation length can be much larger (up to $\sim$20 spectral pixels) for the azimuthally averaged spectra extracted from the innermost beam. This is probably because the individual spectra that were azimuthally averaged are all spatially correlated. This spatial correlation is transformed to spectral correlation by the Keplerian shifting procedure.

The correlation length is difficult to determine for spectra extracted from inner disk regions because the line emission becomes very broad. Thus, no line-free regions is available. To circumvent this problem, for the innermost regions (within 1.5 beams), we adopt the correlation lengths determined from the weakest lines: H$^{13}$CN~1--0 for HCN 1--0, HCN 3--2 and DCN 3--2, and \ntdp~3--2 for \nthp 3--2.

As we move outward in radius, the upper boundaries of the priors on $\mathrm{FWHM}_\mathrm{B3}$ and $\mathrm{FWHM}_\mathrm{B6}$ are adjusted according to the following scheme. First, we calculate the SNR of $\mathrm{FWHM}_\mathrm{Bx}$, defined as the median of the MCMC ('signal') divided by the mean of the 16th and 84th percentiles ('error'). If the SNR is higher than a set threshold (we choose between 3 and 8), then the upper boundary of the prior for the next radial bin is set to the median plus five times the error. This procedure reflects the observed sharp decrease of the line width with increasing radius and allows us to set tighter priors in the outer regions that show little signal. In the case where only the B6 or B3 FWHM exceeds the SNR required for adjustment, we still adjust both FWHM by using the higher SNR FWHM as a basis to estimate a conservative upper boundary for the lower SNR FWHM, taking into account the difference in channel size if necessary. Figure \ref{fig:FWHM_fit_example} shows an example of the fitted FWHM together with the radially adjusted prior boundaries.

\begin{figure*}
\plotone{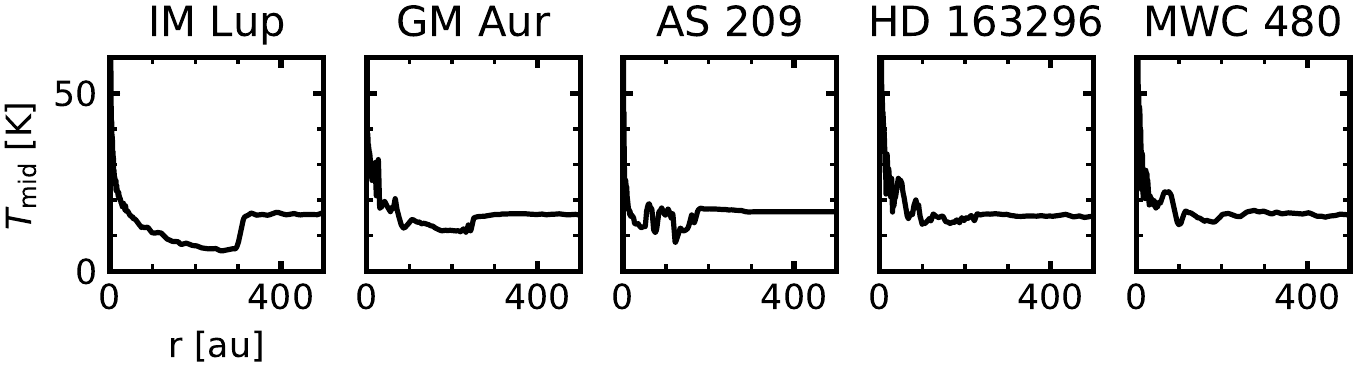}
\caption{Profiles of the midplane gas temperature $T_\mathrm{mid}$. These profiles were extracted from the MAPS reference models \citep{Zhang21_MAPS} and are used for fitting the \nthp and \ntdp column densities under the assumption that $T_\mathrm{ex}=T_\mathrm{mid}$ (Section \ref{sec:N2H+_N2D+_fitting}).\label{fig:radial_temp_model}}
\end{figure*}

\begin{figure}
\plotone{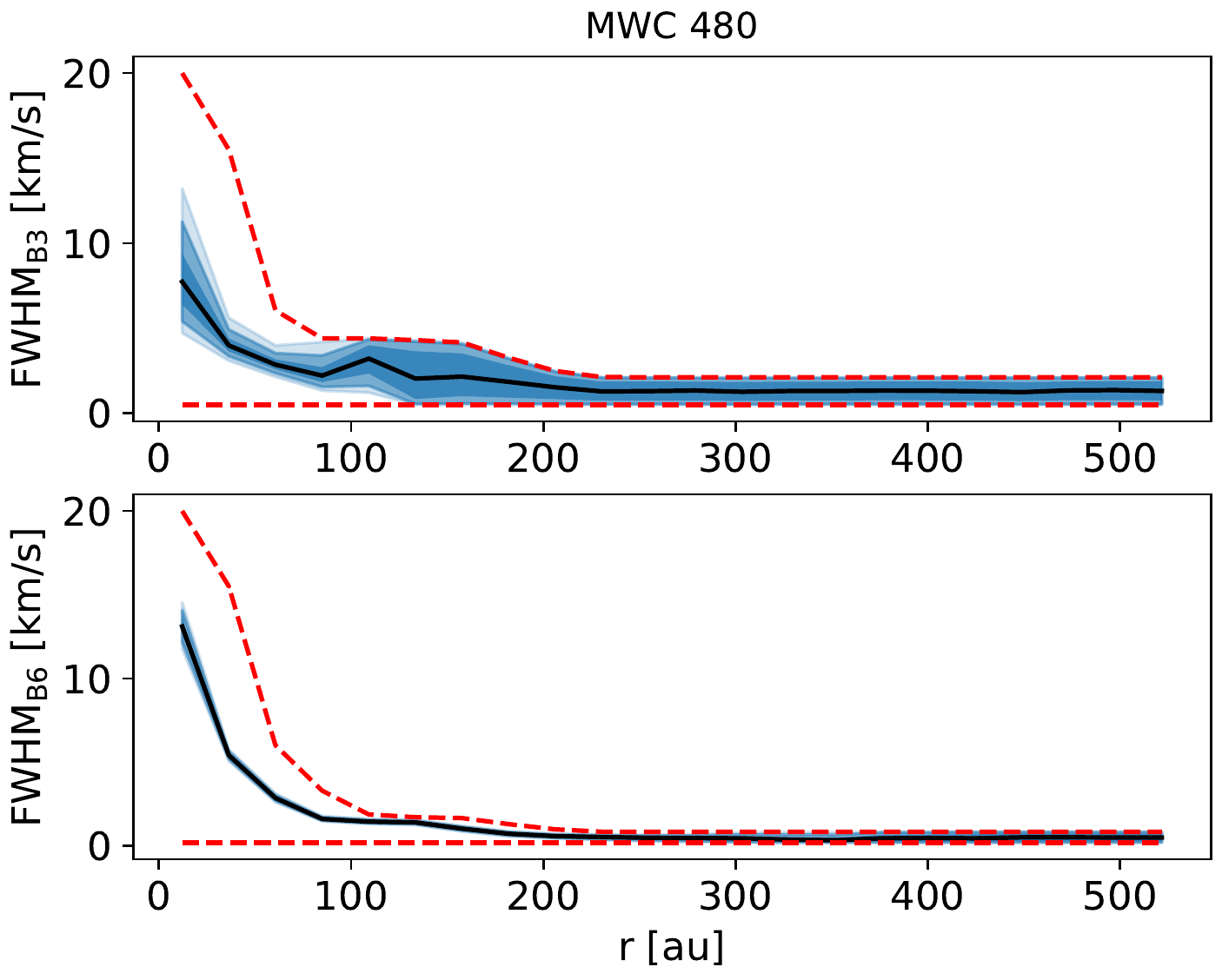}
\caption{Example of the fitted FWHM of the Gaussian kernel used to convolve the model spectrum for MWC~480. The black line marks the median, while the shaded regions encompass the 16th to 84th, 2.3th to 97.7th, and 0.15th to 99.85th percentile regions. The red dashed lines show the upper and lower bound of the flat prior. Sequentially fitting larger and larger radii, the upper bound is adjusted by using the fit results from the previous radius (see text for details). \label{fig:FWHM_fit_example}}
\end{figure}

To derive the column densities of \ce{N2H+} and \ce{N2D+}, we used an analogous approach and simultaneously fitted the 3--2 transitions. However, the excitation temperature was fixed to 20\,K. In addition, to test the dependence on the assumed temperature, we also run fits with the excitation temperature fixed to the midplane temperature extracted from the MAPS reference models by \citet[][Fig.\ \ref{fig:radial_temp_model}]{Zhang21_MAPS}. These source-specific thermochemical models fit the spectral energy distribution (SED), (sub-)millimeter continuum images, and the vertical locations of the CO-emitting layers reported by \citet{Law21_MAPS_surfaces_vertical_distributions}. We note that the increase of temperature in the outer disk (particularly clearly seen for IM~Lup at $\sim$300\,au) coincides with the truncation radius of the pebble disk, which makes the midplane warmer.

Since the \ce{N2H+} and \ce{N2D+} data come from different observations, two separate parameters for the FWHM of the Gaussian kernels were used, one for each line. The free parameters and their priors are listed in Table \ref{tab:N2H+_free_parameters_overview}.

The method described here is essentially the same as employed by \citet{Bergner21_MAPS} to calculate HCN and CN column densities and by \citet{Guzman21_MAPS} to calculate HCN and \ce{C2H} column densities. One difference is that \citet{Bergner21_MAPS} and \citet{Guzman21_MAPS} do not convolve the intrinsic spectrum but simply redistribute the total flux of the intrinsic spectrum in a model spectrum with broader components. They also do not include H$^{13}$CN in their fits of the HCN column density. Furthermore, \citet{Guzman21_MAPS} employ a two-step fitting method, where they first fit HCN 1--0 and 3--2 at 0.3$\arcsec$ resolution to constrain the excitation temperature and then fit HCN 3--2 at higher 0.15$\arcsec$ resolution, using the constraints on $T_\mathrm{ex}$ from the first step. The HCN column densities derived by \citet{Bergner21_MAPS} and \citet{Guzman21_MAPS} are generally consistent with our results, with differences only seen inward of 50--100\,au \citep[][Appendix C]{Guzman21_MAPS}.

\subsection{Non-LTE}\label{appendix:HCN_DCN_fitting_nonLTE}
We repeated the calculation of the HCN and DCN column densities described in Appendix \ref{appendix:HCN_DCN_fitting_LTE} without the assumption of LTE, that is, $T_\mathrm{ex}$ is not necessarily equal to $T_\mathrm{kin}$ and can now take different values for each of the emission lines. We now fit the kinetic gas temperature $T_\mathrm{kin}$ and the \ce{H2} number density $n_\mathrm{H_2}$, with priors detailed in Table \ref{tab:HCN_free_parameters_overview}. Instead of using equations \ref{eq:x_u} and \ref{eq:x_l}, the fractional level populations are now determined by solving the statistical equilibrium of the excitation and deexcitation processes. To this end, we use the radiative transfer code \texttt{pythonradex}\footnote{\url{https://github.com/gica3618/pythonradex}.} For the purpose of the radiative transfer, we ignore the hyperfine structure: we assume that all hyperfine lines of a given emission line share the same excitation temperature \citep{Teague20} and we do not consider cross-excitation. The excitation temperature for a given transition is then computed from the fractional level populations by using
\begin{equation}
    \frac{x_u}{x_l} = \frac{g_u}{g_l}e^{-\Delta E/(kT_\mathrm{ex})}
\end{equation}
where $\Delta E = E_u-E_l$. The rest of the calculation is identical to the LTE case. In practice, we precompute a grid of fractional level populations and excitation temperatures in the $T_\mathrm{kin}$-$n_\mathrm{H_2}$-space and interpolate on that grid during the MCMC run for computational efficiency. We adopt the scaled HCN-He collision rates by \citet{Dumouchel10} available in the LAMDA database\footnote{\url{https://home.strw.leidenuniv.nl/~moldata/}, accessed 2021 March 1.} for all three species (HCN, DCN, H$^{13}$CN). We find that the calculation of the non-LTE level populations fails for column densities above $\sim10^{16}$\,cm$^{-2}$. Thus, we reduce the upper boundary of the column density priors compared to the LTE case (Table \ref{tab:HCN_free_parameters_overview}). Because we find column densities below $10^{16}$\,cm$^{-2}$ with both the LTE and non-LTE runs, this should not have a significant effect on our results.

\begin{table*}
\centering
\caption{Free parameters for the fitting of azimuthally averaged spectra to derive HCN and DCN column densities.}\label{tab:HCN_free_parameters_overview}
\begin{tabular}{Ccccc}
\hline
\mathrm{Parameter} & prior low\tablenotemark{a} & prior high\tablenotemark{b} & unit & LTE/non-LTE\tablenotemark{c}\\
\hline
T_\mathrm{ex} & 10 & 100 & [K] & LTE \\
\log_{10} N_\mathrm{HCN} & 3 & 18 (LTE), 16 (non-LTE) & $\log_{10}(\mathrm{[cm^{-2}]})$ &\\
\log_{10} N_\mathrm{DCN} & 3 & 18 (LTE), 16 (non-LTE) & $\log_{10}(\mathrm{[cm^{-2}]})$ &\\
\mathrm{FWHM_{B3}} & 0.5\tablenotemark{d} & variable\tablenotemark{e} & [km\,s$^{-1}$] &\\
\mathrm{FWHM_{B6}} & 0.2\tablenotemark{d} & variable\tablenotemark{e} & [km\,s$^{-1}$] &\\
\Delta v_\mathrm{HCN 1-0} & $-$0.3 & 0.3 & [km\,s$^{-1}$] &\\
\Delta v_\mathrm{H13CN 1-0} & $-$0.3 & 0.3 & [km\,s$^{-1}$] &\\
\Delta v_\mathrm{HCN 3-2} & $-$0.2 & 0.2 & [km\,s$^{-1}$] &\\
\Delta v_\mathrm{DCN 3-2} & $-$0.2 & 0.2 & [km\,s$^{-1}$] &\\
T_\mathrm{kin} & 10 & 90 & [K] & non-LTE \\
\log_{10} n_\mathrm{H_2} & 3 & 10 & $\log_{10}(\mathrm{[cm^{-3}]})$ & non-LTE \\
\hline
\end{tabular}
\tablenotetext{a}{Lower bound of flat prior.}
\tablenotetext{b}{Upper bound of flat prior.}
\tablenotetext{c}{Marks parameters used exclusively in the LTE or the non-LTE run.}
\tablenotetext{d}{Equal to the channel width.}
\tablenotetext{e}{Initial value for innermost radial bin is 20\,km\,s$^{-1}$. Dynamically adjusted as larger and larger radii are fitted (see Figure \ref{fig:FWHM_fit_example} and text).}
\end{table*}

\begin{table*}
\centering
\caption{Free parameters for the fitting of azimuthally averaged spectra to derive \ce{N2H+} and \ce{N2D+} column densities.}\label{tab:N2H+_free_parameters_overview}
\begin{tabular}{Cccc}
\hline
\mathrm{Parameter} & prior low\tablenotemark{a} & prior high\tablenotemark{b} & unit\\
\hline
\log_{10} N_\mathrm{N_2H^+} & 3 & 18 & $\log_{10}(\mathrm{[cm^{-2}]})$\\
\log_{10} N_\mathrm{N_2D^+} & 3 & 18 & $\log_{10}(\mathrm{[cm^{-2}]})$\\
\mathrm{FWHM_{N_2H^+ 3-2}} & variable\tablenotemark{c} & variable\tablenotemark{d} & [km\,s$^{-1}$]\\
\mathrm{FWHM_{N_2D^+ 3-2}} & 0.2\tablenotemark{c} & variable\tablenotemark{d} & [km\,s$^{-1}$]\\
\Delta v_\mathrm{N_2H^+ 3-2} & $-$0.2 & 0.2 & [km\,s$^{-1}$]\\
\Delta v_\mathrm{N_2D^+ 3-2} & $-$0.2 & 0.2 & [km\,s$^{-1}$]\\
\hline
\end{tabular}
\tablenotetext{a}{Lower bound of flat prior.}
\tablenotetext{b}{Upper bound of flat prior.}
\tablenotetext{c}{Equal to the channel width (see Table \ref{tab:N2Hp_image_params}).}
\tablenotetext{d}{Initial value for innermost radial bin is 20\,km\,s$^{-1}$. Dynamically adjusted as larger and larger radii are fitted (see Figure \ref{fig:FWHM_fit_example} and text).}
\end{table*}

\section{Molecular data}\label{appendix:molecular_data}
We used spectral line data from the CDMS database \citep{Muller01,Muller05,Endres16}, downloaded with the \texttt{SPECTCOL} software. CDMS gives the following sources for the data:
\begin{itemize}
    \item HCN: \citet{Ebenstein84,Maki00,Ahrens02,Thorwirth03,Lapinov06}
    \item DCN: \citet{DeLeon84,Mollmann02,Brunken04}
    \item H$^{13}$CN: \citet{Winnewisser78,Preusser93,Maki00,Maiwald00,Fuchs04,Cazzoli05}
    \item \nthp: \citet{Verhoeve90,Caselli95,Amano05,Cazzoli12}
    \item \ntdp: \citet{Dore04,Amano05,Pagani09}
\end{itemize}
Table \ref{tab:atomic_parameters_HCN} lists the molecular data for HCN 1--0 and 3--2, DCN 3--2 and H$^{13}$CN 1--0. Tables \ref{tab:atomic_parameters_N2H+} and \ref{tab:atomic_parameters_N2D+} show corresponding data for \ce{N2H+} 3--2 and \ce{N2D+} 3--2, respectively.

\begin{table*}
\centering
\caption{Spectral line data for HCN 1--0 and 3--2, DCN 3--2 and H$^{13}$CN 1--0 used in this study.}\label{tab:atomic_parameters_HCN}
\begin{tabular}{cccccccC}
\hline
\hline
Molecule & $J$ & $F$ & $\nu_0$ & $g_u$ & $A_{ul}$ & $r_i$\tablenotemark{a} & \delta v\tablenotemark{b}\\
& & & [GHz] & & [s$^{-1}$] & & [\mathrm{km\,s}^{-1}] \\
\hline
\multicolumn{8}{c}{hyperfine-resolved transitions}\\
\hline
HCN & 1--0 & 0--1 & 88.6339357 & 1 & $2.41\times10^{-5}$ & 0.1111 & -7.063\\
HCN & 1--0 & 2--1 & 88.6318475 & 5 & $2.41\times10^{-5}$ & 0.5555 & 0.000\\
HCN & 1--0 & 1--1 & 88.6304156 & 3 & $2.41\times10^{-5}$ & 0.3333 & 4.843\\
\hline
HCN & 3--2 & 2--2 & 265.8885221 & 5 & $1.30\times10^{-4}$ & 0.0370 & -2.354\\
HCN & 3--2 & 2--3 & 265.8869793 & 5 & $3.71\times10^{-6}$ & 0.0011 & -0.615\\
HCN & 3--2 & 4--3 & 265.8864999 & 9 & $8.36\times10^{-4}$ & 0.4286 & -0.074\\
HCN & 3--2 & 3--2 & 265.8864339 & 7 & $7.43\times10^{-4}$ & 0.2963 & 0.000\\
HCN & 3--2 & 2--1 & 265.8861886 & 5 & $7.02\times10^{-4}$ & 0.2000 & 0.277\\
HCN & 3--2 & 3--3 & 265.8848912 & 7 & $9.29\times10^{-5}$ & 0.0370 & 1.739\\
\hline
DCN & 3--2 & 2--2 & 217.240622 & 5 & $7.12\times10^{-5}$ & 0.0370 & -2.876\\
DCN & 3--2 & 2--3 & 217.239079 & 5 & $2.03\times10^{-6}$ & 0.0011 & -0.747\\
DCN & 3--2 & 4--3 & 217.238612 & 9 & $4.57\times10^{-4}$ & 0.4286 & -0.102\\
DCN & 3--2 & 3--2 & 217.238555 & 7 & $4.07\times10^{-4}$ & 0.2963 & -0.024\\
DCN & 3--2 & 2--1 & 217.238300 & 5 & $3.84\times10^{-4}$ & 0.2000 & 0.328\\
DCN & 3--2 & 3--3 & 217.236999 & 7 & $5.08\times10^{-5}$ & 0.0370 & 2.124\\
\hline
H$^{13}$CN & 1--0 & 0--1 & 86.3422543 & 1 & $2.23\times10^{-5}$ & 0.11 & -7.25\\
H$^{13}$CN & 1--0 & 2--1 & 86.3401666 & 5 & $2.23\times10^{-5}$ & 0.56 & 0.00\\
H$^{13}$CN & 1--0 & 1--1 & 86.3387352 & 3 & $2.23\times10^{-5}$ & 0.33 & 4.97\\
\hline
\multicolumn{8}{c}{hyperfine-unresolved transitions}\\
\hline
HCN & 1--0 & & 88.6318475 & 9 & $2.41\times10^{-5}$ & & \\
HCN & 3--2 & & 265.8864339 & 21 & $8.36\times10^{-4}$ & & \\
DCN & 3--2 & & 217.2385378 & 21 & $4.57\times10^{-4}$ & & \\
H$^{13}$CN & 1--0 & & 86.3401666 & 9 & $2.23\times10^{-5}$ & & \\
\hline
\end{tabular}
\tablenotetext{a}{Relative strength of hyperfine component.}
\tablenotetext{b}{Velocity offset with respect to the hyperfine-unresolved transition.}
\end{table*}

\begin{table*}
\centering
\caption{Spectral line data of \nthp 3--2 used in this study.}\label{tab:atomic_parameters_N2H+}
\begin{tabular}{cccccccC}
\hline
\hline
$J$ & $F_1$ & $F$ & $\nu_0$ & $g_u$ & $A_{ul}$ & $r_i$\tablenotemark{a} & \delta v\tablenotemark{b} \\
& & & [GHz] & & [s$^{-1}$] & & [\mathrm{km\,s}^{-1}] \\
\hline
\multicolumn{8}{c}{hyperfine-resolved transitions}\\
\hline
3--2 & 2--2 & 1--2 & 279.5147240 & 3 & $4.88\times10^{-5}$ & 0.0018 & -3.191\\
3--2 & 2--2 & 3--2 & 279.5145858 & 7 & $2.06\times10^{-5}$ & 0.0018 & -3.043\\
3--2 & 2--2 & 1--1 & 279.5143988 & 3 & $9.32\times10^{-5}$ & 0.0035 & -2.842\\
3--2 & 2--2 & 3--3 & 279.5143328 & 7 & $1.60\times10^{-4}$ & 0.0141 & -2.771\\
3--2 & 2--2 & 2--2 & 279.5142143 & 5 & $1.82\times10^{-4}$ & 0.0115 & -2.644\\
3--2 & 2--2 & 2--3 & 279.5139612 & 5 & $3.50\times10^{-5}$ & 0.0022 & -2.373\\
3--2 & 2--2 & 2--1 & 279.5138890 & 5 & $3.42\times10^{-5}$ & 0.0022 & -2.295\\
3--2 & 4--3 & 3--3 & 279.5123144 & 7 & $1.16\times10^{-4}$ & 0.0103 & -0.606\\
3--2 & 3--2 & 2--2 & 279.5121160 & 5 & $1.18\times10^{-4}$ & 0.0074 & -0.394\\
3--2 & 4--3 & 5--4 & 279.5118523 & 11 & $1.26\times10^{-3}$ & 0.1752 & -0.111\\
3--2 & 2--1 & 1--1 & 279.5118447 & 3 & $4.95\times10^{-4}$ & 0.0188 & -0.103\\
3--2 & 4--3 & 4--3 & 279.5118028 & 9 & $1.20\times10^{-3}$ & 0.1360 & -0.058\\
3--2 & 4--3 & 3--2 & 279.5118022 & 7 & $1.14\times10^{-3}$ & 0.1008 & -0.057\\
3--2 & 3--2 & 4--3 & 279.5117986 & 9 & $1.10\times10^{-3}$ & 0.1253 & -0.053\\
3--2 & 3--2 & 2--1 & 279.5117908 & 5 & $9.68\times10^{-4}$ & 0.0612 & -0.045\\
3--2 & 3--2 & 3--2 & 279.5116770 & 7 & $1.00\times10^{-3}$ & 0.0888 & 0.077\\
3--2 & 2--1 & 1--2 & 279.5116372 & 3 & $2.87\times10^{-5}$ & 0.0011 & 0.120\\
3--2 & 2--1 & 3--2 & 279.5114990 & 7 & $1.07\times10^{-3}$ & 0.0948 & 0.268\\
3--2 & 3--2 & 3--3 & 279.5114240 & 7 & $1.59\times10^{-4}$ & 0.0141 & 0.349\\
3--2 & 4--3 & 4--4 & 279.5114066 & 9 & $6.35\times10^{-5}$ & 0.0072 & 0.367\\
3--2 & 2--1 & 1--0 & 279.5113769 & 3 & $5.88\times10^{-4}$ & 0.0223 & 0.399\\
3--2 & 2--1 & 2--1 & 279.5113349 & 5 & $7.60\times10^{-4}$ & 0.0481 & 0.444\\
3--2 & 2--1 & 2--2 & 279.5111274 & 5 & $2.44\times10^{-4}$ & 0.0155 & 0.667\\
3--2 & 3--3 & 4--3 & 279.5102649 & 9 & $1.21\times10^{-5}$ & 0.0014 & 1.592\\
3--2 & 3--3 & 3--3 & 279.5098903 & 7 & $8.05\times10^{-5}$ & 0.0071 & 1.994\\
3--2 & 3--3 & 4--4 & 279.5098686 & 9 & $1.46\times10^{-4}$ & 0.0167 & 2.017\\
3--2 & 3--3 & 2--2 & 279.5098171 & 5 & $1.55\times10^{-4}$ & 0.0098 & 2.072\\
3--2 & 3--3 & 3--4 & 279.5094940 & 7 & $1.13\times10^{-5}$ & 0.0010 & 2.419\\
\hline
\multicolumn{8}{c}{hyperfine-unresolved transition}\\
\hline
3--2 & & & 279.5117491 & 63 & $1.26\times10^{-3}$ & & \\
\hline
\end{tabular}
\tablenotetext{a}{Relative strength of hyperfine component.}
\tablenotetext{b}{Velocity offset with respect to the hyperfine-unresolved transition.}
\end{table*}

\begin{table*}
\centering
\caption{Spectral line data of \ntdp 3--2 used in this study.}\label{tab:atomic_parameters_N2D+}
\begin{tabular}{cccccccC}
\hline
\hline
$J$ & $F_1$ & $F$ & $\nu_0$ & $g_u$ & $A_{ul}$ & $r_i$\tablenotemark{a} & \delta v\tablenotemark{b}\\
& & & [GHz] & & [s$^{-1}$] & & [\mathrm{km\,s}^{-1}] \\
\hline
\multicolumn{8}{c}{hyperfine-resolved transitions}\\
\hline
3--2 & 2--2 & 1--2 & 231.3248131 & 3 & $2.76\times10^{-5}$ & 0.0019 & -3.868\\
3--2 & 2--2 & 3--2 & 231.3246669 & 7 & $1.17\times10^{-5}$ & 0.0018 & -3.679\\
3--2 & 2--2 & 1--1 & 231.3244843 & 3 & $5.25\times10^{-5}$ & 0.0035 & -3.442\\
3--2 & 2--2 & 3--3 & 231.3244147 & 7 & $9.04\times10^{-5}$ & 0.0142 & -3.352\\
3--2 & 2--2 & 2--2 & 231.3242977 & 5 & $1.03\times10^{-4}$ & 0.0115 & -3.200\\
3--2 & 2--2 & 2--3 & 231.3240455 & 5 & $1.99\times10^{-5}$ & 0.0022 & -2.873\\
3--2 & 2--2 & 2--1 & 231.3239689 & 5 & $1.94\times10^{-5}$ & 0.0022 & -2.774\\
3--2 & 4--3 & 3--3 & 231.3223976 & 7 & $6.57\times10^{-5}$ & 0.0103 & -0.738\\
3--2 & 3--2 & 2--2 & 231.3222005 & 5 & $6.64\times10^{-5}$ & 0.0074 & -0.482\\
3--2 & 2--1 & 1--1 & 231.3219314 & 3 & $2.81\times10^{-4}$ & 0.0188 & -0.134\\
3--2 & 4--3 & 5--4 & 231.3219281 & 11 & $7.14\times10^{-4}$ & 0.1758 & -0.129\\
3--2 & 4--3 & 3--2 & 231.3218798 & 7 & $6.46\times10^{-4}$ & 0.1012 & -0.067\\
3--2 & 4--3 & 4--3 & 231.3218794 & 9 & $6.78\times10^{-4}$ & 0.1365 & -0.066\\
3--2 & 3--2 & 4--3 & 231.3218770 & 9 & $6.24\times10^{-4}$ & 0.1257 & -0.063\\
3--2 & 3--2 & 2--1 & 231.3218717 & 5 & $5.49\times10^{-4}$ & 0.0614 & -0.056\\
3--2 & 3--2 & 3--2 & 231.3217560 & 7 & $5.69\times10^{-4}$ & 0.0891 & 0.094\\
3--2 & 2--1 & 3--2 & 231.3215801 & 7 & $6.07\times10^{-4}$ & 0.0951 & 0.322\\
3--2 & 3--2 & 3--3 & 231.3215039 & 7 & $9.01\times10^{-5}$ & 0.0141 & 0.420\\
3--2 & 4--3 & 4--4 & 231.3214836 & 9 & $3.59\times10^{-5}$ & 0.0072 & 0.447\\
3--2 & 2--1 & 1--0 & 231.3214595 & 3 & $3.33\times10^{-4}$ & 0.0224 & 0.478\\
3--2 & 2--1 & 2--1 & 231.3214160 & 5 & $4.31\times10^{-4}$ & 0.0482 & 0.534\\
3--2 & 2--1 & 2--2 & 231.3212109 & 5 & $1.39\times10^{-4}$ & 0.0155 & 0.800\\
3--2 & 3--3 & 3--3 & 231.3199797 & 7 & $4.55\times10^{-5}$ & 0.0071 & 2.396\\
3--2 & 3--3 & 4--4 & 231.3199570 & 9 & $8.30\times10^{-5}$ & 0.0167 & 2.425\\
3--2 & 3--3 & 2--2 & 231.3199064 & 5 & $8.82\times10^{-5}$ & 0.0099 & 2.491\\
\hline
\multicolumn{8}{c}{hyperfine-unresolved transition}\\
\hline
3--2 & & & 231.3218283 & 63 & $4.38\times10^{-4}$ & & \\
\hline
\end{tabular}
\tablenotetext{a}{Relative strength of hyperfine component.}
\tablenotetext{b}{Velocity offset with respect to the hyperfine-unresolved transition.}
\end{table*}

\section{Data available for Download}\label{appendix:data_download}
The MAPS data products (calibrated visibilities, image cubes, moment maps, radial emission profiles and emission surfaces) as well as the scripts used to generate the data products are available for download from the ALMA archive (\url{https://almascience.nrao.edu/alma-data/lp/maps}) and from the MAPS website (\url{http://www.alma-maps.info}). In addition, the following data, specific to this paper, are also available for download:
\begin{itemize}
    \item image cubes of HCN 1--0 and 3--2 as well as H$^{13}$CN 1--0 covering all hyperfine transitions (Section \ref{sec:adopted_MAPS_imaging_products})
    \item visibilities, (smoothed) image cubes, zeroth moment maps and radial emission profiles of the archival \nthp 3--2 data (Sections \ref{sec:archival_N2H+_data}, \ref{sec:disk_integrated_flux_mom0}, \ref{sec:N2H+_radial_emission_profiles})
    \item azimuthally averaged spectra of the HCN, DCN, H$^{13}$CN, \nthp and \ntdp emission lines analysed in this paper (Section \ref{sec:az_averaged_spectra})
    \item radial profiles of column density, optical depth, temperature and deuteration fraction (Section \ref{sec:analysis})
    \item image cubes, zeroth moment maps and radial emission profiles of the archival DCO$^+$ data (Section \ref{sec:DCN_N2D+_DCO+_comparison})
    \item visibilities, image cubes, radial emission profiles and associated scripts used for studying the negative emission of \ntdp 3--2 (Appendix \ref{appendix:negative_flux})
    \item disk-integrated spectra of \nthp 3--2 (Appendix \ref{appendix:az_averaged_spectra_gallery})
    \item molecular data and partition functions (Appendix \ref{appendix:molecular_data})
\end{itemize}

\bibliography{bibliography}{}

\begin{thebibliography}{}
\expandafter\ifx\csname natexlab\endcsname\relax\def\natexlab#1{#1}\fi
\providecommand{\url}[1]{\href{#1}{#1}}
\providecommand{\dodoi}[1]{doi:~\href{http://doi.org/#1}{\nolinkurl{#1}}}
\providecommand{\doeprint}[1]{\href{http://ascl.net/#1}{\nolinkurl{http://ascl.net/#1}}}
\providecommand{\doarXiv}[1]{\href{https://arxiv.org/abs/#1}{\nolinkurl{https://arxiv.org/abs/#1}}}

\bibitem[{{Ahrens} {et~al.}(2002){Ahrens}, {Lewen}, {Takano}, {Winnewisser},
  {Urban}, {Negirev}, \& {Koroliev}}]{Ahrens02}
{Ahrens}, V., {Lewen}, F., {Takano}, S., {et~al.} 2002, Zeitschrift
  Naturforschung Teil A, 57, 669, \dodoi{10.1515/zna-2002-0806}

\bibitem[{{Aikawa} {et~al.}(2018){Aikawa}, {Furuya}, {Hincelin}, \&
  {Herbst}}]{Aikawa18}
{Aikawa}, Y., {Furuya}, K., {Hincelin}, U., \& {Herbst}, E. 2018, \apj, 855,
  119, \dodoi{10.3847/1538-4357/aaad6c}

\bibitem[{{Aikawa} {et~al.}(2015){Aikawa}, {Furuya}, {Nomura}, \&
  {Qi}}]{Aikawa15}
{Aikawa}, Y., {Furuya}, K., {Nomura}, H., \& {Qi}, C. 2015, \apj, 807, 120,
  \dodoi{10.1088/0004-637X/807/2/120}

\bibitem[{{Aikawa} {et~al.}(1999){Aikawa}, {Umebayashi}, {Nakano}, \&
  {Miyama}}]{Aikawa99}
{Aikawa}, Y., {Umebayashi}, T., {Nakano}, T., \& {Miyama}, S.~M. 1999, \apj,
  519, 705, \dodoi{10.1086/307400}

\bibitem[{{Aikawa} {et~al.}(2021){Aikawa}, {Cataldi}, {Yamato}, {Zhang},
  {Booth}, {Furuya}, {Andrews}, {Bae}, {Bergin}, {Bergner}, {Bosman},
  {Cleeves}, {Czekala}, {Guzm{\'a}n}, {Huang}, {Ilee}, {Law}, {Le Gal},
  {Loomis}, {M{\'e}nard}, {Nomura}, {{\"O}berg}, {Qi}, {Schwarz}, {Teague},
  {Tsukagoshi}, {Walsh}, \& {Wilner}}]{Aikawa21_MAPS}
{Aikawa}, Y., {Cataldi}, G., {Yamato}, Y., {et~al.} 2021, \apjs, 257, 13,
  \dodoi{10.3847/1538-4365/ac143c}

\bibitem[{{Alarc{\'o}n} {et~al.}(2021){Alarc{\'o}n}, {Bosman}, {Bergin},
  {Zhang}, {Teague}, {Bae}, {Aikawa}, {Andrews}, {Booth}, {Calahan}, {Cataldi},
  {Czekala}, {Huang}, {Ilee}, {Law}, {Le Gal}, {Liu}, {Long}, {Loomis},
  {M{\'e}nard}, {{\"O}berg}, {Schwarz}, {van't Hoff}, {Walsh}, \&
  {Wilner}}]{Alarcon21_MAPS}
{Alarc{\'o}n}, F., {Bosman}, A.~D., {Bergin}, E.~A., {et~al.} 2021, \apjs, 257,
  8, \dodoi{10.3847/1538-4365/ac22ae}

\bibitem[{{Alexander}(2017)}]{Alexander17}
{Alexander}, C. M.~O. 2017, Philosophical Transactions of the Royal Society of
  London Series A, 375, 20150384, \dodoi{10.1098/rsta.2015.0384}

\bibitem[{{Amano} {et~al.}(2005){Amano}, {Hirao}, \& {Takano}}]{Amano05}
{Amano}, T., {Hirao}, T., \& {Takano}, J. 2005, Journal of Molecular
  Spectroscopy, 234, 170, \dodoi{10.1016/j.jms.2005.09.004}

\bibitem[{{Andrews} {et~al.}(2018){Andrews}, {Huang}, {P{\'e}rez}, {Isella},
  {Dullemond}, {Kurtovic}, {Guzm{\'a}n}, {Carpenter}, {Wilner}, {Zhang}, {Zhu},
  {Birnstiel}, {Bai}, {Benisty}, {Hughes}, {{\"O}berg}, \& {Ricci}}]{Andrews18}
{Andrews}, S.~M., {Huang}, J., {P{\'e}rez}, L.~M., {et~al.} 2018, \apjl, 869,
  L41, \dodoi{10.3847/2041-8213/aaf741}

\bibitem[{{Astropy Collaboration} {et~al.}(2013){Astropy Collaboration},
  {Robitaille}, {Tollerud}, {Greenfield}, {Droettboom}, {Bray}, {Aldcroft},
  {Davis}, {Ginsburg}, {Price-Whelan}, {Kerzendorf}, {Conley}, {Crighton},
  {Barbary}, {Muna}, {Ferguson}, {Grollier}, {Parikh}, {Nair}, {Unther},
  {Deil}, {Woillez}, {Conseil}, {Kramer}, {Turner}, {Singer}, {Fox}, {Weaver},
  {Zabalza}, {Edwards}, {Azalee Bostroem}, {Burke}, {Casey}, {Crawford},
  {Dencheva}, {Ely}, {Jenness}, {Labrie}, {Lim}, {Pierfederici}, {Pontzen},
  {Ptak}, {Refsdal}, {Servillat}, \& {Streicher}}]{astropy13}
{Astropy Collaboration}, {Robitaille}, T.~P., {Tollerud}, E.~J., {et~al.} 2013,
  \aap, 558, A33, \dodoi{10.1051/0004-6361/201322068}

\bibitem[{{Bergin} {et~al.}(2003){Bergin}, {Calvet}, {D'Alessio}, \&
  {Herczeg}}]{Bergin03}
{Bergin}, E., {Calvet}, N., {D'Alessio}, P., \& {Herczeg}, G.~J. 2003, \apjl,
  591, L159, \dodoi{10.1086/377148}

\bibitem[{{Bergin} {et~al.}(2001){Bergin}, {Ciardi}, {Lada}, {Alves}, \&
  {Lada}}]{Bergin01}
{Bergin}, E.~A., {Ciardi}, D.~R., {Lada}, C.~J., {Alves}, J., \& {Lada}, E.~A.
  2001, \apj, 557, 209, \dodoi{10.1086/321625}

\bibitem[{{Bergin} {et~al.}(2016){Bergin}, {Du}, {Cleeves}, {Blake}, {Schwarz},
  {Visser}, \& {Zhang}}]{Bergin16}
{Bergin}, E.~A., {Du}, F., {Cleeves}, L.~I., {et~al.} 2016, \apj, 831, 101,
  \dodoi{10.3847/0004-637X/831/1/101}

\bibitem[{{Bergner} {et~al.}(2019){Bergner}, {{\"O}berg}, {Bergin}, {Loomis},
  {Pegues}, \& {Qi}}]{Bergner19}
{Bergner}, J.~B., {{\"O}berg}, K.~I., {Bergin}, E.~A., {et~al.} 2019, \apj,
  876, 25, \dodoi{10.3847/1538-4357/ab141e}

\bibitem[{{Bergner} {et~al.}(2020{\natexlab{a}}){Bergner}, {{\"O}berg},
  {Bergin}, {Loomis}, {Pegues}, \& {Qi}}]{Bergner20_erratum}
---. 2020{\natexlab{a}}, \apj, 896, 176, \dodoi{10.3847/1538-4357/ab98f7}

\bibitem[{{Bergner} {et~al.}(2020{\natexlab{b}}){Bergner}, {{\"O}berg},
  {Bergin}, {Andrews}, {Blake}, {Carpenter}, {Cleeves}, {Guzm{\'a}n}, {Huang},
  {J{\o}rgensen}, {Qi}, {Schwarz}, {Williams}, \&
  {Wilner}}]{Bergner20_volatile_evolution}
---. 2020{\natexlab{b}}, \apj, 898, 97, \dodoi{10.3847/1538-4357/ab9e71}

\bibitem[{{Bergner} {et~al.}(2021){Bergner}, {{\"O}berg}, {Guzm{\'a}n}, {Law},
  {Loomis}, {Cataldi}, {Bosman}, {Aikawa}, {Andrews}, {Bergin}, {Booth},
  {Cleeves}, {Czekala}, {Huang}, {Ilee}, {Le Gal}, {Long}, {Nomura},
  {M{\'e}nard}, {Qi}, {Schwarz}, {Teague}, {Tsukagoshi}, {Walsh}, {Wilner}, \&
  {Yamato}}]{Bergner21_MAPS}
{Bergner}, J.~B., {{\"O}berg}, K.~I., {Guzm{\'a}n}, V.~V., {et~al.} 2021,
  \apjs, 257, 11, \dodoi{10.3847/1538-4365/ac143a}

\bibitem[{{Bockel{\'e}e-Morvan} {et~al.}(1998){Bockel{\'e}e-Morvan}, {Gautier},
  {Lis}, {Young}, {Keene}, {Phillips}, {Owen}, {Crovisier}, {Goldsmith},
  {Bergin}, {Despois}, \& {Wootten}}]{Bockelee-Morvan98}
{Bockel{\'e}e-Morvan}, D., {Gautier}, D., {Lis}, D.~C., {et~al.} 1998, \icarus,
  133, 147, \dodoi{10.1006/icar.1998.5916}

\bibitem[{{Bockel{\'e}e-Morvan} {et~al.}(2015){Bockel{\'e}e-Morvan},
  {Calmonte}, {Charnley}, {Duprat}, {Engrand}, {Gicquel}, {H{\"a}ssig},
  {Jehin}, {Kawakita}, {Marty}, {Milam}, {Morse}, {Rousselot}, {Sheridan}, \&
  {Wirstr{\"o}m}}]{Bockelee-Morvan15}
{Bockel{\'e}e-Morvan}, D., {Calmonte}, U., {Charnley}, S., {et~al.} 2015, \ssr,
  197, 47, \dodoi{10.1007/s11214-015-0156-9}

\bibitem[{{Booth} {et~al.}(2017){Booth}, {Dent}, {Jord{\'a}n}, {Lestrade},
  {Hales}, {Wyatt}, {Casassus}, {Ertel}, {Greaves}, {Kennedy}, {Matr{\`a}},
  {Augereau}, \& {Villard}}]{Booth17}
{Booth}, M., {Dent}, W. R.~F., {Jord{\'a}n}, A., {et~al.} 2017, \mnras, 469,
  3200, \dodoi{10.1093/mnras/stx1072}

\bibitem[{{Bosman} {et~al.}(2021{\natexlab{a}}){Bosman}, {Alarc{\'o}n},
  {Bergin}, {Zhang}, {van't Hoff}, {{\"O}berg}, {Guzm{\'a}n}, {Walsh},
  {Aikawa}, {Andrews}, {Bergner}, {Booth}, {Cataldi}, {Cleeves}, {Czekala},
  {Furuya}, {Huang}, {Ilee}, {Law}, {Le Gal}, {Liu}, {Long}, {Loomis},
  {M{\'e}nard}, {Nomura}, {Qi}, {Schwarz}, {Teague}, {Tsukagoshi}, {Yamato}, \&
  {Wilner}}]{Bosman21_MAPS_CtoO}
{Bosman}, A.~D., {Alarc{\'o}n}, F., {Bergin}, E.~A., {et~al.}
  2021{\natexlab{a}}, \apjs, 257, 7, \dodoi{10.3847/1538-4365/ac1435}

\bibitem[{{Bosman} {et~al.}(2021{\natexlab{b}}){Bosman}, {Bergin}, {Loomis},
  {Andrews}, {van't Hoff}, {Teague}, {{\"O}berg}, {Guzm{\'a}n}, {Walsh},
  {Aikawa}, {Alarc{\'o}n}, {Bae}, {Bergner}, {Booth}, {Cataldi}, {Cleeves},
  {Czekala}, {Huang}, {Ilee}, {Law}, {Le Gal}, {Liu}, {Long}, {M{\'e}nard},
  {Nomura}, {P{\'e}rez}, {Qi}, {Schwarz}, {Sierra}, {Tsukagoshi}, {Yamato},
  {Wilner}, \& {Zhang}}]{Bosman21_MAPS_inner20au}
{Bosman}, A.~D., {Bergin}, E.~A., {Loomis}, R.~A., {et~al.} 2021{\natexlab{b}},
  \apjs, 257, 15, \dodoi{10.3847/1538-4365/ac1433}

\bibitem[{{Brasser} \& {Morbidelli}(2013)}]{Brasser13}
{Brasser}, R., \& {Morbidelli}, A. 2013, \icarus, 225, 40,
  \dodoi{10.1016/j.icarus.2013.03.012}

\bibitem[{{Br{\"u}nken} {et~al.}(2004){Br{\"u}nken}, {Fuchs}, {Lewen}, {Urban},
  {Giesen}, \& {Winnewisser}}]{Brunken04}
{Br{\"u}nken}, S., {Fuchs}, U., {Lewen}, F., {et~al.} 2004, Journal of
  Molecular Spectroscopy, 225, 152, \dodoi{10.1016/j.jms.2004.02.021}

\bibitem[{{Carney} {et~al.}(2017){Carney}, {Hogerheijde}, {Loomis}, {Salinas},
  {{\"O}berg}, {Qi}, \& {Wilner}}]{Carney17}
{Carney}, M.~T., {Hogerheijde}, M.~R., {Loomis}, R.~A., {et~al.} 2017, \aap,
  605, A21, \dodoi{10.1051/0004-6361/201629342}

\bibitem[{{Carney} {et~al.}(2018){Carney}, {Fedele}, {Hogerheijde}, {Favre},
  {Walsh}, {Bruderer}, {Miotello}, {Murillo}, {Klaassen}, {Henning}, \& {van
  Dishoeck}}]{Carney18}
{Carney}, M.~T., {Fedele}, D., {Hogerheijde}, M.~R., {et~al.} 2018, \aap, 614,
  A106, \dodoi{10.1051/0004-6361/201732384}

\bibitem[{{Caselli} {et~al.}(1995){Caselli}, {Myers}, \&
  {Thaddeus}}]{Caselli95}
{Caselli}, P., {Myers}, P.~C., \& {Thaddeus}, P. 1995, \apjl, 455, L77,
  \dodoi{10.1086/309805}

\bibitem[{{Cazzoli} {et~al.}(2012){Cazzoli}, {Cludi}, {Buffa}, \&
  {Puzzarini}}]{Cazzoli12}
{Cazzoli}, G., {Cludi}, L., {Buffa}, G., \& {Puzzarini}, C. 2012, \apjs, 203,
  11, \dodoi{10.1088/0067-0049/203/1/11}

\bibitem[{{Cazzoli} \& {Puzzarini}(2005)}]{Cazzoli05}
{Cazzoli}, G., \& {Puzzarini}, C. 2005, Journal of Molecular Spectroscopy, 233,
  280, \dodoi{10.1016/j.jms.2005.07.009}

\bibitem[{{Ceccarelli} {et~al.}(2014){Ceccarelli}, {Caselli},
  {Bockel{\'e}e-Morvan}, {Mousis}, {Pizzarello}, {Robert}, \&
  {Semenov}}]{Ceccarelli14}
{Ceccarelli}, C., {Caselli}, P., {Bockel{\'e}e-Morvan}, D., {et~al.} 2014, in
  Protostars and Planets VI, ed. H.~{Beuther}, R.~S. {Klessen}, C.~P.
  {Dullemond}, \& T.~{Henning}, 859,
  \dodoi{10.2458/azu_uapress_9780816531240-ch037}

\bibitem[{{Chapillon} {et~al.}(2012){Chapillon}, {Guilloteau}, {Dutrey},
  {Pi{\'e}tu}, \& {Gu{\'e}lin}}]{Chapillon12}
{Chapillon}, E., {Guilloteau}, S., {Dutrey}, A., {Pi{\'e}tu}, V., \&
  {Gu{\'e}lin}, M. 2012, \aap, 537, A60, \dodoi{10.1051/0004-6361/201116762}

\bibitem[{{Cleeves}(2016)}]{Cleeves16_COThermalInversion}
{Cleeves}, L.~I. 2016, \apjl, 816, L21, \dodoi{10.3847/2041-8205/816/2/L21}

\bibitem[{{Cleeves} {et~al.}(2016){Cleeves}, {{\"O}berg}, {Wilner}, {Huang},
  {Loomis}, {Andrews}, \& {Czekala}}]{Cleeves16}
{Cleeves}, L.~I., {{\"O}berg}, K.~I., {Wilner}, D.~J., {et~al.} 2016, \apj,
  832, 110, \dodoi{10.3847/0004-637X/832/2/110}

\bibitem[{{Cleeves} {et~al.}(2018){Cleeves}, {{\"O}berg}, {Wilner}, {Huang},
  {Loomis}, {Andrews}, \& {Guzman}}]{Cleeves18}
---. 2018, \apj, 865, 155, \dodoi{10.3847/1538-4357/aade96}

\bibitem[{{Crovisier} {et~al.}(2004){Crovisier}, {Bockel{\'e}e-Morvan},
  {Colom}, {Biver}, {Despois}, {Lis}, \& {Teamtarget-of-opportunity radio
  observations of comets}}]{Crovisier04}
{Crovisier}, J., {Bockel{\'e}e-Morvan}, D., {Colom}, P., {et~al.} 2004, \aap,
  418, 1141, \dodoi{10.1051/0004-6361:20035688}

\bibitem[{{Cuppen} {et~al.}(2017){Cuppen}, {Walsh}, {Lamberts}, {Semenov},
  {Garrod}, {Penteado}, \& {Ioppolo}}]{Cuppen17}
{Cuppen}, H.~M., {Walsh}, C., {Lamberts}, T., {et~al.} 2017, \ssr, 212, 1,
  \dodoi{10.1007/s11214-016-0319-3}

\bibitem[{{Cyburt} {et~al.}(2016){Cyburt}, {Fields}, {Olive}, \&
  {Yeh}}]{Cyburt16}
{Cyburt}, R.~H., {Fields}, B.~D., {Olive}, K.~A., \& {Yeh}, T.-H. 2016, Reviews
  of Modern Physics, 88, 015004, \dodoi{10.1103/RevModPhys.88.015004}

\bibitem[{{Czekala} {et~al.}(2021){Czekala}, {Loomis}, {Teague}, {Booth},
  {Huang}, {Cataldi}, {Ilee}, {Law}, {Walsh}, {Bosman}, {Guzm{\'a}n}, {Gal},
  {{\"O}berg}, {Yamato}, {Aikawa}, {Andrews}, {Bae}, {Bergin}, {Bergner},
  {Cleeves}, {Kurtovic}, {M{\'e}nard}, {Nomura}, {P{\'e}rez}, {Qi}, {Schwarz},
  {Tsukagoshi}, {Waggoner}, {Wilner}, \& {Zhang}}]{Czekala21_MAPS}
{Czekala}, I., {Loomis}, R.~A., {Teague}, R., {et~al.} 2021, \apjs, 257, 2,
  \dodoi{10.3847/1538-4365/ac1430}

\bibitem[{{DeLeon} \& {Muenter}(1984)}]{DeLeon84}
{DeLeon}, R.~L., \& {Muenter}, J.~S. 1984, \jcp, 80, 6092,
  \dodoi{10.1063/1.446708}

\bibitem[{{Dore} {et~al.}(2004){Dore}, {Caselli}, {Beninati}, {Bourke},
  {Myers}, \& {Cazzoli}}]{Dore04}
{Dore}, L., {Caselli}, P., {Beninati}, S., {et~al.} 2004, \aap, 413, 1177,
  \dodoi{10.1051/0004-6361:20034025}

\bibitem[{{Dumouchel} {et~al.}(2010){Dumouchel}, {Faure}, \&
  {Lique}}]{Dumouchel10}
{Dumouchel}, F., {Faure}, A., \& {Lique}, F. 2010, \mnras, 406, 2488,
  \dodoi{10.1111/j.1365-2966.2010.16826.x}

\bibitem[{{Dutrey} {et~al.}(1997){Dutrey}, {Guilloteau}, \&
  {Guelin}}]{Dutrey97}
{Dutrey}, A., {Guilloteau}, S., \& {Guelin}, M. 1997, \aap, 317, L55

\bibitem[{{Ebenstein} \& {Muenter}(1984)}]{Ebenstein84}
{Ebenstein}, W.~L., \& {Muenter}, J.~S. 1984, \jcp, 80, 3989,
  \dodoi{10.1063/1.447269}

\bibitem[{{Endres} {et~al.}(2016){Endres}, {Schlemmer}, {Schilke}, {Stutzki},
  \& {M{\"u}ller}}]{Endres16}
{Endres}, C.~P., {Schlemmer}, S., {Schilke}, P., {Stutzki}, J., \&
  {M{\"u}ller}, H. S.~P. 2016, Journal of Molecular Spectroscopy, 327, 95,
  \dodoi{10.1016/j.jms.2016.03.005}

\bibitem[{{Facchini} {et~al.}(2017){Facchini}, {Birnstiel}, {Bruderer}, \& {van
  Dishoeck}}]{Facchini17}
{Facchini}, S., {Birnstiel}, T., {Bruderer}, S., \& {van Dishoeck}, E.~F. 2017,
  \aap, 605, A16, \dodoi{10.1051/0004-6361/201630329}

\bibitem[{{Facchini} {et~al.}(2021){Facchini}, {Teague}, {Bae}, {Benisty},
  {Keppler}, \& {Isella}}]{Facchini21}
{Facchini}, S., {Teague}, R., {Bae}, J., {et~al.} 2021, \aj, 162, 99,
  \dodoi{10.3847/1538-3881/abf0a4}

\bibitem[{{Favre} {et~al.}(2015){Favre}, {Bergin}, {Cleeves}, {Hersant}, {Qi},
  \& {Aikawa}}]{Favre15}
{Favre}, C., {Bergin}, E.~A., {Cleeves}, L.~I., {et~al.} 2015, \apjl, 802, L23,
  \dodoi{10.1088/2041-8205/802/2/L23}

\bibitem[{{Favre} {et~al.}(2019){Favre}, {Fedele}, {Maud}, {Booth}, {Tazzari},
  {Miotello}, {Testi}, {Semenov}, \& {Bruderer}}]{Favre19}
{Favre}, C., {Fedele}, D., {Maud}, L., {et~al.} 2019, \apj, 871, 107,
  \dodoi{10.3847/1538-4357/aaf80c}

\bibitem[{{Feng} {et~al.}(2019){Feng}, {Caselli}, {Wang}, {Lin}, {Beuther}, \&
  {Sipil{\"a}}}]{Feng19}
{Feng}, S., {Caselli}, P., {Wang}, K., {et~al.} 2019, \apj, 883, 202,
  \dodoi{10.3847/1538-4357/ab3a42}

\bibitem[{{Flaherty} {et~al.}(2017){Flaherty}, {Hughes}, {Rose}, {Simon}, {Qi},
  {Andrews}, {K{\'o}sp{\'a}l}, {Wilner}, {Chiang}, {Armitage}, \&
  {Bai}}]{Flaherty17}
{Flaherty}, K.~M., {Hughes}, A.~M., {Rose}, S.~C., {et~al.} 2017, \apj, 843,
  150, \dodoi{10.3847/1538-4357/aa79f9}

\bibitem[{{Foreman-Mackey}(2016)}]{Foreman-Mackey16}
{Foreman-Mackey}, D. 2016, The Journal of Open Source Software, 1, 24,
  \dodoi{10.21105/joss.00024}

\bibitem[{{Foreman-Mackey} {et~al.}(2013){Foreman-Mackey}, {Hogg}, {Lang}, \&
  {Goodman}}]{Foreman-Mackey13}
{Foreman-Mackey}, D., {Hogg}, D.~W., {Lang}, D., \& {Goodman}, J. 2013, \pasp,
  125, 306, \dodoi{10.1086/670067}

\bibitem[{{Fuchs} {et~al.}(2004){Fuchs}, {Bruenken}, {Fuchs}, {Thorwirth},
  {Ahrens}, {Lewen}, {Urban}, {Giesen}, \& {Winnewisser}}]{Fuchs04}
{Fuchs}, U., {Bruenken}, S., {Fuchs}, G.~W., {et~al.} 2004, Zeitschrift
  Naturforschung Teil A, 59, 861, \dodoi{10.1515/zna-2004-1123}

\bibitem[{{Furuya} {et~al.}(2019){Furuya}, {Aikawa}, {Hama}, \&
  {Watanabe}}]{Furuya19}
{Furuya}, K., {Aikawa}, Y., {Hama}, T., \& {Watanabe}, N. 2019, \apj, 882, 172,
  \dodoi{10.3847/1538-4357/ab3790}

\bibitem[{{Gerner} {et~al.}(2015){Gerner}, {Shirley}, {Beuther}, {Semenov},
  {Linz}, {Albertsson}, \& {Henning}}]{Gerner15}
{Gerner}, T., {Shirley}, Y.~L., {Beuther}, H., {et~al.} 2015, \aap, 579, A80,
  \dodoi{10.1051/0004-6361/201423989}

\bibitem[{{Gicquel} {et~al.}(2014){Gicquel}, {Milam}, {Villanueva}, {Remijan},
  {Coulson}, {Chuang}, {Charnley}, {Cordiner}, \& {Kuan}}]{Gicquel14}
{Gicquel}, A., {Milam}, S.~N., {Villanueva}, G.~L., {et~al.} 2014, \apj, 794,
  1, \dodoi{10.1088/0004-637X/794/1/1}

\bibitem[{{Guzm{\'a}n} {et~al.}(2018){Guzm{\'a}n}, {Huang}, {Andrews},
  {Isella}, {P{\'e}rez}, {Carpenter}, {Dullemond}, {Ricci}, {Birnstiel},
  {Zhang}, {Zhu}, {Bai}, {Benisty}, {{\"O}berg}, \& {Wilner}}]{Guzman18}
{Guzm{\'a}n}, V.~V., {Huang}, J., {Andrews}, S.~M., {et~al.} 2018, \apjl, 869,
  L48, \dodoi{10.3847/2041-8213/aaedae}

\bibitem[{{Guzm{\'a}n} {et~al.}(2021){Guzm{\'a}n}, {Bergner}, {Law},
  {{\"O}berg}, {Walsh}, {Cataldi}, {Aikawa}, {Bergin}, {Czekala}, {Huang},
  {Andrews}, {Loomis}, {Zhang}, {Le Gal}, {Alarc{\'o}n}, {Ilee}, {Teague},
  {Cleeves}, {Wilner}, {Long}, {Schwarz}, {Bosman}, {P{\'e}rez}, {M{\'e}nard},
  \& {Liu}}]{Guzman21_MAPS}
{Guzm{\'a}n}, V.~V., {Bergner}, J.~B., {Law}, C.~J., {et~al.} 2021, \apjs, 257,
  6, \dodoi{10.3847/1538-4365/ac1440}

\bibitem[{{H{\'e}brard} {et~al.}(2005){H{\'e}brard}, {Tripp}, {Chayer},
  {Friedman}, {Dupuis}, {Sonnentrucker}, {Williger}, \& {Moos}}]{Hebrard05}
{H{\'e}brard}, G., {Tripp}, T.~M., {Chayer}, P., {et~al.} 2005, \apj, 635,
  1136, \dodoi{10.1086/497574}

\bibitem[{{Henning} \& {Semenov}(2013)}]{Henning13}
{Henning}, T., \& {Semenov}, D. 2013, Chemical Reviews, 113, 9016,
  \dodoi{10.1021/cr400128p}

\bibitem[{{Hily-Blant} {et~al.}(2019){Hily-Blant}, {Magalhaes de Souza},
  {Kastner}, \& {Forveille}}]{Hily-Blant19}
{Hily-Blant}, P., {Magalhaes de Souza}, V., {Kastner}, J., \& {Forveille}, T.
  2019, \aap, 632, L12, \dodoi{10.1051/0004-6361/201936750}

\bibitem[{{H{\"o}gbom}(1974)}]{Hogbom74}
{H{\"o}gbom}, J.~A. 1974, \aaps, 15, 417

\bibitem[{{Huang} \& {{\"O}berg}(2015)}]{Huang15}
{Huang}, J., \& {{\"O}berg}, K.~I. 2015, \apjl, 809, L26,
  \dodoi{10.1088/2041-8205/809/2/L26}

\bibitem[{{Huang} {et~al.}(2016){Huang}, {{\"O}berg}, \& {Andrews}}]{Huang16}
{Huang}, J., {{\"O}berg}, K.~I., \& {Andrews}, S.~M. 2016, \apjl, 823, L18,
  \dodoi{10.3847/2041-8205/823/1/L18}

\bibitem[{{Huang} {et~al.}(2017){Huang}, {{\"O}berg}, {Qi}, {Aikawa},
  {Andrews}, {Furuya}, {Guzm{\'a}n}, {Loomis}, {van Dishoeck}, \&
  {Wilner}}]{Huang17}
{Huang}, J., {{\"O}berg}, K.~I., {Qi}, C., {et~al.} 2017, \apj, 835, 231,
  \dodoi{10.3847/1538-4357/835/2/231}

\bibitem[{{Huang} {et~al.}(2018){Huang}, {Andrews}, {Dullemond}, {Isella},
  {P{\'e}rez}, {Guzm{\'a}n}, {{\"O}berg}, {Zhu}, {Zhang}, {Bai}, {Benisty},
  {Birnstiel}, {Carpenter}, {Hughes}, {Ricci}, {Weaver}, \& {Wilner}}]{Huang18}
{Huang}, J., {Andrews}, S.~M., {Dullemond}, C.~P., {et~al.} 2018, \apjl, 869,
  L42, \dodoi{10.3847/2041-8213/aaf740}

\bibitem[{{Huang} {et~al.}(2020){Huang}, {Andrews}, {Dullemond}, {{\"O}berg},
  {Qi}, {Zhu}, {Birnstiel}, {Carpenter}, {Isella}, {Mac{\'\i}as}, {McClure},
  {P{\'e}rez}, {Teague}, {Wilner}, \& {Zhang}}]{Huang20}
---. 2020, \apj, 891, 48, \dodoi{10.3847/1538-4357/ab711e}

\bibitem[{Hunter(2007)}]{Hunter07}
Hunter, J.~D. 2007, Computing in Science \& Engineering, 9, 90,
  \dodoi{10.1109/MCSE.2007.55}

\bibitem[{{Jorsater} \& {van Moorsel}(1995)}]{Jorsater95}
{Jorsater}, S., \& {van Moorsel}, G.~A. 1995, \aj, 110, 2037,
  \dodoi{10.1086/117668}

\bibitem[{{Koda} {et~al.}(2019){Koda}, {Teuben}, {Sawada}, {Plunkett}, \&
  {Fomalont}}]{Koda19}
{Koda}, J., {Teuben}, P., {Sawada}, T., {Plunkett}, A., \& {Fomalont}, E. 2019,
  \pasp, 131, 054505, \dodoi{10.1088/1538-3873/ab047e}

\bibitem[{{Lapinov}(2006)}]{Lapinov06}
{Lapinov}, A.~V. 2006, in Society of Photo-Optical Instrumentation Engineers
  (SPIE) Conference Series, Vol. 6580, Society of Photo-Optical Instrumentation
  Engineers (SPIE) Conference Series, ed. Y.~N. {Ponomarev}, S.~N.
  {Mikhailenko}, \& L.~N. {Sinitsa}, 658001--658001--12,
  \dodoi{10.1117/12.724761}

\bibitem[{{Law} {et~al.}(2021{\natexlab{a}}){Law}, {Loomis}, {Teague},
  {{\"O}berg}, {Czekala}, {Andrews}, {Huang}, {Aikawa}, {Alarc{\'o}n}, {Bae},
  {Bergin}, {Bergner}, {Boehler}, {Booth}, {Bosman}, {Calahan}, {Cataldi},
  {Cleeves}, {Furuya}, {Guzm{\'a}n}, {Ilee}, {Le Gal}, {Liu}, {Long},
  {M{\'e}nard}, {Nomura}, {Qi}, {Schwarz}, {Sierra}, {Tsukagoshi}, {Yamato},
  {van't Hoff}, {Walsh}, {Wilner}, \& {Zhang}}]{Law21_MAPS_radial_profiles}
{Law}, C.~J., {Loomis}, R.~A., {Teague}, R., {et~al.} 2021{\natexlab{a}},
  \apjs, 257, 3, \dodoi{10.3847/1538-4365/ac1434}

\bibitem[{{Law} {et~al.}(2021{\natexlab{b}}){Law}, {Teague}, {Loomis}, {Bae},
  {{\"O}berg}, {Czekala}, {Andrews}, {Aikawa}, {Alarc{\'o}n}, {Bergin},
  {Bergner}, {Booth}, {Bosman}, {Calahan}, {Cataldi}, {Cleeves}, {Furuya},
  {Guzm{\'a}n}, {Huang}, {Ilee}, {Le Gal}, {Liu}, {Long}, {M{\'e}nard},
  {Nomura}, {P{\'e}rez}, {Qi}, {Schwarz}, {Soto}, {Tsukagoshi}, {Yamato},
  {van't Hoff}, {Walsh}, {Wilner}, \&
  {Zhang}}]{Law21_MAPS_surfaces_vertical_distributions}
{Law}, C.~J., {Teague}, R., {Loomis}, R.~A., {et~al.} 2021{\natexlab{b}},
  \apjs, 257, 4, \dodoi{10.3847/1538-4365/ac1439}

\bibitem[{{Le Gal} {et~al.}(2020){Le Gal}, {{\"O}berg}, {Huang}, {Law},
  {M{\'e}nard}, {Lefloch}, {Vastel}, {Lopez-Sepulcre}, {Favre}, {Bianchi}, \&
  {Ceccarelli}}]{LeGal20}
{Le Gal}, R., {{\"O}berg}, K.~I., {Huang}, J., {et~al.} 2020, \apj, 898, 131,
  \dodoi{10.3847/1538-4357/ab9ebf}

\bibitem[{{Le Gal} {et~al.}(2021){Le Gal}, {{\"O}berg}, {Teague}, {Loomis},
  {Law}, {Walsh}, {Bergin}, {M{\'e}nard}, {Wilner}, {Andrews}, {Aikawa},
  {Booth}, {Cataldi}, {Bergner}, {Bosman}, {Cleeves}, {Czekala}, {Furuya},
  {Guzm{\'a}n}, {Huang}, {Ilee}, {Nomura}, {Qi}, {Schwarz}, {Tsukagoshi},
  {Yamato}, \& {Zhang}}]{LeGal21_MAPS}
{Le Gal}, R., {{\"O}berg}, K.~I., {Teague}, R., {et~al.} 2021, \apjs, 257, 12,
  \dodoi{10.3847/1538-4365/ac2583}

\bibitem[{{Linsky} {et~al.}(2006){Linsky}, {Draine}, {Moos}, {Jenkins}, {Wood},
  {Oliveira}, {Blair}, {Friedman}, {Gry}, {Knauth}, {Kruk}, {Lacour}, {Lehner},
  {Redfield}, {Shull}, {Sonneborn}, \& {Williger}}]{Linsky06}
{Linsky}, J.~L., {Draine}, B.~T., {Moos}, H.~W., {et~al.} 2006, \apj, 647,
  1106, \dodoi{10.1086/505556}

\bibitem[{{Lis} {et~al.}(2019){Lis}, {Bockel{\'e}e-Morvan}, {G{\"u}sten},
  {Biver}, {Stutzki}, {Delorme}, {Dur{\'a}n}, {Wiesemeyer}, \& {Okada}}]{Lis19}
{Lis}, D.~C., {Bockel{\'e}e-Morvan}, D., {G{\"u}sten}, R., {et~al.} 2019, \aap,
  625, L5, \dodoi{10.1051/0004-6361/201935554}

\bibitem[{{Liu} {et~al.}(2019){Liu}, {Dipierro}, {Ragusa}, {Lodato}, {Herczeg},
  {Long}, {Harsono}, {Boehler}, {Menard}, {Johnstone}, {Pascucci}, {Pinilla},
  {Salyk}, {van der Plas}, {Cabrit}, {Fischer}, {Hendler}, {Manara}, {Nisini},
  {Rigliaco}, {Avenhaus}, {Banzatti}, \& {Gully-Santiago}}]{Liu19}
{Liu}, Y., {Dipierro}, G., {Ragusa}, E., {et~al.} 2019, \aap, 622, A75,
  \dodoi{10.1051/0004-6361/201834157}

\bibitem[{{Long} {et~al.}(2018){Long}, {Pinilla}, {Herczeg}, {Harsono},
  {Dipierro}, {Pascucci}, {Hendler}, {Tazzari}, {Ragusa}, {Salyk}, {Edwards},
  {Lodato}, {van de Plas}, {Johnstone}, {Liu}, {Boehler}, {Cabrit}, {Manara},
  {Menard}, {Mulders}, {Nisini}, {Fischer}, {Rigliaco}, {Banzatti}, {Avenhaus},
  \& {Gully-Santiago}}]{Long18}
{Long}, F., {Pinilla}, P., {Herczeg}, G.~J., {et~al.} 2018, \apj, 869, 17,
  \dodoi{10.3847/1538-4357/aae8e1}

\bibitem[{{Loomis} {et~al.}(2018){Loomis}, {{\"O}berg}, {Andrews}, {Walsh},
  {Czekala}, {Huang}, \& {Rosenfeld}}]{Loomis18}
{Loomis}, R.~A., {{\"O}berg}, K.~I., {Andrews}, S.~M., {et~al.} 2018, \aj, 155,
  182, \dodoi{10.3847/1538-3881/aab604}

\bibitem[{{Loomis} {et~al.}(2020){Loomis}, {{\"O}berg}, {Andrews}, {Bergin},
  {Bergner}, {Blake}, {Cleeves}, {Czekala}, {Huang}, {Le Gal}, {M{\'e}nard},
  {Pegues}, {Qi}, {Walsh}, {Williams}, \& {Wilner}}]{Loomis20}
---. 2020, \apj, 893, 101, \dodoi{10.3847/1538-4357/ab7cc8}

\bibitem[{{Mac{\'\i}as} {et~al.}(2018){Mac{\'\i}as}, {Espaillat}, {Ribas},
  {Schwarz}, {Anglada}, {Osorio}, {Carrasco-Gonz{\'a}lez}, {G{\'o}mez}, \&
  {Robinson}}]{Macias18}
{Mac{\'\i}as}, E., {Espaillat}, C.~C., {Ribas}, {\'A}., {et~al.} 2018, \apj,
  865, 37, \dodoi{10.3847/1538-4357/aad811}

\bibitem[{{Maiwald} {et~al.}(2000){Maiwald}, {Lewen}, {Ahrens}, {Beaky},
  {Gendriesch}, {Koroliev}, {Negirev}, {Paveljev}, {Vowinkel}, \&
  {Winnewisser}}]{Maiwald00}
{Maiwald}, F., {Lewen}, F., {Ahrens}, V., {et~al.} 2000, Journal of Molecular
  Spectroscopy, 202, 166, \dodoi{10.1006/jmsp.2000.8118}

\bibitem[{{Maki} {et~al.}(2000){Maki}, {Mellau}, {Klee}, {Winnewisser}, \&
  {Quapp}}]{Maki00}
{Maki}, A.~G., {Mellau}, G.~C., {Klee}, S., {Winnewisser}, M., \& {Quapp}, W.
  2000, Journal of Molecular Spectroscopy, 202, 67,
  \dodoi{10.1006/jmsp.2000.8113}

\bibitem[{{Matr{\`a}} {et~al.}(2017){Matr{\`a}}, {MacGregor}, {Kalas}, {Wyatt},
  {Kennedy}, {Wilner}, {Duchene}, {Hughes}, {Pan}, {Shannon}, {Clampin},
  {Fitzgerald}, {Graham}, {Holland }, {Pani{\'c}}, \& {Su}}]{Matra17}
{Matr{\`a}}, L., {MacGregor}, M.~A., {Kalas}, P., {et~al.} 2017, \apj, 842, 9,
  \dodoi{10.3847/1538-4357/aa71b4}

\bibitem[{{Mawet} {et~al.}(2012){Mawet}, {Absil}, {Montagnier}, {Riaud},
  {Surdej}, {Ducourant}, {Augereau}, {R{\"o}ttinger}, {Girard}, {Krist}, \&
  {Stapelfeldt}}]{Mawet12}
{Mawet}, D., {Absil}, O., {Montagnier}, G., {et~al.} 2012, \aap, 544, A131,
  \dodoi{10.1051/0004-6361/201219662}

\bibitem[{{McMullin} {et~al.}(2007){McMullin}, {Waters}, {Schiebel}, {Young},
  \& {Golap}}]{McMullin07}
{McMullin}, J.~P., {Waters}, B., {Schiebel}, D., {Young}, W., \& {Golap}, K.
  2007, in Astronomical Society of the Pacific Conference Series, Vol. 376,
  Astronomical Data Analysis Software and Systems XVI, ed. R.~A. {Shaw},
  F.~{Hill}, \& D.~J. {Bell}, 127

\bibitem[{{Meier} {et~al.}(1998){Meier}, {Owen}, {Jewitt}, {Matthews}, {Senay},
  {Biver}, {Bockelee-Morvan}, {Crovisier}, \& {Gautier}}]{Meier98}
{Meier}, R., {Owen}, T.~C., {Jewitt}, D.~C., {et~al.} 1998, Science, 279, 1707,
  \dodoi{10.1126/science.279.5357.1707}

\bibitem[{{Milam} {et~al.}(2005){Milam}, {Savage}, {Brewster}, {Ziurys}, \&
  {Wyckoff}}]{Milam05}
{Milam}, S.~N., {Savage}, C., {Brewster}, M.~A., {Ziurys}, L.~M., \& {Wyckoff},
  S. 2005, \apj, 634, 1126, \dodoi{10.1086/497123}

\bibitem[{{Millar} {et~al.}(1989){Millar}, {Bennett}, \& {Herbst}}]{Millar89}
{Millar}, T.~J., {Bennett}, A., \& {Herbst}, E. 1989, \apj, 340, 906,
  \dodoi{10.1086/167444}

\bibitem[{{M{\"o}llmann} {et~al.}(2002){M{\"o}llmann}, {Maki}, {Winnewisser},
  {Winnewisser}, \& {Quapp}}]{Mollmann02}
{M{\"o}llmann}, E., {Maki}, A.~G., {Winnewisser}, M., {Winnewisser}, B.~P., \&
  {Quapp}, W. 2002, Journal of Molecular Spectroscopy, 212, 22,
  \dodoi{10.1006/jmsp.2001.8519}

\bibitem[{{Montesinos} {et~al.}(2009){Montesinos}, {Eiroa}, {Mora}, \&
  {Mer{\'\i}n}}]{Montesinos09}
{Montesinos}, B., {Eiroa}, C., {Mora}, A., \& {Mer{\'\i}n}, B. 2009, \aap, 495,
  901, \dodoi{10.1051/0004-6361:200810623}

\bibitem[{{Mousis} {et~al.}(2000){Mousis}, {Gautier}, {Bockel{\'e}e-Morvan},
  {Robert}, {Dubrulle}, \& {Drouart}}]{Mousis00}
{Mousis}, O., {Gautier}, D., {Bockel{\'e}e-Morvan}, D., {et~al.} 2000, \icarus,
  148, 513, \dodoi{10.1006/icar.2000.6499}

\bibitem[{{M{\"u}ller} {et~al.}(2005){M{\"u}ller}, {Schl{\"o}der}, {Stutzki},
  \& {Winnewisser}}]{Muller05}
{M{\"u}ller}, H. S.~P., {Schl{\"o}der}, F., {Stutzki}, J., \& {Winnewisser}, G.
  2005, Journal of Molecular Structure, 742, 215,
  \dodoi{10.1016/j.molstruc.2005.01.027}

\bibitem[{{M{\"u}ller} {et~al.}(2001){M{\"u}ller}, {Thorwirth}, {Roth}, \&
  {Winnewisser}}]{Muller01}
{M{\"u}ller}, H.~S.~P., {Thorwirth}, S., {Roth}, D.~A., \& {Winnewisser}, G.
  2001, \aap, 370, L49, \dodoi{10.1051/0004-6361:20010367}

\bibitem[{{Noble} {et~al.}(2013){Noble}, {Theule}, {Borget}, {Danger},
  {Chomat}, {Duvernay}, {Mispelaer}, \& {Chiavassa}}]{Noble13}
{Noble}, J.~A., {Theule}, P., {Borget}, F., {et~al.} 2013, \mnras, 428, 3262,
  \dodoi{10.1093/mnras/sts272}

\bibitem[{{Nyman} \& {Yu}(2019)}]{Nyman19}
{Nyman}, G., \& {Yu}, H.-G. 2019, AIP Advances, 9, 095017,
  \dodoi{10.1063/1.5114693}

\bibitem[{{\"O}berg \& Bergin(2021)}]{Oberg21}
{\"O}berg, K.~I., \& Bergin, E.~A. 2021, Physics Reports, 893, 1 ,
  \dodoi{https://doi.org/10.1016/j.physrep.2020.09.004}

\bibitem[{{{\"O}berg} {et~al.}(2015){{\"O}berg}, {Furuya}, {Loomis}, {Aikawa},
  {Andrews}, {Qi}, {van Dishoeck}, \& {Wilner}}]{Oberg15}
{{\"O}berg}, K.~I., {Furuya}, K., {Loomis}, R., {et~al.} 2015, \apj, 810, 112,
  \dodoi{10.1088/0004-637X/810/2/112}

\bibitem[{{{\"O}berg} {et~al.}(2012){{\"O}berg}, {Qi}, {Wilner}, \&
  {Hogerheijde}}]{Oberg12}
{{\"O}berg}, K.~I., {Qi}, C., {Wilner}, D.~J., \& {Hogerheijde}, M.~R. 2012,
  \apj, 749, 162, \dodoi{10.1088/0004-637X/749/2/162}

\bibitem[{{{\"O}berg} {et~al.}(2009){{\"O}berg}, {van Dishoeck}, \&
  {Linnartz}}]{Oberg09}
{{\"O}berg}, K.~I., {van Dishoeck}, E.~F., \& {Linnartz}, H. 2009, \aap, 496,
  281, \dodoi{10.1051/0004-6361/200810207}

\bibitem[{{{\"O}berg} {et~al.}(2011){{\"O}berg}, {Qi}, {Fogel}, {Bergin},
  {Andrews}, {Espaillat}, {Wilner}, {Pascucci}, \& {Kastner}}]{Oberg11_SMA}
{{\"O}berg}, K.~I., {Qi}, C., {Fogel}, J. K.~J., {et~al.} 2011, \apj, 734, 98,
  \dodoi{10.1088/0004-637X/734/2/98}

\bibitem[{{{\"O}berg} {et~al.}(2021{\natexlab{a}}){{\"O}berg}, {Cleeves},
  {Bergner}, {Cavanaro}, {Teague}, {Huang}, {Loomis}, {Bergin}, {Blake},
  {Calahan}, {Cazzoletti}, {Guzm{\'a}n}, {Hogerheijde}, {Kama}, {Terwisscha van
  Scheltinga}, {Qi}, {van Dishoeck}, {Walsh}, \& {Wilner}}]{Oberg21_TWHya}
{{\"O}berg}, K.~I., {Cleeves}, L.~I., {Bergner}, J.~B., {et~al.}
  2021{\natexlab{a}}, \aj, 161, 38, \dodoi{10.3847/1538-3881/abc74d}

\bibitem[{{{\"O}berg} {et~al.}(2021{\natexlab{b}}){{\"O}berg}, {Guzm{\'a}n},
  {Walsh}, {Aikawa}, {Bergin}, {Law}, {Loomis}, {Alarc{\'o}n}, {Andrews},
  {Bae}, {Bergner}, {Boehler}, {Booth}, {Bosman}, {Calahan}, {Cataldi},
  {Cleeves}, {Czekala}, {Furuya}, {Huang}, {Ilee}, {Kurtovic}, {Le Gal}, {Liu},
  {Long}, {M{\'e}nard}, {Nomura}, {P{\'e}rez}, {Qi}, {Schwarz}, {Sierra},
  {Teague}, {Tsukagoshi}, {Yamato}, {van't Hoff}, {Waggoner}, {Wilner}, \&
  {Zhang}}]{Oberg21_MAPS}
{{\"O}berg}, K.~I., {Guzm{\'a}n}, V.~V., {Walsh}, C., {et~al.}
  2021{\natexlab{b}}, \apjs, 257, 1, \dodoi{10.3847/1538-4365/ac1432}

\bibitem[{{O'Brien} {et~al.}(2018){O'Brien}, {Izidoro}, {Jacobson}, {Raymond},
  \& {Rubie}}]{OBrien18}
{O'Brien}, D.~P., {Izidoro}, A., {Jacobson}, S.~A., {Raymond}, S.~N., \&
  {Rubie}, D.~C. 2018, \ssr, 214, 47, \dodoi{10.1007/s11214-018-0475-8}

\bibitem[{{Pagani} {et~al.}(2009){Pagani}, {Daniel}, \& {Dubernet}}]{Pagani09}
{Pagani}, L., {Daniel}, F., \& {Dubernet}, M.~L. 2009, \aap, 494, 719,
  \dodoi{10.1051/0004-6361:200810570}

\bibitem[{{Pagani} {et~al.}(2011){Pagani}, {Roueff}, \& {Lesaffre}}]{Pagani11}
{Pagani}, L., {Roueff}, E., \& {Lesaffre}, P. 2011, \apjl, 739, L35,
  \dodoi{10.1088/2041-8205/739/2/L35}

\bibitem[{{Pegues} {et~al.}(2020){Pegues}, {{\"O}berg}, {Bergner}, {Loomis},
  {Qi}, {Le Gal}, {Cleeves}, {Guzm{\'a}n}, {Huang}, {J{\o}rgensen}, {Andrews},
  {Blake}, {Carpenter}, {Schwarz}, {Williams}, \& {Wilner}}]{Pegues20}
{Pegues}, J., {{\"O}berg}, K.~I., {Bergner}, J.~B., {et~al.} 2020, \apj, 890,
  142, \dodoi{10.3847/1538-4357/ab64d9}

\bibitem[{{Pinte} {et~al.}(2018){Pinte}, {M{\'e}nard}, {Duch{\^e}ne}, {Hill},
  {Dent}, {Woitke}, {Maret}, {van der Plas}, {Hales}, {Kamp}, {Thi}, {de
  Gregorio-Monsalvo}, {Rab}, {Quanz}, {Avenhaus}, {Carmona}, \&
  {Casassus}}]{Pinte18}
{Pinte}, C., {M{\'e}nard}, F., {Duch{\^e}ne}, G., {et~al.} 2018, \aap, 609,
  A47, \dodoi{10.1051/0004-6361/201731377}

\bibitem[{{Pinte} {et~al.}(2020){Pinte}, {Price}, {M{\'e}nard}, {Duch{\^e}ne},
  {Christiaens}, {Andrews}, {Huang}, {Hill}, {van der Plas}, {Perez}, {Isella},
  {Boehler}, {Dent}, {Mentiplay}, \& {Loomis}}]{Pinte20}
{Pinte}, C., {Price}, D.~J., {M{\'e}nard}, F., {et~al.} 2020, \apjl, 890, L9,
  \dodoi{10.3847/2041-8213/ab6dda}

\bibitem[{{Preusser} \& {Maki}(1993)}]{Preusser93}
{Preusser}, J., \& {Maki}, A.~G. 1993, Journal of Molecular Spectroscopy, 162,
  484, \dodoi{10.1006/jmsp.1993.1300}

\bibitem[{{Price-Whelan} {et~al.}(2018){Price-Whelan}, {Sip{\H{o}}cz},
  {G{\"u}nther}, {Lim}, {Crawford}, {Conseil}, {Shupe}, {Craig}, {Dencheva},
  {Ginsburg}, {VanderPlas}, {Bradley}, {P{\'e}rez-Su{\'a}rez}, {de Val-Borro},
  {Paper Contributors}, {Aldcroft}, {Cruz}, {Robitaille}, {Tollerud},
  {Coordination Committee}, {Ardelean}, {Babej}, {Bach}, {Bachetti}, {Bakanov},
  {Bamford}, {Barentsen}, {Barmby}, {Baumbach}, {Berry}, {Biscani}, {Boquien},
  {Bostroem}, {Bouma}, {Brammer}, {Bray}, {Breytenbach}, {Buddelmeijer},
  {Burke}, {Calderone}, {Cano Rodr{\'\i}guez}, {Cara}, {Cardoso}, {Cheedella},
  {Copin}, {Corrales}, {Crichton}, {D{\textquoteright}Avella}, {Deil},
  {Depagne}, {Dietrich}, {Donath}, {Droettboom}, {Earl}, {Erben}, {Fabbro},
  {Ferreira}, {Finethy}, {Fox}, {Garrison}, {Gibbons}, {Goldstein}, {Gommers},
  {Greco}, {Greenfield}, {Groener}, {Grollier}, {Hagen}, {Hirst}, {Homeier},
  {Horton}, {Hosseinzadeh}, {Hu}, {Hunkeler}, {Ivezi{\'c}}, {Jain}, {Jenness},
  {Kanarek}, {Kendrew}, {Kern}, {Kerzendorf}, {Khvalko}, {King}, {Kirkby},
  {Kulkarni}, {Kumar}, {Lee}, {Lenz}, {Littlefair}, {Ma}, {Macleod},
  {Mastropietro}, {McCully}, {Montagnac}, {Morris}, {Mueller}, {Mumford},
  {Muna}, {Murphy}, {Nelson}, {Nguyen}, {Ninan}, {N{\"o}the}, {Ogaz}, {Oh},
  {Parejko}, {Parley}, {Pascual}, {Patil}, {Patil}, {Plunkett}, {Prochaska},
  {Rastogi}, {Reddy Janga}, {Sabater}, {Sakurikar}, {Seifert}, {Sherbert},
  {Sherwood-Taylor}, {Shih}, {Sick}, {Silbiger}, {Singanamalla}, {Singer},
  {Sladen}, {Sooley}, {Sornarajah}, {Streicher}, {Teuben}, {Thomas},
  {Tremblay}, {Turner}, {Terr{\'o}n}, {van Kerkwijk}, {de la Vega}, {Watkins},
  {Weaver}, {Whitmore}, {Woillez}, {Zabalza}, \& {Contributors}}]{astropy18}
{Price-Whelan}, A.~M., {Sip{\H{o}}cz}, B.~M., {G{\"u}nther}, H.~M., {et~al.}
  2018, \aj, 156, 123, \dodoi{10.3847/1538-3881/aabc4f}

\bibitem[{{Qi} {et~al.}(2011){Qi}, {D'Alessio}, {{\"O}berg}, {Wilner},
  {Hughes}, {Andrews}, \& {Ayala}}]{Qi11}
{Qi}, C., {D'Alessio}, P., {{\"O}berg}, K.~I., {et~al.} 2011, \apj, 740, 84,
  \dodoi{10.1088/0004-637X/740/2/84}

\bibitem[{{Qi} {et~al.}(2015){Qi}, {{\"O}berg}, {Andrews}, {Wilner}, {Bergin},
  {Hughes}, {Hogherheijde}, \& {D'Alessio}}]{Qi15}
{Qi}, C., {{\"O}berg}, K.~I., {Andrews}, S.~M., {et~al.} 2015, \apj, 813, 128,
  \dodoi{10.1088/0004-637X/813/2/128}

\bibitem[{{Qi} {et~al.}(2013){Qi}, {{\"O}berg}, {Wilner}, {D'Alessio},
  {Bergin}, {Andrews}, {Blake}, {Hogerheijde}, \& {van Dishoeck}}]{Qi13}
{Qi}, C., {{\"O}berg}, K.~I., {Wilner}, D.~J., {et~al.} 2013, Science, 341,
  630, \dodoi{10.1126/science.1239560}

\bibitem[{{Qi} {et~al.}(2019){Qi}, {{\"O}berg}, {Espaillat}, {Robinson},
  {Andrews}, {Wilner}, {Blake}, {Bergin}, \& {Cleeves}}]{Qi19}
{Qi}, C., {{\"O}berg}, K.~I., {Espaillat}, C.~C., {et~al.} 2019, \apj, 882,
  160, \dodoi{10.3847/1538-4357/ab35d3}

\bibitem[{{Remijan} {et~al.}(2021){Remijan}, {Biggs}, {Cortes}, {Dent}, {Di
  Francesco}, {Fomalont}, {Hales}, {Kameno}, {Mason}, {Philips}, {Saini},
  {Stoehr}, {Vila Vilaro}, \& {Villard}}]{ALMA_THB_Cycle8}
{Remijan}, A., {Biggs}, A., {Cortes}, P., {et~al.} 2021, ALMA Technical
  Handbook, ALMA Doc. 8.5

\bibitem[{{Roberts} \& {Millar}(2000)}]{Roberts00}
{Roberts}, H., \& {Millar}, T.~J. 2000, \aap, 361, 388

\bibitem[{{Roueff} {et~al.}(2013){Roueff}, {Gerin}, {Lis}, {Wootten},
  {Marcelino}, {Cernicharo}, \& {Tercero}}]{Roueff13}
{Roueff}, E., {Gerin}, M., {Lis}, D.~C., {et~al.} 2013, Journal of Physical
  Chemistry A, 117, 9959, \dodoi{10.1021/jp400119a}

\bibitem[{{Rubin} {et~al.}(2020){Rubin}, {Engrand}, {Snodgrass}, {Weissman},
  {Altwegg}, {Busemann}, {Morbidelli}, \& {Mumma}}]{Rubin20}
{Rubin}, M., {Engrand}, C., {Snodgrass}, C., {et~al.} 2020, \ssr, 216, 102,
  \dodoi{10.1007/s11214-020-00718-2}

\bibitem[{{Salinas} {et~al.}(2017){Salinas}, {Hogerheijde}, {Mathews},
  {{\"O}berg}, {Qi}, {Williams}, \& {Wilner}}]{Salinas17}
{Salinas}, V.~N., {Hogerheijde}, M.~R., {Mathews}, G.~S., {et~al.} 2017, \aap,
  606, A125, \dodoi{10.1051/0004-6361/201731223}

\bibitem[{{Salinas} {et~al.}(2018){Salinas}, {Hogerheijde}, {Murillo},
  {Mathews}, {Qi}, {Williams}, \& {Wilner}}]{Salinas18}
{Salinas}, V.~N., {Hogerheijde}, M.~R., {Murillo}, N.~M., {et~al.} 2018, \aap,
  616, A45, \dodoi{10.1051/0004-6361/201731745}

\bibitem[{{Schwarz} {et~al.}(2016){Schwarz}, {Bergin}, {Cleeves}, {Blake},
  {Zhang}, {{\"O}berg}, {van Dishoeck}, \& {Qi}}]{Schwarz16}
{Schwarz}, K.~R., {Bergin}, E.~A., {Cleeves}, L.~I., {et~al.} 2016, \apj, 823,
  91, \dodoi{10.3847/0004-637X/823/2/91}

\bibitem[{{Sierra} {et~al.}(2021){Sierra}, {P{\'e}rez}, {Zhang}, {Law},
  {Guzm{\'a}n}, {Qi}, {Bosman}, {{\"O}berg}, {Andrews}, {Long}, {Teague},
  {Booth}, {Walsh}, {Wilner}, {M{\'e}nard}, {Cataldi}, {Czekala}, {Bae},
  {Huang}, {Bergner}, {Ilee}, {Benisty}, {Le Gal}, {Loomis}, {Tsukagoshi},
  {Liu}, {Yamato}, \& {Aikawa}}]{Sierra21_MAPS}
{Sierra}, A., {P{\'e}rez}, L.~M., {Zhang}, K., {et~al.} 2021, \apjs, 257, 14,
  \dodoi{10.3847/1538-4365/ac1431}

\bibitem[{{Simon} {et~al.}(2000){Simon}, {Dutrey}, \& {Guilloteau}}]{Simon00}
{Simon}, M., {Dutrey}, A., \& {Guilloteau}, S. 2000, \apj, 545, 1034,
  \dodoi{10.1086/317838}

\bibitem[{{Simon} {et~al.}(2019){Simon}, {Guilloteau}, {Beck}, {Chapillon}, {Di
  Folco}, {Dutrey}, {Feiden}, {Grosso}, {Pi{\'e}tu}, {Prato}, \&
  {Schaefer}}]{Simon19}
{Simon}, M., {Guilloteau}, S., {Beck}, T.~L., {et~al.} 2019, \apj, 884, 42,
  \dodoi{10.3847/1538-4357/ab3e3b}

\bibitem[{Teague(2019)}]{Teague19_gofish}
Teague, R. 2019, The Journal of Open Source Software, 4, 1632,
  \dodoi{10.21105/joss.01632}

\bibitem[{{Teague} {et~al.}(2018{\natexlab{a}}){Teague}, {Bae}, {Birnstiel}, \&
  {Bergin}}]{Teague18_AS209}
{Teague}, R., {Bae}, J., {Birnstiel}, T., \& {Bergin}, E.~A.
  2018{\natexlab{a}}, \apj, 868, 113, \dodoi{10.3847/1538-4357/aae836}

\bibitem[{{Teague} \& {Foreman-Mackey}(2018)}]{Teague18_bettermoments}
{Teague}, R., \& {Foreman-Mackey}, D. 2018, Research Notes of the American
  Astronomical Society, 2, 173, \dodoi{10.3847/2515-5172/aae265}

\bibitem[{{Teague} \& {Loomis}(2020)}]{Teague20}
{Teague}, R., \& {Loomis}, R. 2020, \apj, 899, 157,
  \dodoi{10.3847/1538-4357/aba956}

\bibitem[{{Teague} {et~al.}(2016){Teague}, {Guilloteau}, {Semenov}, {Henning},
  {Dutrey}, {Pi{\'e}tu}, {Birnstiel}, {Chapillon}, {Hollenbach}, \&
  {Gorti}}]{Teague16}
{Teague}, R., {Guilloteau}, S., {Semenov}, D., {et~al.} 2016, \aap, 592, A49,
  \dodoi{10.1051/0004-6361/201628550}

\bibitem[{{Teague} {et~al.}(2018{\natexlab{b}}){Teague}, {Henning},
  {Guilloteau}, {Bergin}, {Semenov}, {Dutrey}, {Flock}, {Gorti}, \&
  {Birnstiel}}]{Teague18_CS}
{Teague}, R., {Henning}, T., {Guilloteau}, S., {et~al.} 2018{\natexlab{b}},
  \apj, 864, 133, \dodoi{10.3847/1538-4357/aad80e}

\bibitem[{{Teague} {et~al.}(2021){Teague}, {Bae}, {Aikawa}, {Andrews},
  {Bergin}, {Bergner}, {Boehler}, {Booth}, {Bosman}, {Cataldi}, {Czekala},
  {Guzm{\'a}n}, {Huang}, {Ilee}, {Law}, {Le Gal}, {Long}, {Loomis},
  {M{\'e}nard}, {{\"O}berg}, {P{\'e}rez}, {Schwarz}, {Sierra}, {Walsh},
  {Wilner}, {Yamato}, \& {Zhang}}]{Teague21_MAPS}
{Teague}, R., {Bae}, J., {Aikawa}, Y., {et~al.} 2021, \apjs, 257, 18,
  \dodoi{10.3847/1538-4365/ac1438}

\bibitem[{{Thorwirth} {et~al.}(2003){Thorwirth}, {M{\"u}ller}, {Lewen},
  {Br{\"u}nken}, {Ahrens}, \& {Winnewisser}}]{Thorwirth03}
{Thorwirth}, S., {M{\"u}ller}, H.~S.~P., {Lewen}, F., {et~al.} 2003, \apjl,
  585, L163, \dodoi{10.1086/374327}

\bibitem[{{Turner}(2001)}]{Turner01}
{Turner}, B.~E. 2001, \apjs, 136, 579, \dodoi{10.1086/322536}

\bibitem[{{van der Walt} {et~al.}(2011){van der Walt}, {Colbert}, \&
  {Varoquaux}}]{vanderWalt11}
{van der Walt}, S., {Colbert}, S.~C., \& {Varoquaux}, G. 2011, Computing in
  Science and Engineering, 13, 22, \dodoi{10.1109/MCSE.2011.37}

\bibitem[{{van 't Hoff} {et~al.}(2017){van 't Hoff}, {Walsh}, {Kama},
  {Facchini}, \& {van Dishoeck}}]{vantHoff17}
{van 't Hoff}, M.~L.~R., {Walsh}, C., {Kama}, M., {Facchini}, S., \& {van
  Dishoeck}, E.~F. 2017, \aap, 599, A101, \dodoi{10.1051/0004-6361/201629452}

\bibitem[{{Verhoeve} {et~al.}(1990){Verhoeve}, {Zwart}, {Versluis}, {Drabbels},
  {ter Meulen}, {Meerts}, {Dymanus}, \& {McLay}}]{Verhoeve90}
{Verhoeve}, P., {Zwart}, E., {Versluis}, M., {et~al.} 1990, Review of
  Scientific Instruments, 61, 1612, \dodoi{10.1063/1.1141123}

\bibitem[{{Virtanen} {et~al.}(2020){Virtanen}, {Gommers}, {Oliphant},
  {Haberland}, {Reddy}, {Cournapeau}, {Burovski}, {Peterson}, {Weckesser},
  {Bright}, {van der Walt}, {Brett}, {Wilson}, {Millman}, {Mayorov}, {Nelson},
  {Jones}, {Kern}, {Larson}, {Carey}, {Polat}, {Feng}, {Moore}, {VanderPlas},
  {Laxalde}, {Perktold}, {Cimrman}, {Henriksen}, {Quintero}, {Harris},
  {Archibald}, {Ribeiro}, {Pedregosa}, {van Mulbregt}, \& {SciPy 1. 0
  Contributors}}]{Virtanen20}
{Virtanen}, P., {Gommers}, R., {Oliphant}, T.~E., {et~al.} 2020, Nature
  Methods, 17, 261, \dodoi{10.1038/s41592-019-0686-2}

\bibitem[{{Walsh} {et~al.}(2012){Walsh}, {Nomura}, {Millar}, \&
  {Aikawa}}]{Walsh12}
{Walsh}, C., {Nomura}, H., {Millar}, T.~J., \& {Aikawa}, Y. 2012, \apj, 747,
  114, \dodoi{10.1088/0004-637X/747/2/114}

\bibitem[{{Walter} \& {Brinks}(1999)}]{Walter99}
{Walter}, F., \& {Brinks}, E. 1999, \aj, 118, 273, \dodoi{10.1086/300906}

\bibitem[{{Walter} {et~al.}(2008){Walter}, {Brinks}, {de Blok}, {Bigiel},
  {Kennicutt}, {Thornley}, \& {Leroy}}]{Walter08}
{Walter}, F., {Brinks}, E., {de Blok}, W.~J.~G., {et~al.} 2008, \aj, 136, 2563,
  \dodoi{10.1088/0004-6256/136/6/2563}

\bibitem[{{Weaver} {et~al.}(2018){Weaver}, {Isella}, \& {Boehler}}]{Weaver18}
{Weaver}, E., {Isella}, A., \& {Boehler}, Y. 2018, \apj, 853, 113,
  \dodoi{10.3847/1538-4357/aaa481}

\bibitem[{{Willacy}(2007)}]{Willacy07}
{Willacy}, K. 2007, \apj, 660, 441, \dodoi{10.1086/512796}

\bibitem[{{Willacy} \& {Langer}(2000)}]{Willacy00}
{Willacy}, K., \& {Langer}, W.~D. 2000, \apj, 544, 903, \dodoi{10.1086/317236}

\bibitem[{{Willacy} \& {Woods}(2009)}]{Willacy09}
{Willacy}, K., \& {Woods}, P.~M. 2009, \apj, 703, 479,
  \dodoi{10.1088/0004-637X/703/1/479}

\bibitem[{{Willacy} {et~al.}(2015){Willacy}, {Alexander}, {Ali-Dib},
  {Ceccarelli}, {Charnley}, {Doronin}, {Ellinger}, {Gast}, {Gibb}, {Milam},
  {Mousis}, {Pauzat}, {Tornow}, {Wirstr{\"o}m}, \& {Zicler}}]{Willacy15}
{Willacy}, K., {Alexander}, C., {Ali-Dib}, M., {et~al.} 2015, \ssr, 197, 151,
  \dodoi{10.1007/s11214-015-0167-6}

\bibitem[{{Winnewisser} \& {Vogt}(1978)}]{Winnewisser78}
{Winnewisser}, M., \& {Vogt}, J. 1978, Zeitschrift Naturforschung Teil A, 33,
  1323, \dodoi{10.1515/zna-1978-1110}

\bibitem[{{Yen} {et~al.}(2016){Yen}, {Koch}, {Liu}, {Puspitaningrum}, {Hirano},
  {Lee}, \& {Takakuwa}}]{Yen16}
{Yen}, H.-W., {Koch}, P.~M., {Liu}, H.~B., {et~al.} 2016, \apj, 832, 204,
  \dodoi{10.3847/0004-637X/832/2/204}

\bibitem[{{Zhang} {et~al.}(2021){Zhang}, {Booth}, {Law}, {Bosman}, {Schwarz},
  {Bergin}, {{\"O}berg}, {Andrews}, {Guzm{\'a}n}, {Walsh}, {Qi}, {van't Hoff},
  {Long}, {Wilner}, {Huang}, {Czekala}, {Ilee}, {Cataldi}, {Bergner}, {Aikawa},
  {Teague}, {Bae}, {Loomis}, {Calahan}, {Alarc{\'o}n}, {M{\'e}nard}, {Le Gal},
  {Sierra}, {Yamato}, {Nomura}, {Tsukagoshi}, {P{\'e}rez}, {Trapman}, {Liu}, \&
  {Furuya}}]{Zhang21_MAPS}
{Zhang}, K., {Booth}, A.~S., {Law}, C.~J., {et~al.} 2021, \apjs, 257, 5,
  \dodoi{10.3847/1538-4365/ac1580}

\bibitem[{{Zhu} {et~al.}(2019){Zhu}, {Zhang}, {Jiang}, {Kataoka}, {Birnstiel},
  {Dullemond}, {Andrews}, {Huang}, {P{\'e}rez}, {Carpenter}, {Bai}, {Wilner},
  \& {Ricci}}]{zhu19}
{Zhu}, Z., {Zhang}, S., {Jiang}, Y.-F., {et~al.} 2019, \apjl, 877, L18,
  \dodoi{10.3847/2041-8213/ab1f8c}

\end{thebibliography}
\bibliographystyle{aasjournal}



\end{document}